%% file: arXiv1.tex
\documentclass[aps,prl,reprint,superscriptaddress]{revtex4-2}

\usepackage{graphicx}
\usepackage{bm}
\usepackage{float}
\usepackage{amsmath}
\usepackage{amssymb}
\usepackage[dvipsnames]{xcolor}
\usepackage{hyperref}

\begin{document}
\title{Hubbard models with ultracold shielded molecules in optical lattices}
\author{Kevin P{\'e}rez}
\affiliation{Institute for Molecules and Materials, Radboud University, 6525 AJ Nijmegen, The Netherlands}
\author{Joseph W. Desroches}
\affiliation{Department of Physics and Astronomy, Rice University, Houston, Texas 77005-1892, USA}
\author{Kaden R. A. Hazzard}
\affiliation{Department of Physics and Astronomy, Rice University, Houston, Texas 77005-1892, USA}
\affiliation{Smalley-Curl Institute, Rice University, Houston, Texas 77005-1892, USA}
\author{Tijs Karman}
\email{t.karman@science.ru.nl}
\affiliation{Institute for Molecules and Materials, Radboud University, 6525 AJ Nijmegen, The Netherlands}
\date{\today}

\begin{abstract}
We show that, in appropriate regimes, ultracold collisionally shielded molecules in an optical lattice realize a controlled extended Hubbard model, despite their large-radius hard-core repulsion. We compute the Hubbard parameters and show    lattice depth and microwave-induced dipolar interactions provide flexible, independent control of   on- and off-site interactions. This makes shielded molecules a powerful platform for  extended Hubbard models and enables an  \textit{interaction microscope} that can   distinguish  inter-hole interactions  in doped Mott insulators --  crucial to  superconductivity and strange metallicity -- that  are indistinguishable with quantum gas microscopy alone.
\end{abstract}

\maketitle

Ultracold atoms in optical lattices are a powerful platform for quantum simulation of the Hubbard model,
a paradigmatic model for strongly correlated electrons in condensed matter \cite{bloch2005ultracold,bloch2012quantum,Bloch2008,gross2017quantum,BOHRDT2021168651}.
The optical lattice can realize diverse geometries, control tunneling rates, and enable site-resolved imaging,
while on-site interactions are flexibly tuned by magnetic field using Feshbach resonances or lattice depth. 
Rydberg atoms \cite{weckesser2025realization}, magnetic atoms \cite{baier2016extended,su2023dipolar}, and polar molecules in optical lattices can realize extended Hubbard models with off-site interactions realized by their long-range interactions.
Molecules' long coherence times and strong interactions, as well as the rapid development of molecule quantum gas microscopy \cite{rosenberg2022observation,christakis2023probing,mortlock2025multi,cornish2024quantum}, make them an excellent platform to study extended Hubbard models that augment the usual Hubbard model with off-site  interactions~\cite{Hirsch1984ExtendedHubbard,Zhang1989ExtendedHubbard,Huang2014ExtendedDMFT,barnett:quantum_2006,Dutta2015NonstandardHubbard,Trefzger2011DipolarLattices,CapogrossoSansone2010PolarMolecules,Sengupta2005Supersolids,gorshkov:quantum_2011,gorshkov:tunable_2011,carroll:observation_2025}. 

While two molecules in a molecular Hubbard simulator undergo reactive loss or loss from sticky collisions \cite{mayle2012statistical,christianen2019photoinduced} when they tunnel into the same site,
collisional shielding can essentially eliminate these losses \cite{yuan2025extreme}.
Shielding has enabled evaporative cooling to quantum degenerate gases of fermionic ~\cite{valtolina2020dipolar,li2021tuning,schindewolf2022evaporation} and bosonic molecules~\cite{bigagli2024observation, shi2025bose},
and dense but stable quantum droplets~\cite{Zhang_2026}.
There exist various schemes for shielding \cite{gorshkov2008suppression,gonzalez2017adimensional,augustovivcova2019ultracold,karam2023two,karman2018microwave,Lassabliere2018,karman2025double}.
We focus on double microwave shielding \cite{karman2025double},
which uses 
$\sigma^+$ circular and $\pi$ linear polarization microwaves.
Interaction between shielded molecules can be thought of as a repulsive shielding core,
plus long-range dipole-dipole interactions that can be flexibly tuned in magnitude and sign, or even turned off.

\begin{figure}
    \centering
    \includegraphics[]{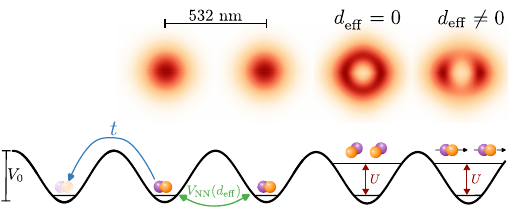}
    \caption{{\bf Hubbard model with shielded molecules}. Molecules can tunnel between lattice sites with hopping amplitude $t$  and interact on-site with strength $U$. The induced dipole moment $d_\mathrm{eff}$ gives rise to dipolar off-site interactions, with nearest-neighbor strength $V_{\mathrm{NN}}$. The wave functions shown illustrate how repulsive shielding and dipolar interactions deform the two-molecule wave function on length scales comparable to the lattice spacing.
    } 
    \label{fig:cartoon}
\end{figure}

Because the repulsive shielding interactions
are strong and
have a range on the scale of a hundred nanometers, which is
comparable to 
optical lattice spacings,
standard methods to compute Hubbard parameters  for atoms are not necessarily accurate \cite{jaksch1998cold, buchler2010microscopic}.  
Zero-range approximations replacing the microscopic molecule-molecule by the Fermi pseudopotential, typically made for ultracold atoms, need not be valid for shielded molecules.
Similarly, the even-longer-ranged dipolar interaction acts on the micron scale,
and its contribution to the on-site interaction cannot be treated perturbatively,
as is usually done for magnetic atoms~\cite{baier2016extended}.
Beyond this technical challenge of how to meaningfully compute Hubbard parameters,
the strong on-site repulsion raises a more fundamental question about 
whether shielded molecules can fully quantum simulate an extended Hubbard model. 
The strong long-ranged repulsive interactions
naturally suggest that shielded molecules effectively realize hard-core interactions,
completely inhibiting double occupancy.
Therefore the outlook has rather pessimistically been that shielded molecules can  operate only deeply in the Mott insulator regime, with effectively no tunability of the interactions.

In this Letter, we non-perturbatively calculate on-site interactions between shielded ultracold  molecules in a 3D optical lattice, as illustrated in Fig.~\ref{fig:cartoon}. These can be bosonic or  fermions (necessarily with multiple components). We find that 
by simultaneously tuning the lattice depth and dipolar interaction,
one can flexibly and independently tune on-site and off-site interactions,
between roughly $\pm 10$ times the tunneling.
Due to the strong repulsion this is realized in shallow lattices where the absolute tunneling rate is large,
an advantage for 
reaching the lowest, most interesting temperatures in strongly interacting Hubbard models 
a key requirement for  state-of-the-art quantum simulation \cite{xu2025neutral}.
Furthermore, we show that the 
tunability of the off-site interaction can be used as an \emph{interaction microscope,}
a novel diagnostic of effective hole-hole (or other dopant) interactions in the doped Hubbard model.

We consider  ultracold molecules loaded in a three-dimensional cubic optical lattice with lattice spacing $a=\lambda/2=532$~nm, where $\lambda$ is the optical wavelength.
Explicitly, the lattice potential is given by the sum of $V_0\sin^2(k\rho)$ for $\rho=x,y,z$.
The lattice depth $V_0$ is experimentally adjusted by the laser intensity and this controls the tunneling $t$, which is a universal function of $V_0/E_r$ that is identical to the well-known atomic case \cite{jaksch1998cold},
where $E_r = \hbar^2 k^2/2m$ is the recoil energy.
The molecules experience on-site interactions $U$ and off-site nearest  interactions, whose strength for nearest-neighbors displaced along the quantization axis we denote  $V_{\mathrm{NN}}$. Under the usual conditions -- a deep enough lattice for tunneling to be described by a tight-binding model, interaction range small compared to the lattice spacing, and interaction energy small compared to the bandgap -- the system is described by an extended Hubbard model. Writing this explicitly for fermions, 
\begin{align}
{\hat H} &= -t \sum_{\langle i,j \rangle, \sigma} \left( {\hat c}_{i \sigma}^\dagger {\hat c}_{j \sigma}^{\phantom{\dagger}} + \mathrm{h.c.} \right) + \frac{U}{2} \sum_{i,\sigma, \tau} {\hat n}_{i \sigma} {\hat n}_{i \tau} \nonumber \\
&\hspace{0.25in}{}+ \sum_{i\ne j;\sigma\tau} \frac{V_{ij}}{2} n_{i\sigma} n_{j\tau},
\end{align}
where $c_{i\sigma}^{\phantom\dagger}$ ($c_{i\sigma}^{\dagger}$) is a fermionic annihilation (creation) operator for site $i$ and nuclear spin $\sigma$~\footnote{We do not include the -- possibly non-degenerate -- single-molecule nuclear spin state energies here, as these can be eliminated by moving to the rotating frame due to the SU($N$) symmetry of the model~\cite{Mukherjee:SUN_2025}. }, $t$ is the tunneling, and to a good approximation 
$V_{ij}=V_\mathrm{NN} (a/r_{ij})^3 (3\cos^2\theta_{ij}-1)/2$
where $r_{ij}$ is the distance from site $i$ to site $j$, $\theta_{ij}$ is the angle between the line connecting $i$ to $j$ and the quantization axis.
Small deviations from this dependence occur due to the non-point-like nature of the Wannier functions (percent level), as well as from the extension of the shielding core across lattice sites    (small for bialkalis).
The bosonic case can be written similarly. We note that this model enjoys an SU($N$) symmetry that opens up a wealth of interesting physics~\cite{Mukherjee:SUN_2025,mukherjee:SUN_alkali_2025,stepp:trion_2026}, some of which may not be obtained with SU($N$) systems of alkaline-earth-like atoms~\cite{wu:exact_2003,Cazalilla_2009,gorshkov2010two,cazalilla2014ultracold,ibarra2025many}. 
Computing effective $U$ and $V_{\mathrm{NN}}$ that result from the microscopic shielding and dipolar molecule-molecule interactions, and their dependence on $V_0$ and $d_\mathrm{eff}$, is the goal of the paper.

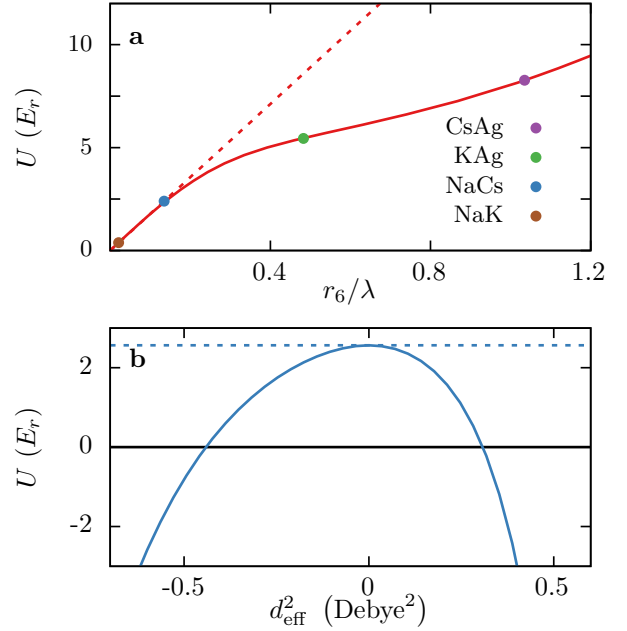
\begin{figure}
    \input{u_r6_3d}
        \caption{{\bf On-site interaction energy at lattice depth $V_0=6E_R$}.
        \textbf{a} as a function of the shielding core repulsion for compensated dipolar interactions ($d_{\text{eff}}=0$).
Markers indicate representative molecular species considered in this work. 
        \textbf{b} On-site interaction as a function of the squared effective dipole moment $d_{\mathrm{eff}}^{2}$ for NaCs molecules. 
        Perturbative results are shown as dashed lines.
        }
    \label{fig:scaling}
\end{figure}

The microscopic interaction between shielded molecules consists of a repulsive shielding core plus long-range dipolar interactions.
The dipolar interaction is given by $d_\mathrm{eff}^2 (1-3\cos^2\theta) / 4\pi \epsilon_0 r^3$,
where $r$ is the distance between the molecules and $\theta$ the angle between intermolecular axis and the microwave quantization axis.
The microwave fields control the effective dipole moment $d_\mathrm{eff}$,
enabling tuning between dipolar ($d^2_\mathrm{eff}>0$), anti-dipolar ($d^2_\mathrm{eff}<0$), and compensated ($d_\mathrm{eff}=0$) interactions.
The shielding core can be accurately modeled as a $+C_6 r^{-6}$ van der Waals repulsion that results from the microwave shielding \cite{deng2023effective}.
The repulsive shielding core has a long range, with characteristic size $r_6 \equiv (mC_6/\hbar^2)^{1/4}$. For NaCs with compensated dipolar interactions, $r_6=160$~nm is not small compared to the $\lambda/2 \approx 532$~nm lattice spacing in a typical optical lattice.
The range of dipolar interactions can be on the micron scale,
and hence be appreciable over several sites of a typical optical lattice.
Due to the universality of microwave shielding \cite{dutta2025universality} both length scales are molecule dependent and scale with $r_d = m d^2 / 4\pi\epsilon_0\hbar^2$. For  relatively less polar and lighter NaK, $r_6=22$~nm, while both length scales 
may  be substantially enhanced for ``ultrapolar'' molecules of assembled alkali-metal-coinage-metal atoms~\cite{Smialkowski}, 
to 510~nm for KAg and up to 1100~nm for CsAg.

\begin{figure*}
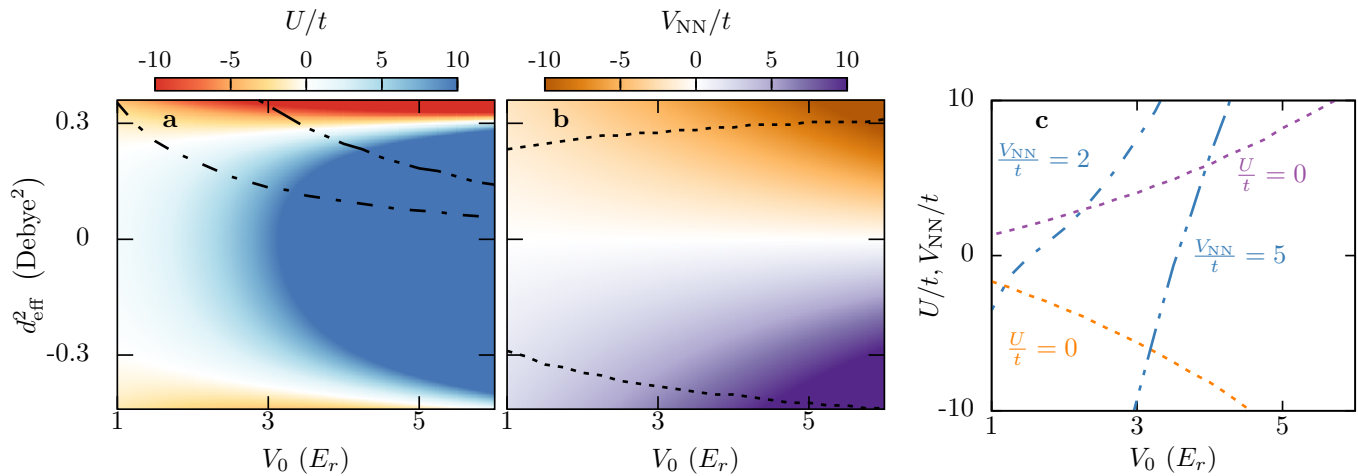

    \centering
     \include{heatmaps}
    \caption{{\bf Hubbard parameters for NaCs molecules} controlled by the effective squared dipole moment $d_{\mathrm{eff}}^2$ and lattice depth $V_{0}$. (a)  $U/t$. Dash-dotted and dash--double-dotted contours indicate constant $V_{\mathrm{NN}}/t = 2$ and $V_{\mathrm{NN}}/t = 5$, respectively. (b)  $V_{\mathrm{NN}}/t$. Dashed curves indicate $U/t = 0$. (c) One-dimensional cuts along the  trajectories indicated in (a) and (b), i.e.\ $U/t$ for fixed-$V_{\mathrm{NN}}/t$ (blue) and $V_{\mathrm{NN}}/t$ along the $U/t = 0$ trajectories (orange and purple). 
    }
    \label{Fig:heatmaps}
\end{figure*}

To calculate the on-site interaction  non-perturbatively, we  numerically solve the Schr\"odinger equation for two molecules in the optical lattice using a limited basis of \emph{contractions} that are localized on a single site.
These contractions are determined by first solving two molecules in a harmonic trap.
In a truncated basis of these reference solutions, we then solve the full problem in the actual lattice potential including the resulting coupling between center-of-mass and relative coordinates.
The on-site interaction $U$ is computed as the difference in ground-state energy between calculations for interacting and non-interacting molecules.
We also compute off-site interactions $V_{\mathrm{NN}}$ between molecules in nearest neighbor sites as matrix elements in the Wannier basis,
and give estimates of density-assisted tunneling.
The on-site interaction here is computed differently from  typical methods for ultracold atoms,
where one replaces the microscopic interaction by a Fermi pseudo-potential $g\delta(\bm{r})$ and treats this perturbatively
in the Wannier basis, $U=g\int |w_0(\bm{r})|^4~d\bm{r}$.
The philosophy of our calculation is that the localized basis leads to localized approximations of eigenstates, and hence are close to the Wannier description,
while our approach can still be applied where the interaction is non-perturbative.
In the Supplemental Material (SM), we extensively validate this and show that with suitable contractions one recovers the usual results calculated in the Wannier basis in the special cases where the Wannier approach is  applicable~\cite{supplement}.

Figure~\ref{fig:scaling}\textbf{a} shows the on-site interaction energy for compensated dipolar interactions ($d_\mathrm{eff}=0$) as a function of the repulsive shielding interactions' $r_6$,
for a fixed lattice depth $V_0/E_r = 6.$
For $d_\mathrm{eff}=0$  the interaction in recoil units is a universal function of $r_6/\lambda$ and $V_0/E_r$.
The markers indicate parameters corresponding to typical ultracold molecules NaK~\cite{park2015ultracold} and NaCs~\cite{Cairncross2021, Stevenson2023} that have been realized experimentally, and ultrapolar KAg and CsAg that are actively pursued~\cite{Vayninger2025}.
With the exception of NaK, $r_6$ exceeds the harmonic oscillator length,
and the perturbative pseudo-potential description is invalid, see the SM~\cite{supplement}.

Figure~\ref{fig:scaling}\textbf{b} shows how the on-site interaction $U$ for NaCs depends on the induced dipolar interactions, proportional to $d_\mathrm{eff}^2$.
The dipolar interaction  non-perturbatively modifies the on-site interaction already for a very small induced dipole moment.
An induced dipole moment of $d_\mathrm{eff}\approx0.5$~Debye -- a small fraction of the limiting 4.7~Debye dipole -- can compensate the strongly repulsive on-site interaction and tune $U$ close to zero,
for example over a range where $|U| \approx t$.
Tuning of the on-site interaction to these small values using the induced dipole moment is  a powerful feature that enables accessing regimes beyond the  hard-core, infinitely repulsive on-site interactions, and effectively reinstating quantum fluctuations into doubly occupied configurations.

Next we show that one may independently tune  the on-site $U$ and off-site $V_{\mathrm{NN}}$ interactions, relative to tunneling $t$, by controlling the lattice depth $V_0$ and the induced dipolar interaction, proportional to $d_{\mathrm{eff}}^2$.
Figure~\ref{Fig:heatmaps}\textbf{a} shows $U/t$ with superimposed paths along which the off-site interaction is fixed at $V_{\mathrm{NN}}/t = 2$ and $5$, demonstrating that $U/t$ may be tuned at fixed $V_{\mathrm{NN}}/t$.
Figure~\ref{Fig:heatmaps}\textbf{b} shows $V_{\mathrm{NN}}/t$ where the superimposed curves mark $U/t = 0$.
Figure~\ref{Fig:heatmaps}\textbf{c} shows line plots of the interactions along these paths,
demonstrating that $U/t$ can be flexibly tuned from weakly to strongly interacting at fixed $V_{\mathrm{NN}}/t$ paths and vice versa.
This establishes comprehensive control of on and off-site interactions. Excitingly, this control is possible at shallow lattices where $t$ is large, advantageous experimentally for accessing low temperatures $T/t$. 

\begin{figure}
    \centering
    \includegraphics[width=\linewidth]{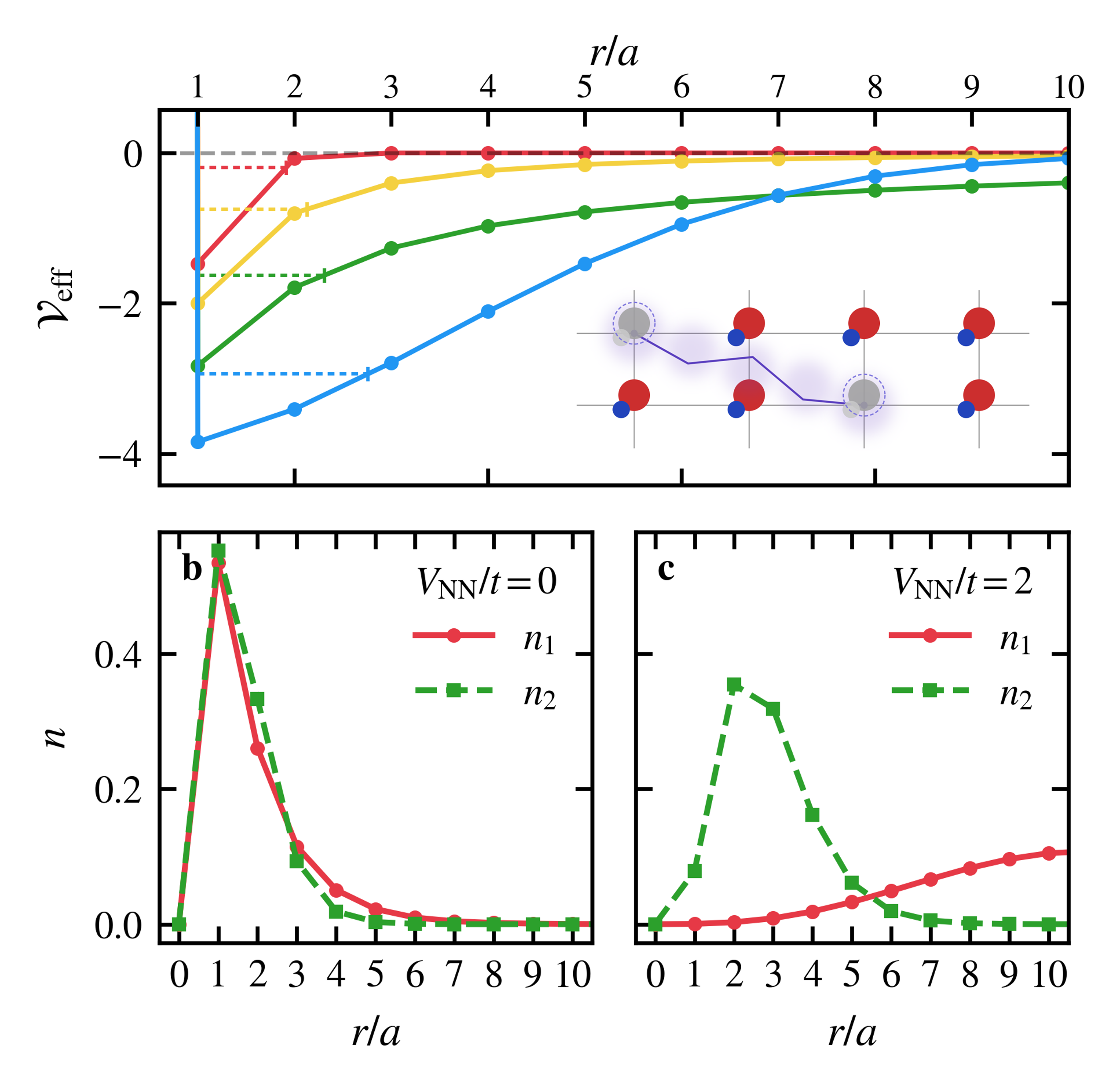}
    \caption{\textbf{An interaction microscope: using dipolar interactions to distinguish inter-hole interactions.} \textbf{(a)} Four toy effective 1D interaction potentials potentials between holes in a doped Hubbard model (inset) that are essentially indistinguishable via snapshots of hole locations in a quantum gas microscope. \textbf{(b, c)} Probability to find two holes at separation $r$ in the ground state, for the red ($n_1$) and green ($n_2$) potentials in  (a).  Results (b) without ($V_{\mathrm{NN}}/t = 0$) and (c) with ($V_{\mathrm{NN}}/t = 2$) dipolar interactions, demonstrating that the tunable $V_{\text{NN}}$ allows one to discriminate otherwise indistinguishable interactions. Results are for a 20-site lattice and potential coefficients $\mathcal{W}_\mathrm{eff}/t = 4$ (see the SM~\cite{supplement}).}
    \label{fig:interaction_microscope}
\end{figure}
 
In addition to flexibility for quantum simulations, the independent tunability of on- and off-site interactions enables an \textit{interaction microscope} -- a probe of the effective inter-dopant interactions, which are fundamental for understanding doped Mott insulators and the myriad associated phenomena in them, such as pseudogaps, strange metals, stripes, and, most notably, high-temperature superconductivity \cite{keimerQuantumMatterHightemperature2015}. Notable proposed inter-hole interaction mechanisms and associated potentials (or more generally interaction vertices)  include geometric string theory and parton constructions \cite{bohrdtDichotomyHeavyLight2023, grusdtMesonFormationMixeddimensional2018, grusdtMicroscopicSpinonchargonTheory2019, grusdtPairingHolesConfining2023, grusdtPartonTheoryMagnetic2018, alamProgrammableDigitalQuantum2025, alamFermionicDynamicsTrappedion2025, homeierFeshbachHypothesisHighTc2025}, the spin-bag mechanism \cite{schriefferDynamicSpinFluctuations1989, kampfPseudogapsSpinbagApproach1990, schriefferSpinbagMechanismHightemperature1988}, and magnon-exchange mediated pairing \cite{scalapino_d-wave_1986, monthoux_spin-fluctuation-induced_1992}. 

To demonstrate the principle of this interaction microscope, we consider the simplified problem of two holes doped into a Mott insulator (inset in Fig. \ref{fig:interaction_microscope}\textbf{a}) and show that adding $V_{\text{NN}}$ allows experiments to discriminate inter-dopant interactions that would be otherwise  indistinguishable using density and correlation measurements in quantum gas microscopes. For illustrative purposes, we take a simplified model where the holes interact via an effective one-dimensional potential  $\mathcal{V}_{\mathrm{eff}}(r)$ that depends only on their  separation $r$. All the potentials in Fig.~\ref{fig:interaction_microscope}\textbf{(a)} produce ground state  probability distributions of the distance between holes, $n(r)$, that are highly similar and therefore  difficult to distinguish experimentally, as illustrated in Fig.~\ref{fig:interaction_microscope}\textbf{b} for the red and green potentials of Fig~\ref{fig:interaction_microscope}\textbf{a}. However, Fig.~\ref{fig:interaction_microscope}\textbf{(c)} shows that an easily accessible off-site interaction strength ($V_{\text{NN}}/t=2$) modifies the $n(r)$ so that the two potentials are easily distinguished. More quantitatively the maximum difference in the two $n(r)$ increases from 0.07 to  0.4 after adding dipolar interactions, see the SM~\cite{supplement}. 
The mechanism of this distinguishability for the shown pair of potentials  is  that despite  the similar binding length and inter-hole wavefunctions, the binding energies are quite different and therefore adding $V_{\text{NN}}$  causes the states   of one potential to unbind while the other stays strongly bound. Even if the binding energies were similar, potentials are still generally distinguishable because the bound state wavefunction's characteristic size changes differently with $V_{\text{NN}}$ for different potential shapes.

This ability of the off-site interactions to distinguish potentials at reasonable $V_{\text{NN}}$ is not an artifact of the two potentials compared in Fig.~\ref{fig:interaction_microscope}\textbf{(b--c)}, but is broadly true for different potentials. In the SM, we show that this holds for all four potentials illustrated in Fig.~\ref{fig:interaction_microscope}\textbf{(a)}, and that furthermore the interaction microscope's effectiveness survives to experimentally relevant temperatures for Hubbard model quantum simulation $(T/t \lesssim 1)$~\cite{supplement}. The SM also more generally quantifies the distinguishability of the $n(r)$ via the $\ell^2$ norm for these potentials and how this evolves with $V_{\text{NN}}$ and $T$. 
Finally, we note that our interaction microscope is not restricted to hole-hole interactions, but can also be applied to other inter-dopant interactions hole-doublon and doublon-doublon interactions. 
 
We have studied the tunable extended Hubbard model realized with collisionally shielded ultracold molecules in an optical lattice.
The strong, long-range interactions between typical molecules cannot be described by a Fermi pseudo-potential, and the on-site interactions Hubbard $U$ are not perturbative.
Nevertheless, shielded molecules do not necessarily realize hard-core interactions that forbid double occupancy. In fact, we find that
by simultaneously tuning the lattice depth and dipolar interaction in double microwave shielding\cite{yuan2025extreme,karman2025double} we can flexibly and independently tune on-site and off-site interactions,  between, say, $\pm 10$ times the tunneling.
Due to the strong repulsion this is realized in shallow lattices where the absolute tunneling rate is high, creating advantages for simulating strongly interacting Hubbard physics at low $T/t$, a key bottleneck in state-of-the-art quantum simulation.
Additional tunability of on- and off-site interactions is possible using the microwave quantization axis, anisotropy and microwave ellipticity\cite{chen2023field,Zhang_2026},
and Floquet engineering of the microwave-induced interactions.
Parallel work shows in single-microwave shielding that coherent coupling of a neighboring monomer pair to a long-lived field-linked on-site dimer provides independently tunable SU($N$)-symmetric on-site and off-site interactions with complementary theoretical techniques, and shows that the microwave orientation axis is a powerful tuning parameter~\cite{austrians}.
Furthermore, we show that the tunable long-range off-site interaction can be used as an \emph{interaction microscope,}
a  tool to characterize effective hole-hole and other interdopant interactions in the doped Hubbard model that are predicted by various theoretical frameworks, but which could  otherwise be indistinguishable.

\textit{Acknowledgments ---} We thank Sebastian Will and Lin Su for useful discussions.
K.P.~is supported by NWO ENW-M grant OCENW.M.21.359.
J.W.D and K.R.A.H. are supported in part by the W. M. Keck Foundation (Grant No. 995764) and K.R.A.H. is supported in part by National Science Foundation (NSF DGE-2346014). T.K.~is supported by NWO VIDI (grant ID 10.61686/AKJWK33335).
 T.K. and K.R.A.H. performed part of this work at the Aspen Center for Physics, which is supported in part by the National Science Foundation (PHY-1066293).

\bibliographystyle{apsrev4-2}
\bibliography{biblio}

\clearpage
\begin{widetext}
\begin{center}
\textbf{\large Supplemental Material for: ``Hubbard models with ultracold shielded molecules in optical lattices''}
\end{center}
\end{widetext}

\setcounter{equation}{0}
\setcounter{figure}{0}
\setcounter{table}{0}
\setcounter{page}{1}
\makeatletter
\renewcommand{\theequation}{S\arabic{equation}}
\renewcommand{\thefigure}{S\arabic{figure}}
\renewcommand{\thetable}{S\arabic{table}}

\section{One molecule in an optical lattice \label{sec:onemolecule}}

We first consider a single particle in a periodic optical lattice in one dimension $V(x)=V_0\sin^2(kx)$,
where $k=2\pi/\lambda$ with $\lambda$ the wavelength of the light,
and $\lambda/2$ the period of the resulting optical lattice.
The Bloch wavefunction $\psi_q(x) = u_q(x) \exp(iqx)$ where cell-periodic function $u_q(x)$ statisfies
\begin{align}
    \left[ \frac{\hbar^2}{2m} \left(-i\frac{d}{dx} + q\right)^2 + V(x) \right] u_q(x) = e_q u_q(x),
\end{align}
with periodic boundary conditions.
We solve the Schr\"odinger equation for $\psi_q(x)$ using sinc-function discrete variable representation (DVR) \cite{colbert1992novel} subject to the boundary condition $\psi_q(x+\lambda/2) = \exp(iq\lambda/2) \psi_q(x)$ for each quasi-momentum $q$.
Fourier transformation of $u_q(x)$ yields the Wannier function $w_i(x) = \int \exp(iqx_i) u_q(x) dq$ localized at site $i$.
The tunneling is obtained directly from the bandwidth as {$t = (e_{q=k}-e_0)/4$}.

For a single molecule in a three-dimensional cubic optical lattice we consider the separable potential $V(\bm{r}) = V_0 [\sin^2(kx)+\sin^2(ky) + \sin^2(kz)]$.
The solutions for the Wannier functions are products of one-dimensional Wannier functions $w_{i}(\bm{r}) = w(x-x_i)w(y-y_i)w(z-z_i)$.

Given two nearest neighbor sites $i$ and $j$, the tunneling can also be obtained as
\begin{equation}
t = -\int w_{i}^{\ast}(\bm{r})\, \hat{h}_{0}\, w_{j}(\bm{r}) d\bm{r},
\end{equation}
where $\hat{h}_{0}$ is the single-particle Hamiltonian.
The tunneling calculated in this way agrees closely with that obtained from the bandwidth as described above.
On-site interaction are written as
\begin{equation}
U = \int \int \left|w_{i}(\bm{r}_1)\right|^{2} V_{\mathrm{int}}(\bm{r}_1-\bm{r}_2) \left|w_{i}(\bm{r}_2)\right|^{2}\ d\bm{r}_1\ d\bm{r}_2.
\label{eq:Uphys}
\end{equation}
Computing the on-site interaction in the basis of the single-particle Wannier function for a single band implicitly assumes the interactions are perturbatively weak and do not modify the wavefunction.
For atoms, one first replaces the physical interaction potential with a pseudopotential $V = \frac{4\pi\hbar^2 a_s}{m} \delta(\bm{r-\bm{r}'})$ with the same scattering length, $a_s$, which gives
\begin{equation}
U_\mathrm{pseudopotential} = \frac{4\pi\hbar^2a_s}{m} \int |w_{i}(\bm{r})|^{4}~d\bm{r}.
\label{eq:Upseudo}
\end{equation}
This replacement is not just convenient,
but necessary as Eq.~\eqref{eq:Uphys} diverges for $V_{\mathrm{int}}(\bm{r}_1-\bm{r}_2) = c_6 r_{12}^{-6}$.
However, the replacement of the physical potential by a pseudopotential which is treated perturbatively as in Eq.~\eqref{eq:Upseudo} is valid only if the range of the physical potential is much shorter than the size of $w_i(\bm{r})$.
For atoms in optical lattices this might typically be valid,
but for microwave shielded molecules that interact over a much longer range, considered in this work, this in general breaks down \cite{buchler2010microscopic}.
Evaluation of $U$ in this case in this regime is the main subject of this paper.

We also evaluate the off-site interaction for nearest-neighbors displaced along the quantization axis, 
\begin{equation}
V_{\mathrm{NN}} = \int \left|w_{i}(\bm{r}_1)\right|^{2} V_{\mathrm{int}}(\bm{r}_1-\bm{r}_2) |w_{j}(\bm{r}_2)|^{2}\ d\bm{r}_1\ d\bm{r}_2,
\label{eq:V}
\end{equation}
where $j$ is a neighbor of site $i$ in the $z$ direction, i.e., $w_{j}(\bm{r}) = w(x-x_i)w(y-y_i)w(z-z_{i+1})$.
We will evaluate this in the case of the dipolar interaction $V_{\mathrm{int}}(\bm{r})= \frac{d_\mathrm{eff}^2}{4\pi\epsilon_0 r^3}(1-3\cos^2\theta)$, where $\theta$ is the polar angle of $\bm{r}$.
We note that all integrals involving the dipolar interaction converge without problems and are computed here by numerical integration.
One might have expected these integrals to diverge due to the $r^{-3}$ scaling of the interaction.
However, the interaction is also completely anisotropic and averages to zero when integrated over the direction of $\bm{r} = \bm{r}_1-\bm{r}_2$.
Hence, the integral $V$ is only sensitive to the anisotropy of the density distribution $|w_{i}(\bm{r}_1)|^{2}|w_{j}(\bm{r}_2)|^{2}$ which vanishes at linearly in ${r}$ around $\bm{r}=0$, leading to a well-defined finite integral \cite{baier2016extended}.

\section{Two molecules in a harmonic potential}

Next we consider two ultracold molecules confined in a three-dimensional isotropic harmonic potential.
We introduce the relative coordinate $\bm{r}=\bm{r}_{1}-\bm{r}_{2}$ and the center-of-mass coordinate $\bm{R}=(\bm{r}_{1}+\bm{r}_{2})/2$,
where $\bm{r}_1$ and $\bm{r}_2$ are the positions of the two molecules.
The Hamiltonian is
\begin{equation}
\hat{H} = - \frac{\hbar^{2}}{2m} \left(\frac{1}{2} \nabla^{2}_{\bm{R}} + 2\nabla^{2}_{\bm{r}} \right) + m\omega^2 R^2 + \frac{1}{4} m\omega^2 r^2 + V_{\mathrm{int}}(\bm{r}),
\end{equation}
where $m$ is the molecular mass and $\omega$ is the harmonic trap frequency. 
The effective interactions between the two shielded molecules with compensated ($d_{\text{eff}}=0$) dipolar interactions are described accurately by a repulsive van der Waals potential\cite{deng2023effective,b8pm-3prn}
\begin{equation} \label{vdw}
V_{\mathrm{int}}(r) = \frac{C_{6}}{r^{6}}.
\end{equation}
Note this is isotropic and independent of the center-of-mass coordinate $\bm{R}$.

Since the Hamiltonian is separable into center-of-mass and relative motion,
and the interaction and external potential are isotropic,
we can write radial Schr\"odinger equations
\begin{align}
\left[  -\frac{1}{4}  \frac{\partial^2}{\partial X^2} + \frac{L(L+1)}{4X^2}  + {X}^{2} - E_{N,L}  \right] \Phi_{N,L}({X}) &= 0, \nonumber\\
\left[- \frac{\partial^2}{\partial x^2} + \frac{l(l+1)}{x^2}  + \frac{1}{4}{x}^{2}  + \left(\frac{r_6}{a_h}\right)^4 x^{-6} - \epsilon_{n,l} \right]\phi_{n,l}({x}) &=0.
\label{eq:harmonicSE}
\end{align}
The corresponding angular solutions are the spherical harmonics $Y_{L,M}(\bm{X})$ and $Y_{l,m}(\bm{x})$, respectively.
Here, we have introduced dimensionless positions $\bm{x}=\bm{r}/a_h$ and $\bm{X}=\bm{R}/a_h$ where $a_h=\sqrt{\frac{\hbar}{m\omega}}$ is the harmonic oscillator length,
and $\epsilon_n$ and $E_n$ are eigenenergies in units of the harmonic oscillator energy $\hbar\omega = \frac{\hbar^2}{ma_h^2}$.
We furthermore introduced a characteristic length of the van der Waals interaction as
\begin{equation} \label{r6}
r_{6} = \left( \frac{m C_{6}}{\hbar^{2}}\right)^{1/4},
\end{equation}
and energy scale $E_6 = \hbar^2/m r_6^2$.

The uncoupled radial Schr\"odinger equations are solved numerically using sinc-function discrete variable representation (DVR) \cite{colbert1992novel}.

For NaCs molecules under conditions that enable effective shielding \cite{bigagli2023collisionally,yuan2025extreme},
$r_6 = 3000~a_0$ (160~nm).
Using the universality of microwave shielding \cite{dutta2025universality} we estimate the relevant length scales for other molecules by assuming $r_6$ scales with the limiting dipolar length $\propto m d^2$.
This results in $r_6 = 9700~a_0$ (513~nm) for KAg,
20\,811~$a_0$ (1101~nm) for CsAg,
423~$a_0$ (22~nm) for NaK,
and 1598~$a_0$ (84~nm) for LiCu, for example\cite{Smialkowski,aymar2005calculation}.
If $r_6\ll a_h$ the van der Waals potential can be safely replaced by a Fermi-pseudopotential $\frac{4\pi\hbar^2a_s}{m} \delta(\bm{r})$, which is typically done for atoms.
The problem of two particles interacting by a contact pseudopotential has been solved exactly \cite{busch1998two}.
For small $a_s/a_h$ the interaction is given by $U/\hbar\omega = \sqrt{\frac{2}{\pi}} \frac{a_s}{a_h}$.
As $a_s/a_h$ increases, the exact interaction energy increases less than suggested by this linear relation,
and saturates at $U=\hbar\omega$ as $a_s \rightarrow \infty$.
In this discussion, $a_s$ is the scattering length associated with the repulsive van der Waals potential,
which we expect to be of order $r_6$ based on dimensional analysis.
More precisely, the $s$-wave Schr\"odinger equation for the repulsive van der Waals potential at zero energy is
\begin{align}
    \left[- \frac{d^2}{dy^2}  + y^{-6} \right]\varphi({y}) = 0,
\end{align}
where $y=r/r_6$.
The exact solution is given by
\begin{align}
    \varphi(y) &= \sqrt{y} K_{1/4}\left(\frac{1}{2y^2}\right) \nonumber \\
    &\simeq \frac{\Gamma(1/4)}{\sqrt{2}}~y + \frac{\Gamma(-1/4)}{\sqrt{2}} + \mathcal{O}(y^{-3}),
\end{align}
where $K_n(y)$ is the irregular Bessel function of the first kind.
We find the scattering length as the root of the asymptotic wavefunction, which yields
\begin{align}
    {a_s} = -\frac{\Gamma(-\frac{1}{4})}{2\Gamma(\frac{1}{4})}~r_6 \approx 0.676~r_6.
    \label{eq:as}
\end{align}

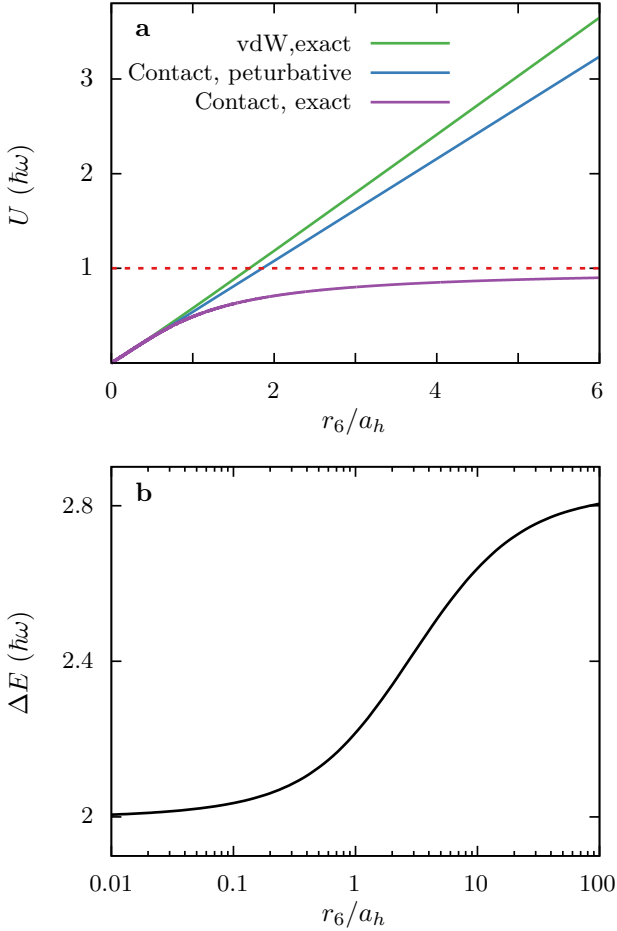
\begin{figure}
    \input{uharm_r6} 
    \input{delta_E_harmonic} 
        \caption{{\bf Two molecules in a harmonic potential }  \textbf{a} Comparison of the on-site interaction energy as a function of $r_{6}/a_{h}$ for two molecules confined in a harmonic potential interacting via a repulsive van der Waals potential (green) and with the Fermi-pseudopotential treated either perturbatively (blue) or exactly (purple) \cite{busch1998two}. The dashed line indicates the limit of $U$ for contact interactions at unitarity, $a_s\rightarrow \infty$.
        \textbf{b} Energy difference between the ground and first excited state of two molecules in a harmonic potential with repulsive van der Waals interactions.}
    \label{fig:r6_ah_vs_u}
\end{figure}

For large $r_6$, the pseudopotential description breaks down.
Instead, we can think of the ground state as a harmonic oscillator state localized at intermediate distances that balance repulsion of the van der Waals and harmonic oscillator potentials, respectively.
The minimum of this potential occurs at $x_\ast = (12)^{1/8} (r_6/a_h)^{1/2}$. 
At $x_\ast$ the value of the potential in harmonic oscillator units is
$\frac{\sqrt{2}}{3^{3/4}} \left(\frac{r_6}{a_h}\right)$,
and the second derivative of the potential in harmonic oscillator units at $x_\ast$ is equal to 4,
which implies a harmonic oscillator frequency of $2\sqrt{2}$ harmonic oscillator units.
For $r_6\gg a_h$, anharmonic terms become negligible and the ground state energy for the relative motion thus becomes  the sum of potential at the minimum and the harmonic zero-point energy,  $\frac{\sqrt{2}}{3^{3/4}} \left(\frac{r_6}{a_h}\right) + \sqrt{2}$ in units of $\hbar\omega$.
This simple result is somewhat remarkable;
the ground state energy in units of $\hbar\omega$ is essentially always linear in $r_6/a_h$,
with a small change of slope from $-\frac{\Gamma(-\frac{1}{4})}{2\Gamma(\frac{1}{4})}\sqrt{\frac{\pi}{2}} \approx 0.539$
for perturbatively weak interactions at $r_6 \ll a_h$ to $\frac{\sqrt{2}}{3^{3/4}}\approx 0.620$ for strong interactions at $r_6\gg a_h$.
This can be seen in Fig.~\ref{fig:r6_ah_vs_u}\textbf{a}.
Treating the pseudopotential interaction perturbatively for $r_6 > a_h$ is inaccurate and deviates strongly from the exact solution for the pseudopotential,
but when compared to the solution for the full interaction this ``double approximation'' is fortuitously accurate.
The other remarkable feature is that the harmonic frequency at large $r_6/a_h$ is independent of $r_6/a_h$.
As a result, the lowest excitation energy for $s$-wave relative motion states changes from $2\hbar\omega$ at $r_6\ll a_h$ to $2\sqrt{2} \hbar\omega$ at $r_6\gg a_h$, shown in Fig.~\ref{fig:r6_ah_vs_u}\textbf{b}.
This change of excitation energy upon interaction could be useful for selective parametric heating of pairs of interacting molecules in a single trap, for example for removing double occupancies or deterministic loading.

\section{Two molecules in an optical lattice}

Next, we consider two ultracold molecules confined in a three-dimensional optical lattice, with Hamiltonian
\begin{equation}\label{total_hamiltonian}
\hat{H} = - \frac{\hbar^{2}}{2m} \left(\frac{1}{2} \nabla^{2}_{\bm{R}} + 2\nabla^{2}_{\bm{r}} \right) + V_{\mathrm{lat}}\left( \bm{r}, \bm{R}\right) + V_{\mathrm{int}}(\bm{r}).
\end{equation}
The lattice potential takes the form
\begin{align}
V_{\mathrm{lat}}(\bm{r},\bm{R}) &= V_{0} \sum_{j=1}^{3} \left[  \sin^2(\bm{k}_j\cdot\bm{r}_1)+\sin^2(\bm{k}_j \cdot\bm{r}_2) \right], \nonumber \\
 &= V_{0} \sum_{j=1}^{3} \left[ 1 -  \cos \left(\bm{k}_{j} \cdot\bm{r}\right) \cos \left(2 \bm{k}_j \cdot \bm{R}\right) \right],
\end{align}
where again the relative coordinate $\bm{r}=\bm{r}_{1}-\bm{r}_{2}$ and the center-of-mass coordinate $\bm{R}=(\bm{r}_{1}+\bm{r}_{2})/2$,
with $\bm{r}_1$ and $\bm{r}_2$ the positions of the two molecules.
We assume a three dimensional cubic lattice where $\bm{k}_j$ point in the $x$, $y$, $z$ directions, respectively,
and $k_j = 2\pi/\lambda$ with $\lambda = 1064$~nm the wavelength of the optical lattice, in this case equal in each of the directions.

Instead of solving the Schr\"odinger equation for the full Hamiltonian directly,
we first solve a ``reference problem'' where the lattice potential is replaced by an isotropic harmonic potential with the same harmonic oscillator frequency $\omega=\sqrt{ \frac{2 V_{0} k^{2}_{\lambda} }{m}}$. 
This leads to uncoupled radial Schr\"odinger equations, Eq.~\eqref{eq:harmonicSE}, which are simpler to solve.
We determine the radial solutions for $l,L=0$,
and use these as contractions to set up a direct product basis set 
\begin{align}
|n\, l\, m\, N\, L\, M\rangle = \phi_n(r/a_h) Y_{l,m}(\bm{r}) \Phi_N(R/a_h) Y_{L,M}(\bm{R}/a_h),
\label{eq:basis}
\end{align}
which is typically truncated as $n,N \le 4$, and $l,L=0,2,4$.
We note that these contractions merely serve to define an efficient basis,
but ultimately we will solve the Schr\"odinger equation for the full Hamiltonian.
We note that the contractions are determined while fully including the interaction between molecules.
The basis that is generated is automatically localized near a particular minimum of the periodic optical lattice potential.
As a result, we expect that eigenstates computed in this localized basis will approximate the Wannier functions,
and we will confirm this numerically in Sec.~\ref{sec:convergence}.

To evaluate the lattice potential in the basis of contractions, we use the Legendre expansion
\begin{equation} \label{coskr}
\cos\left(\bm{k}\cdot\bm{r}\right) =  \sum_{\mathrm{even} \: \ell} i^{\ell}\left( 2 \ell +1 \right) j_{\ell}(k  {r}) P_{\ell}\left( \cos\theta \right),
\end{equation}
where $j_{\ell}(x)$ are the spherical Bessel functions of the first kind,
$P_\ell$ is a Legendre polynomial,
and $\theta$ is the angle between $\bm{k}_j$ and $\bm{r}$.
For the cubic lattice that we consider, we will make use of
\begin{align}
P_{2}(\hat{\bm{x}} \cdot \hat{\bm{r}}) &= -\frac{1}{2}C_{2,0}(\hat{\bm{r}}) + \sqrt{\frac{3}{8}} \left[ C_{2,2}\left(\hat{\bm{r}}\right) + C_{2,-2}(\hat{\bm{r}}) \right] \nonumber \\
P_{2}(\hat{\bm{y}} \cdot \hat{\bm{r}}) &= -\frac{1}{2}C_{2,0}(\hat{\bm{r}}) - \sqrt{\frac{3}{8}} \left[ C_{2,2}\left(\hat{\bm{r}}\right) + C_{2,-2}(\hat{\bm{r}}) \right] \nonumber \\
P_{2}(\hat{\bm{z}} \cdot \hat{\bm{r}}) &= C_{2,0}(\hat{\bm{r}}),
\end{align}
where $C_{\ell,m}$ is a Racah-normalized spherical harmonic.
To evaluate matrix elements of the angular coupling induced by the lattice potential in our direct product basis, we use
\begin{align}
\langle L'M' | &C_{\ell,m}\left(\cos\theta \right) | L M \rangle \nonumber \\&= \sqrt{\frac{2L +1}{2L'+1}}
\langle LM\ell m| L'M' \rangle \langle L0\ell0| L'0 \rangle 
\end{align}
where $\langle LM\ell m| L'M' \rangle$ is a Clebsch-Gordan coefficient.

The total Hamiltonian is evaluated in the basis of Eq.~\eqref{eq:basis} as the sum of the reference Hamiltonian, centrifugal kinetic energy, and the difference in confining potential.
The reference Hamiltonian is solved by the contracted basis of Eq.~\eqref{eq:basis}, and hence is diagonal.
The direct product structure of independent radial and angular functions, used in this basis,
required determining radial contractions without including the centrifugal kinetic energies $\hat{l}^2/2\mu r^2$ and $\hat{L}^2/2MR^2$.
The centrifugal kinetic energy terms are then evaluated in the contracted basis and included in the total Hamiltonian.
Finally, the contracted Hamiltonian uses a harmonic potential rather than the optical lattice potential.
After setting up the total Hamiltonian in this basis, we diagonalize it to determine eigenfunctions and energies.
We determine the on-site interactions $U=E-E_0$ in the lowest band as the difference between the lowest eigenenergies $E$ and $E_0$ obtained including and excluding the molecule-molecule interaction, respectively.

\section{Validation \label{sec:convergence}}

Hubbard parameters are generally defined in terms of Wannier functions \cite{buchler2010microscopic,baier2016extended}.
The computational procedure outlined above is to compute on-site interactions as the difference between the lowest eigenstates for two molecules in an optical lattice, computed with and without interactions.
It is, however, not guarranteed that the on-site interaction energy computed in this way corresponds to the on-site interaction in the Wannier basis.
In fact, the ground state of the problem we solve numerically, two molecules in an extended lattice,
should not correspond to two molecules on the same site, especially not if these molecules are repulsively interacting.
The essence of our approach is to solve the two-molecule problem in a limited localized basis which forces the two molecules onto a single site.
However, one can imagine that if one attempts to converge this calculation by making the localized basis more complete,
at some point the lowest energy solution will delocalize over multiple sites.
We expect that when the lattice is sufficiently deep that the overlap is sufficiently small between solutions to the two-molecule problem on one site and solutions with one or more molecules moved to other sites, one  can obtain meaningful, converged solutions for the on-site interaction before the delocalization happens.
In this section, we first examine the convergence of our calculations,
and numerically establish that the interactions computed in this way agree with integrals over Wannier functions when that approach is justified.

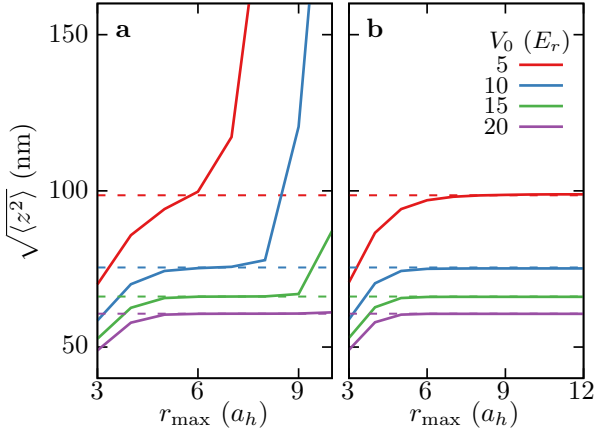
\begin{figure}
\centering
\input{scan_convergence_latex}
\caption{\label{fig:convergence_z2} {\bf Size of the non-interacting wavefunction} $\sqrt{\langle z^{2} \rangle}$, where $z=z_1-z_2$ is the relative z coordinate, as a function of the maximum value of the relative radial grid. Panel \textbf{a} shows results of calculations in which the relative- and center-of-mass cutoffs are increased together, leading to an intermediate plateau before the solutions delocalize at large cutoff. \textbf{b} shows calculations in which the center-of-mass cutoff is fixed at $R_{\mathrm{max}} = 2.5\,a_{h}$ while the relative-coordinate cutoff is varied. Solid lines denote solutions computed with the approach developed here, and dashed lines denote the corresponding Wannier-function results. To ensure full convergence over the entire range of cutoffs, an enlarged basis with ($n=N=l=L=8$) was used for all calculations shown.
}
\end{figure}

First, we consider non-interacting molecules and analyze the size of the wave function $\sqrt{\langle z^2\rangle}$ where $z=z_1-z_2$ is the relative distance in one of the lattice directions.
This expectation value can be computed in the Wannier basis as $\langle z^2\rangle = \int \int w(z_1)^2 (z_1-z_2)^2 w(z_2)^2\ dz_1\ dz_2$.
We compute the same expectation value using our numerical approach,
shown in Fig.~\ref{fig:convergence_z2}\textbf{a} as a function of the size of the radial grid used for computing the contracted basis functions, $r_\mathrm{max}$ for the relative coordinate and $R_\mathrm{max}=2r_\mathrm{max}$ for the center of mass coordinate.
For deep lattices we see that the size of the wavefunction converges for $r_\mathrm{max}> 5~a_h$,
and the value obtained agrees with that from the independent calculation using the Wannier functions, shown as the dashed lines.
If $r_\mathrm{max}$ is made even larger, however, $\sqrt{\langle z^2\rangle}$ starts to grow again to exceed hundreds of nanometer,
as the basis now allows the wavefunction to delocalize over multiple sites.
For shallower lattices this occurs at smaller $r_\mathrm{max}/a_h$ as the harmonic oscillator length grows while the lattice constant is fixed.
Thus, we see a plateau form at intermediate $r_\mathrm{max}/a_h$ where $\sqrt{\langle z^2\rangle}$ is insensitive to $r_\mathrm{max}/a_h$ and in agreement with the result based on integration over products of Wannier functions.
This plateau shrinks with decreasing lattice depth, and essentially disappears around $V_0/E_r = 5.$ For lattice depths around or below this value, our calculations offer only a rough guideline, and quantitative calculations will require new techniques. 

Next we perform a similar calculation of $\sqrt{\langle z^2\rangle}$ as a function of the size of the radial grid for the relative coordinate,
while truncating the center of mass grid at fixed $R_\mathrm{max}=2.5~a_h$.
By contrast, the previous calculation truncated both center of mass and relative coordinates based on $r_\mathrm{max}$.
The idea is that if the center of mass coordinate cannot localize in between lattice sites,
delocalization of the two molecules over two sites requires $r\simeq \lambda$ rather than $r\simeq \lambda/2,$
extending the plateau over which meaningful expectation values are computed, in agreement with the Wannier basis.
As seen in Fig.~\ref{fig:convergence_z2}\textbf{b}, this procedure effectively removes the large-cutoff instability and produces stable convergence over a broader range of relative-coordinate cutoffs.

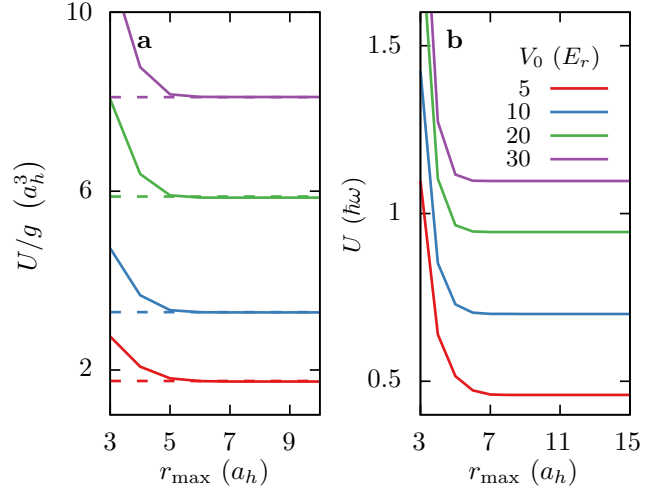
\begin{figure}
\centering
\input{scan_interactions_convergence}
\caption{\label{fig:convergence_interac} 
{\bf Convergence of the on-site interaction energy} as a function of the relative-grid cutoff $r_{\mathrm{max}}$. \textbf{a} pseudopotential, see text, comparing our method (solid lines) with Wannier-function estimates (dashed lines) for several lattice depths. \textbf{b} repulsive van der Waals interactions for NaCs molecules. In both cases we fixed $R_{\mathrm{max}} = 2.5~ a_{h}$. The results show a smooth convergence for all lattice depths.}
\end{figure}

Next, we verify that the same convergence behavior persists in the interacting case.
We first consider a short-range pseudopotential but replace the delta function by a normalized Gaussian, $V(r) = \frac{g}{(\sigma\sqrt{\pi})^3} \exp\left(-\frac{r^{2}}{\sigma^{2}}\right)$,
where we choose the interaction strength $g$ small enough such that this interaction can be treated perturbatively by integrating over Wannier functions
$U = \int \int w(\bm{r}_1)^2 V(\bm{r}_1-\bm{r}_2) w(\bm{r}_2)^2\ d\bm{r}_1\ d\bm{r}_2$.
Figure~\ref{fig:convergence_interac}\textbf{a} shows $U/g$ as a function of the truncation of the relative-coordinate grid $r_{\mathrm{max}}$ for fixed truncation of the center-of-mass grid.
Similar to what we saw in the previous calculations of the size of the wavefunction,
we observe that the interaction energy converges for $r_\mathrm{max}>5~a_h$ to a stable value that agrees with that obtained as matrix elements in the Wannier basis.
This confirms numerically that our method yields the same result as the standard integration over products of Wannier functions where that approach is valid.

Next, we compute the on-site interaction for the repulsive van der Waals potential, shown in Fig.~\ref{fig:convergence_interac}\textbf{b}.
As before, we observe that the computed on-site interactions converge  for  $r_\mathrm{max}>5~a_h$.
Here, we do not include a reference value computed by integrating over products of Wannier functions, because this approach is not applicable here.
As discussed in Sec.~\ref{sec:onemolecule}, the integral over Wannier functions Eq.~\eqref{eq:Uphys} does not converge for the microscopic van der Waals interaction which diverges as $r^{-6}$.
For atoms, the typical approach is to replace this microscopic interaction by a Fermi pseudopotential with the same scattering length.
However, here the range of interactions is so large that the pseudopotential approximation is not valid, nor is its perturbative treatment by computing integrals over the lowest-band Wannier function.
As discussed before, this is flagged by $U$ exceeding $\hbar \omega$. We saw earlier that  the exact result for a harmonic oscillator with a contact pseudopotential deviates from results with a van der Waals interaction already for $U \approx 0.5\hbar \omega$, which we see is exceeded here already for the shallowest lattices, $V_0/E_r =5$, considered.

\section{On-site interactions}
\begin{figure}
    \input{u_r6_3d_supplement}
        \caption{{\bf On-site interaction energy for compensated dipolar interactions} as a function of the dimensionless ratio $r_{6}/a_{h}$ in a three-dimensional optical lattice. Solid curves show results obtained with the full repulsive van der Waals potential at fixed lattice depths $V_{0}=15\,E_{r}$ (orange) and $3\,E_{r}$ (red). Dashed lines show the perturbative scaling of contact-interactions $U/\hbar\omega \propto r_6/a_h$. Markers indicate representative molecular species considered in this work.}
    \label{fig:r6_vs_u}
\end{figure}
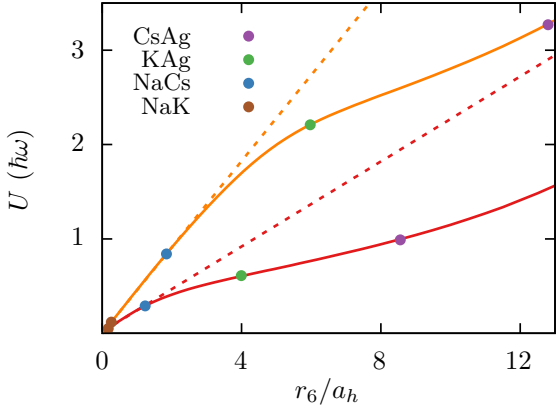

Figure~\ref{fig:r6_vs_u} shows the on-site interaction energy in harmonic-oscillator units, $U/\hbar\omega$, as a function of the dimensionless ratio $r_{6}/a_{h}$ in a three-dimensional optical lattice and compensated ($d_{\text{eff}}=0$) interactions. 
Each solid curve is computed at fixed lattice depth, and hence fixed $a_{h}$, while varying $r_{6}$. For $r_{6}/a_{h} \ll 1$, the interaction is weak, and the repulsive van der Waals potential can be approximated by a contact pseudopotential and  treated perturbatively, leading to the scaling $U/\hbar\omega \propto r_6/a_h$ shown by the dashed lines.
Markers indicate $r_6/a_h$ that correspond to typical polar molecules,
and --except for NaK-- all of them lie in the regime $r_6>a_h$ where we have seen the pseudopotential description breaks down in a harmonic trap.
As before in the case of the harmonic trap, we find that the true $U$ for strong  interactions (with $d_{\text{eff}}=0$) are fortuitously close to what would be expected by extending the perturbative scaling well outside its regime of applicability, the dashed lines.
For the most strongly dipolar molecules such as KAg and CsAg, 
however,
we find that $U$ is no longer linear in $r_6$,
which by construction can never be described by integration of a pseudopotential over products of Wannier functions.

\begin{figure}
    \centering
    \input{U_NaCs} 
    \caption{{\bf On-site interaction energy for NaCs with compensated dipolar interactions} in a three-dimensional optical lattice as a function of the lattice depth $V_{0}$.
    \label{vdw_onsite_interactions}}
\end{figure}
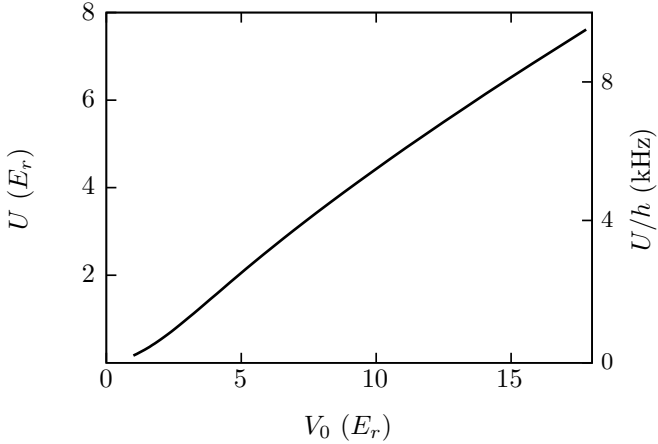

Figure \ref{vdw_onsite_interactions} shows the on-site interaction $U$ for NaCs molecules in a three-dimensional optical lattice as a function of the lattice depth $V_{0}$,
in photon recoil units and in frequency units instead of harmonic oscillator units.

\begin{figure}
\input{u_t_3D} 
\caption{{\bf On-site interaction over tunneling} $U/t$ as a function of the lattice depth $V_{0}$ for several shielded molecular species in a three-dimensional optical lattice with compensated dipolar interactions. $U/t$ is large already at shallow lattices, indicating access to strongly correlated Hubbard regimes in shallow lattices.
}
\label{vdw_U_t}
\end{figure}
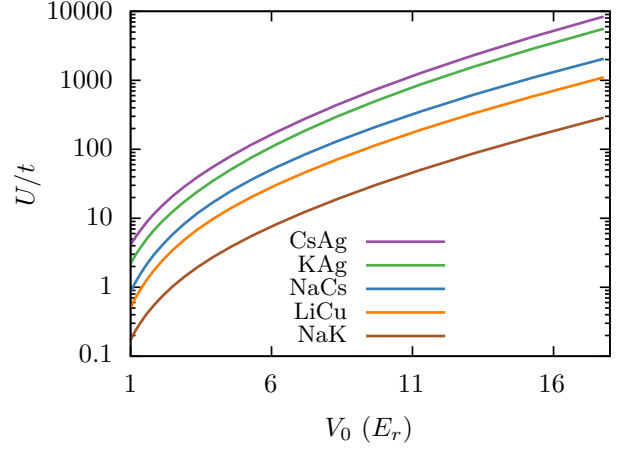

Figure \ref{vdw_U_t} shows the ratio of interaction energy to tunneling $U/t$ as a function of lattice depth $V_{0}$ for several shielded polar molecules in a three-dimensional optical lattice. The differences between molecular species arise primarily from the scaling of the radius of the shield, $r_6$, which for double microwave shielding scales with dipolar length \cite{dutta2025universality}.
Notably, $U/t$ is large already for low lattice depth $V_0$.
This indicates strongly interacting Hubbard physics can be simulated in shallow lattices where $t$ is large,
such that at fixed experimentally realizable temperatures,
the temperature over tunneling $T/t$ can be low.
Achieving large $U/t$ at low $T/t$ is a main challenge for quantum simulation of the Hubbard model with ultracold atoms. 

\section{Dipolar interactions}

Next, we consider turning on long-range dipole-dipole interactions between the molecules.
For ultracold polar molecules with double microwave shielding this can be achieved by tuning the microwaves away from compensation of the dipolar interactions \cite{karman2025double}.
First, we revisit the convergence of the calculated on-site interaction in this case.

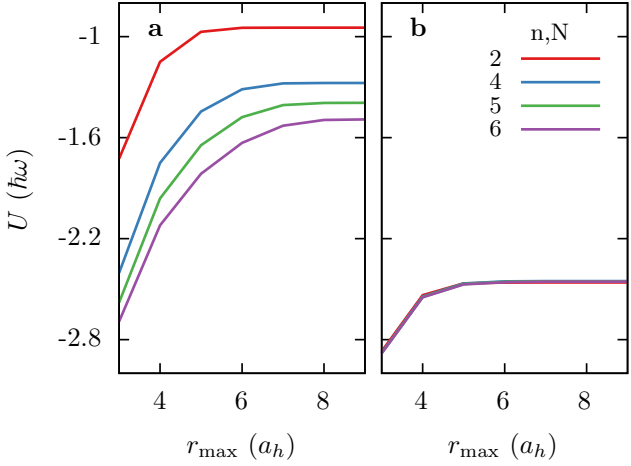
\begin{figure}
    \input{dip_conv_comparison}
    \caption{{\bf Convergence of on-site interaction energies in the presence of dipolar interactions}. The legend indicates the number of contracted basis functions $n=N$ used for the relative-motion ($n$) and center-of-mass ($N$) radial coordinates. In all calculations, ($l=L=4$). Calculations based on contractions generated from the lowest adiabatic potential \textbf{b} exhibit significantly improved numerical stability compared to those obtained using a harmonic reference potential \textbf{a}. The results were obtained for NaCs molecules with a lattice potential of $5 \, E_r$ and $d_{\mathrm{eff}}^2 = -0.67~\mathrm{Debye}^{2}$. 
    }
    \label{Rmaxdip}
\end{figure}

First, we attempt to include dipolar interactions in the same computational procedure as before:
We compute a contracted basis as before, as direct products of angular and radial functions,
where the radial functions are determined by solving a reference problem, the $s$-wave radial Schr\"odinger equation for a harmonic confining potential.
This reference problem accounts for the shielding core, the repulsive van der Waals interaction that results from microwave shielding.
Then, we compute matrix elements of the dipole-dipole interaction in this direct product basis,
and include it in the total Hamiltonian.
Figure~\ref{Rmaxdip}\textbf{a} shows the on-site interaction energy for NaCs including long-range dipolar interactions, as a function of the relative-grid cutoff $r_{\mathrm{max}}$ and for different numbers of radial contractions.
We observe no convergence of the computed on-site interactions,
which is somewhat surprising compared to the previous results.

The essential difference is that in our previous calculations,
the basis functions were optimized while fully accounting for the molecule-molecule interactions.
By contrast, here, the basis functions are not optimized for the dipolar interaction.
The reason for this is that the dipolar interaction is anisotropic and cannot be fully described in a purely radial wave equation, as used to generate our contractions as direct products.
We improve upon this by picking a different reference potential,
which is still a function of the radial relative coordinate alone, but effectively accounts for the dipolar interactions.
This is done by computing the lowest adiabatic potential curve including the dipolar interaction,
by diagonalizing the Hamiltonian for the relative degree of freedom without radial kinetic energy.
When using this reference potential to compute the contracted basis functions,
the on-site interactions converge rapidly as shown in Fig.~\ref{Rmaxdip}\textbf{b}.

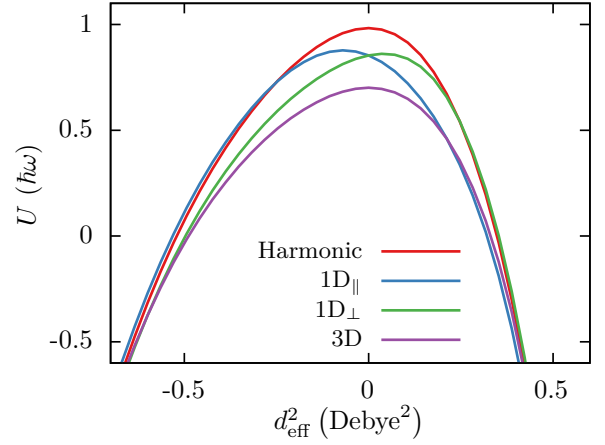
\begin{figure}
    \centering
    \input{deff_v10}
    \caption{{\bf Control of the on-site interaction by tuning dipolar interactions} parameterized by the squared effective dipole moment $d_{\mathrm{eff}}^{2}$ for NaCs molecules in a lattice of depth $10\,E_{r}$. The harmonic curve shows the single-site harmonic approximation, while the lattice curves labeled 1D and 3D include the optical lattice potential in one or three dimensions, respectively. Curve labeled $1\mathrm{D}_{\parallel}$ correspond to dipoles aligned parallel to the lattice direction, and curves labeled $1\mathrm{D}_{\perp}$ correspond to dipoles aligned perpendicular to it. The first-order dipolar correction to $U$ is read off from the slope of the curves at $d_{\mathrm{eff}}^{2}=0$.
    We emphasize that this figure contains the only 1D-lattices results in this work, and all other results are for a 3D cubic lattice.
    }
    \label{d2eff_u}
\end{figure}

The on-site interactions for NaCs molecules as a function of the dipole-dipole interaction are shown in Fig.~\ref{d2eff_u}.
Here, we consider different models for the external trapping potential;
3D corresponds to the three dimensional optical lattice we have considered so far,
whereas harmonic corresponds to an isotropic harmonic approximation to this site potential.
One can see that there is a substantial difference between the 3D optical lattice and the harmonic approximation,
especially at $d_\mathrm{eff}^2 = 0$ where the interactions are repulsive,
whereas the effect of the confining potential becomes smaller at large $d_\mathrm{eff}^2$ when the contact interaction becomes attractive and the molecules bind.

In Fig.~\ref{d2eff_u}, we also show results for an anisotropic situation to illustrate the resulting first-order shifts from the dipolar interaction. We consider a one-dimensional lattice, where the harmonic approximation is made for the two remaining directions. 
In the case labeled $1\mathrm{D}_{\parallel}$, the lattice is in the $z$ direction which is also the polarization direction of the molecular dipole moments.
In the case labeled $1\mathrm{D}_{\perp}$, the lattice is perpendicular to the polarization direction of the molecular dipole moments. 
We emphasize that these are the only 1D-lattices results, and all other results are for a 3D cubic lattice.
The relevance of the 1D lattice here is that the anisotropic potential leads to anisotropic single particle solutions , and therefore  a first-order contribution to the on-site interaction, which can be computed perturbatively as
\begin{equation}
U^\mathrm{dip} = \frac{d_\mathrm{eff}^2}{4\pi\epsilon_0} \int \int  \left|w_{i}(\bm{r}_1)\right|^{2}  \frac{1-3(\hat{\bm{r}}_{12}\cdot \hat{\bm{z}})^2}{r_{12}^3} \left|w_{i}(\bm{r}_2)\right|^{2}\ d\bm{r}_1\ d\bm{r}_2.
\end{equation}
This approximation is frequently used for magnetic atoms \cite{baier2016extended},
and leads to a $U$ that is necessarily linear in $d_\mathrm{eff}^2$.
In the 3D cubic lattice, the first-order dipolar interaction vanishes.
In Fig.~\ref{d2eff_u}, one can indeed see that $U$ versus $d_\mathrm{eff}^2$ has zero slope at $d_\mathrm{eff}=0$ in the isotropic harmonic trap and in the three dimensional optical lattice,
whereas the slope is non-zero for $1\mathrm{D}_{\parallel}$ and $1\mathrm{D}_{\perp}$.
In all cases, however, this perturbative effect is fairly weak in the regime it applies and is quickly overwhelmed by non-perturbative effects:
the on-site interaction deviates strongly from it already for $d_\mathrm{eff}^2=0.5$~Debye$^2$, which corresponds to only a small fraction of the limiting molecular dipole moment.

The non-perturbative effect of the dipolar interaction  already at a small induced dipole moment corresponding to $d_\mathrm{eff}^2\approx0.5$~Debye$^2$ can compensate the strongly repulsive on-site interaction $U\approx \hbar\omega$,
giving weak interactions $U\ll \hbar\omega$ and allowing  $U$ to be tuned on scales comparable to $t$.
Tuning of the on-site interaction to smaller values using the induced dipole moment can be a powerful feature that enables accessing regimes other than hard-core infinitely repulsive on=site interactions,  effectively re-activating quantum fluctuations to doubly occupied sites.

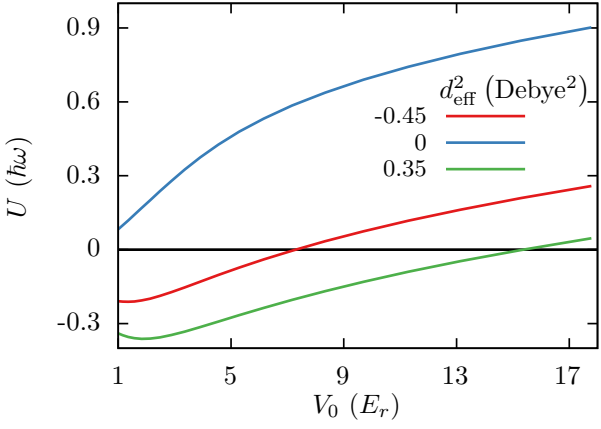
\begin{figure}
    \centering
    \input{U_with_harm} 
    \caption{{\bf On-site interaction} $U$ for NaCs molecules as a function of lattice depth for several fixed values of the squared effective dipole moment $d_{\mathrm{eff}}^{2}$.
    }
    \label{udip_v0}
\end{figure}

Figure~\ref{udip_v0} shows a complementary way of using the dipolar tunability. Instead of scanning $d_{\mathrm{eff}}^{2}$ at fixed lattice depth, we fix several representative values of $d_{\mathrm{eff}}^{2}$ and follow the resulting on-site interaction as a function of the lattice depth. This illustrates that depending on the induced dipole moment, one can obtain either repulsive or attractive on-site interactions, also tunable with lattice depth.
For particular choices of $d_{\mathrm{eff}}^{2}$, this balance can be tuned through $U=0$, corresponding to a cancellation of the net on-site interaction. These zero crossings are indicated by the black lines in Fig.~\ref{udip_v0}.

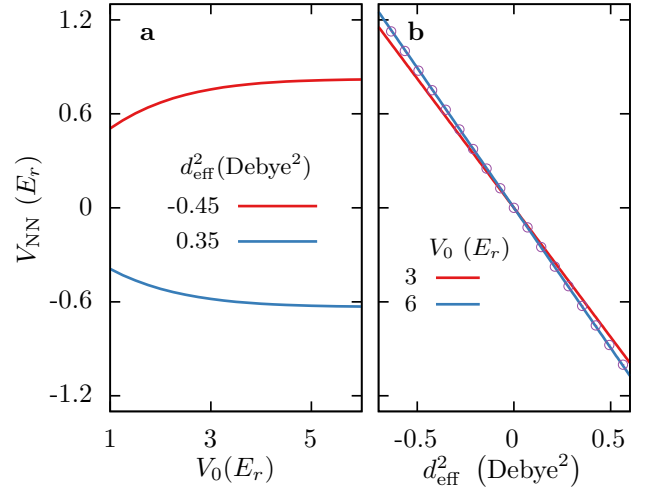
\begin{figure}
    \centering
    \input{Vnn_together} 
    \caption{{\bf Off-site dipolar interaction} $V_{\mathrm{NN}}$ for NaCs molecules as a function of {\bf a} lattice depth for several fixed values of the squared effective dipole moment $d_{\mathrm{eff}}^{2}$ and {\bf b} vice versa. The circles show the point-dipole approximation obtained by fixing the dipoles at the lattice-site centers.}
    \label{Figure:Vnn_dip_v0}
\end{figure}

Figure~\ref{Figure:Vnn_dip_v0} shows the nearest-neighbor off-site dipolar interaction, computed as in Eq.~\eqref{eq:V}.
The neighboring sites are displaced in the $\hat{z}$ direction which is the quantization direction of the molecular dipoles, such that dipolar interactions ($d^2_\mathrm{eff}>0$) are attractive,
and anti-dipolar interactions ($d^2_\mathrm{eff}<0$) are repulsive.
Figure~\ref{Figure:Vnn_dip_v0}{\bf a} shows the off-site interaction for two values of the induced dipole moment as a function of the lattice depth.
Figure~\ref{Figure:Vnn_dip_v0}{\bf b} shows the same interaction as a function of the induced squared dipole for fixed lattice depths.
The dependence on induced squared dipole is necessarily linear, see Eq.~\eqref{eq:V}, and we find the slope increases slightly with lattice depth as the molecular dipoles localize further, see Fig~\ref{Figure:Vnn_dip_v0}.
The interaction is only weakly depth dependent, and close to that expected for two point dipoles separated by $\lambda/2$, shown as the markers.

The point-dipole approximation is more accurate for distant pairs than for nearest neighbors,
so the agreement discussed above implies that to good approximation the off-site interaction beyond nearest neighbors follows the geometric scaling of the dipole-dipole interaction,
$V_{ij}=V_\mathrm{NN}(3\cos^2\theta_{ij}-1) \lambda^3 / 16 r_{ij}^3$,
as discussed in the main text.

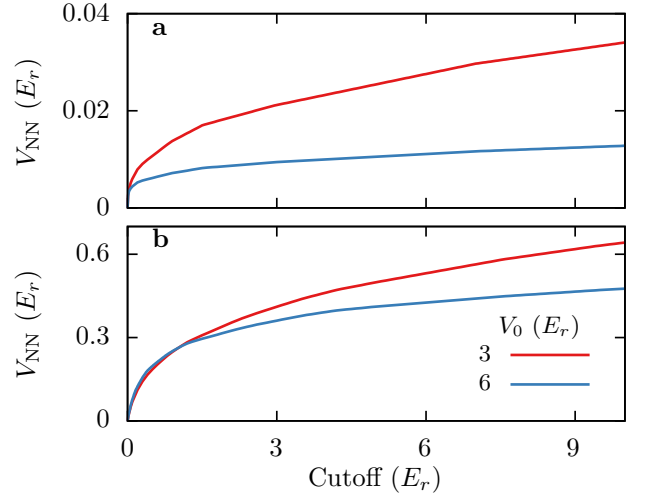
\begin{figure}
    \centering
    \input{Vnn_r6} 
    \caption{{\bf Shielding core contribution to off-site interactions}. Nearest neighbour off-site interaction energies ($V_{\mathrm{NN}}$) produced by the shielding-induced repulsive van der Waals interaction. The horizontal axis shows the energy cutoff applied to the repulsive van der Waals potential, which regularizes the $r^{-6}$ divergence in the numerical evaluation of Eq.~\eqref{eq:V}. Results are shown for lattice depths $V_0=3~E_r$ and $6~E_r$ for \textbf{a} NaCs and \textbf{b} KAg molecules.}
    \label{Figure:Vnn}
\end{figure}

Next we estimate the contribution of the $r^{-6}$ shielding core to the off-site interaction by evaluating Eq.~\eqref{eq:V} for $V_\mathrm{int} = \min(V_\mathrm{cutoff}, C_6 r^{-6})$ as a function of $V_\mathrm{cutoff}$.
The reason for this cutoff is that for any non-zero overlap of the Wannier functions, Eq.~\eqref{eq:V} diverges for the van der Waals interaction.
Physically, short-range interactions far exceeding the lattice depth should not contribute to the off-site interaction,
so one could hope that $V_\mathrm{NN}$ obtains a stable value that depends on $V_\mathrm{cutoff}$ only weakly around $V_\mathrm{cutoff}/ E_r=3$--$6$,
before diverging at large cutoff.
Figure~\ref{Figure:Vnn} show the dependence of the resulting estimate of $V_\mathrm{NN}$ as a function of the cutoff, for two lattice depths $V_0/E_r=3$ and 6.
The resulting $V_\mathrm{NN}$ is not independent of the chosen cutoff,
indicating that a proper determination of the off-site interaction may also require full solution of the two-molecule problem,
which is however beyond the scope of this paper.
Nevertheless, the estimates are useful to show that the contribution of the shielding core to the nearest-neighbor off-site interaction are substantially smaller than the dipolar off-site interactions (Fig.~\ref{Figure:Vnn_dip_v0}) for NaCs, shown in Fig.~\ref{Figure:Vnn}\textbf{a}, and for less dipolar molecules such as NaK.
Conversely, the contribution of the shielding core to the off-site interaction may be strong and comparable to the dipolar one for KAg, shown in Fig.~\ref{Figure:Vnn}\textbf{b}, and for even more dipolar molecules such as CsAg.
Interestingly this could realize new extended models involving hard-core interactions with a radius that encompasses multiple lattice sites.

Having already established that on-site interactions between shielded molecules are non-perturbative, 
we next consider the modification of the interacting two-molecule wavefunction.

First, we analyze the wavefunctions of the two interacting molecules on one lattice site.
We compare the localized two-body wave function obtained with our method to the product of Wannier functions, which is typically the basis used in the Hubbard model.
First, we compute wavefunctions in real space.
To facilitate plotting of the six-dimensional two-molecule wavefunction,
we fix the center of mass coordinate to the center of the lattice site, $\bm{R}=0$,
and the relative coordinate $\bm{r}$ to the $x,z$ plane.
We show the resulting wavefunctions in Fig.~1 of the main text, for NaCs molecules in a 6 recoil lattice for three cases: (i)
 Non-interacting molecules, where the relative-coordinate wavefunction is essentially a Gaussian centered around $\bm{r}$, as expected for a harmonic oscillator. (ii)
 Double microwave shielding with compensated dipolar interactions.
The repulsive shield punches a hole with radius $160$~nm in the wavefunction as the molecules cannot approach within the shield radius.
The length scale of this is comparable to that of the lattice potential,
and for this particular lattice depth encompasses much of the non-interacting wavefunction. (iii) Double microwave shielding with uncompensated dipolar interactions.
This results not only in a hole in the wavefunction due to shielding,
but also an anisotropic amplitude outside as the molecules localize along directions for which the anisotropic dipolar interaction is attractive. 
The length scale of the optical lattice potential is indicated for comparison.

\begin{figure}
    \centering
    \input{overlaps_nodipole} 
    \input{overlaps_with_dipole} 
    \caption{{\bf Wannier decomposition of molecule-molecule wavefunctions}. Projection of the interacting two-body ground-state wave function $\psi_{0}$ onto non-interacting Wannier states $w_i$ as a function of lattice depth for NaCs molecules. \textbf{a} considering the repulsive van der Waals interactions, while dipolar interactions are compensated. The largest overlap is with the lowest band, indicating that the interacting state is predominantly described by the lowest-band Wannier component, with smaller contributions from higher bands. \textbf{b}  Including dipolar interactions. Results are shown for NaCs molecules at fixed $d_{\mathrm{eff}}^{2}= 0.19~\mathrm{Debye}^{2} $. The higher-band overlaps are larger than in the purely van der Waals case, but the dominant contribution remains the lowest Wannier band.}
    \label{Fig:overlaps}
\end{figure}
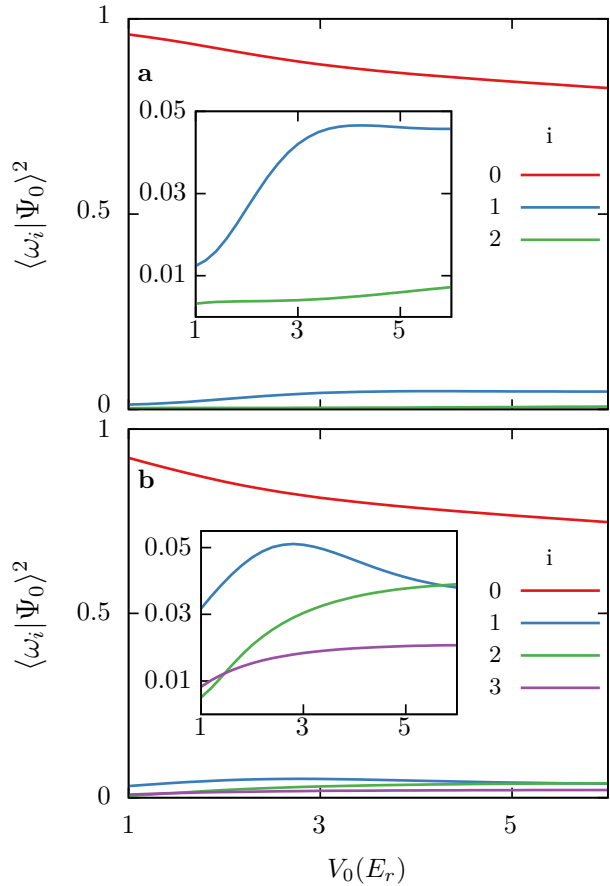

Next, we consider the modification of the interacting two-molecule wavefunction more quantitatively by evaluating the overlap of wavefunctions in the interacting and non-interacting cases.
The non-interacting wavefunctions are ordered by their relative contribution.
For the case of double microwave shielding with compensated dipolar interactions,
the resulting overlaps are shown in Fig.~\ref{Fig:overlaps}\textbf{a}.
The dominant contribution comes from the lowest band over the range of lattice depths considered, while higher-band contributions remain comparatively small but not negligible.
This indicates that, despite the strong modification of the wavefunction inside the radius of the shield, the interacting state is still largely captured by the lowest-band Wannier description.
In the case of uncompensated dipolar interactions, shown in Fig.~\ref{Fig:overlaps}\textbf{b},
we see that the higher-band contributions are somewhat enhanced, but the lowest band remains dominant.

The results above establish the tunability of both $U$ and $V_{NN}$ by tuning induced dipolar interaction and lattice depth.
In the main text Fig~3, we show that indeed comprehensive simultaneous control of $U/t$ and $V_{NN}/t$ is realized.

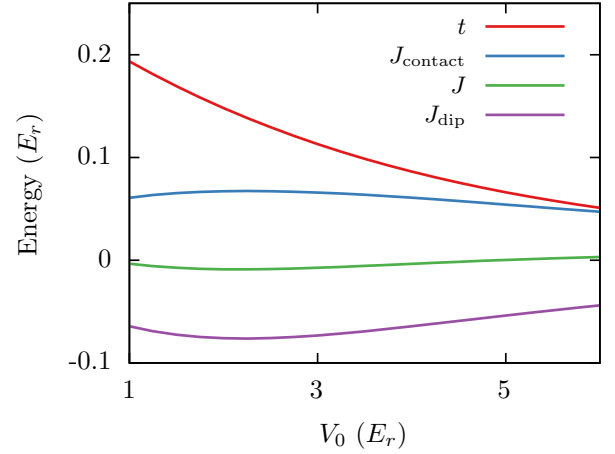
\begin{figure}
     \centering
     \input{induce_t} 
     \caption{{\bf Estimate of the density-induced tunneling} as a function of the lattice depth, split out in contributions of the dipolar interactions, a contact interactions, and compared to the single particle tunneling, see text. Results are shown for NaCs molecules with $d_{\mathrm{eff}}^2 = 0.19$~$\mathrm{Debye}^2$}
     \label{Fig:induce_tunneling}
\end{figure}

We comment on the possibility of density-induced or density-assisted tunneling of the form $-T \sum_{\langle i,j\rangle} a^\dagger_i  a^\dagger_i a_i a_j + h.c.$ in the realization of an extended Hubbard model with ultracold shielded molecules \cite{sowinski,maik2013density}.
The density-induced tunneling can be computed as
\begin{align}
T+t =& -\int \Psi_0(\bm{r}_1,\bm{r}_2) \left[\hat{h}_0(\bm{r}_1)+\hat{h}_0(\bm{r}_2) + V\left(\bm{r}_1-\bm{r}_2\right) \right] \nonumber \\
&\times  w_i(\bm{r}_1) w_{j}(\bm{r}_2)\ d\bm{r}_1\ d\bm{r}_2,
\end{align}
where $i$ and $j$ are nearest-neighbor sites.
The wavefunction analysis above showed that the on-site pair wavefunction, $\Psi_0(\bm{r}_1,\bm{r}_2),$ can be approximated by the wavefunction for non-interacting molecules, $c_0 w_i(\bm{r}_1)w_i(\bm{r}_2)$, with coefficient $c_0=\langle \phi_0 | \Psi_0 \rangle$ close to unity.
The single particle Hamiltonian does not coupled Wannier functions corresponding to different bands, leading to a contribution $(c_0-1)t$ to the density induced tunneling.
For the interaction term
\begin{align}
J = \int w_i(\bm{r}_1)^2\ V_\mathrm{int}(\bm{r}_1-\bm{r}_2) w_i(\bm{r}_2) w_{j}(\bm{r}_2)\ d\bm{r}_1\ d\bm{r}_2,
\end{align}
which neglects corrections of order $1-c_0$ due to contributions of higher bands.
This motivates estimating the density-induced tunneling as $T=(c_0-1) t - J$.
We split the interaction in a dipole-dipole and a contact contribution with $a_s$ given by Eq.~\eqref{eq:as}.
As shown in Figure~\ref{Fig:induce_tunneling}, these contributions are all comparable in the range of lattice depths shown, which is the same range of depths in which we demonstrated tunability of $U/t$ and $V_{\mathrm{NN}}/t$, see Fig.~3 of the main text.
We note that we consider the contribution of the contact interaction to be a useful estimate that indicates the contribution of the shielding core to density-induced tunneling can be substantial.
However, this estimate is likely not quantitatively accurate as we have shown the pseudopotential approximation in applicable to due the long range of the shielding interaction.
Quantitative calculation of the density-induced tunneling therefore would require solving numerically the problem of two molecules in two (or more) sites,
which is beyond the scope of the present paper.

\section{Characterizing the interaction microscope}

In this section, we further investigate the \textit{interaction microscope} enabled by tunable dipolar interactions in ultracold shielded molecules. We first discuss the   effective inter-hole and dipolar potentials we used to demonstrate the technique. We then motivate the $\ell^2$ norm as a metric  to quantify the difference between inter-hole density profiles and efficacy of the interaction microscope, which is then used to illustrate the temperature-dependence of the interaction microscope, showcasing that it survives to experimentally relevant temperatures, and the investigate the robustness of the interaction microscope for all effective inter-hole potentials considered.

\subsection{Effective inter-hole and dipolar potentials}

We consider four effective inter-hole potential functions, shown in Fig. 4\textbf{a},   
\begin{align}
    \mathcal{V}_{\mathrm{eff}}^{(1)}(r) &= -\mathcal{W}_\mathrm{eff}^{(1)} \exp\big( - r^2 \big) + \delta_{r,0} \mathcal{W}_{\mathrm{hc}}
    \label{eq:eff_pot_1} \\
    \mathcal{V}_{\mathrm{eff}}^{(2)}(r) &= - \frac{\mathcal{W}_\mathrm{eff}^{(2)}}{\sqrt{1 + r^2}} + \delta_{r,0} \mathcal{W}_{\mathrm{hc}}
    \label{eq:eff_pot_2} \\
    \mathcal{V}_{\mathrm{eff}}^{(3)}(r) &= - \frac{\mathcal{W}_\mathrm{eff}^{(3)}}{1+r^2 } + \delta_{r,0} \mathcal{W}_{\mathrm{hc}} 
    \label{eq:eff_pot_3} \\
    \mathcal{V}_{\mathrm{eff}}^{(4)}(r) &= -\mathcal{W}_\mathrm{eff}^{(4)} \exp \big( - \frac{r^2}{5^2} \big) + \delta_{r,0} \mathcal{W}_{\mathrm{hc}}
    \label{eq:eff_pot_4}
\end{align}
where $\mathcal{W}_\mathrm{eff}^{(\alpha)}$ sets the strength of the effective potential $\mathcal{V}_{\mathrm{eff}}^{(\alpha)}$, and $\mathcal{W}_{\mathrm{hc}}/t=1000$ is effectively infinite, creating a hardcore repulsion. We specifically choose to consider a long-range potential in $\mathcal{V}_{\mathrm{eff}}^{(4)}$ by re-scaling $r$ by $5$.

For the main text results, $\mathcal{W}_{\mathrm{eff}}^{(\alpha)}/t = 4$ for all $\alpha$.  This  ensures that bound states form, that the potentials are not much deeper than required to form these bound states, and that the resulting bound states are difficult to distinguish without the interaction microscope. Effective potentials one through four presented above and in the main text require a $\mathcal{W}_{\mathrm{eff}}^{(\alpha)}/t$ of approximately 3.1, 0.9, 1.5, and 0.4 respectively to develop  bound states of length $\xi \leq 5$ and 4.3, 3.3, 3.4, and 3.9, respectively, for $\xi \leq 2$, where $\xi \equiv \sqrt{\langle r^2 \rangle}$.

The total inter-hole potential is ${\mathcal V}^{(\alpha)}_\mathrm{eff}+V_{\text{dip}}$ where  
the dipolar off-site interaction is
\begin{equation}
    V_{\mathrm{dip}}(r) = \frac{V_{\mathrm{NN}}}{r^3}
\end{equation}
for $r\neq 0$.

\begin{figure}
    \centering
    \includegraphics[width=\linewidth]{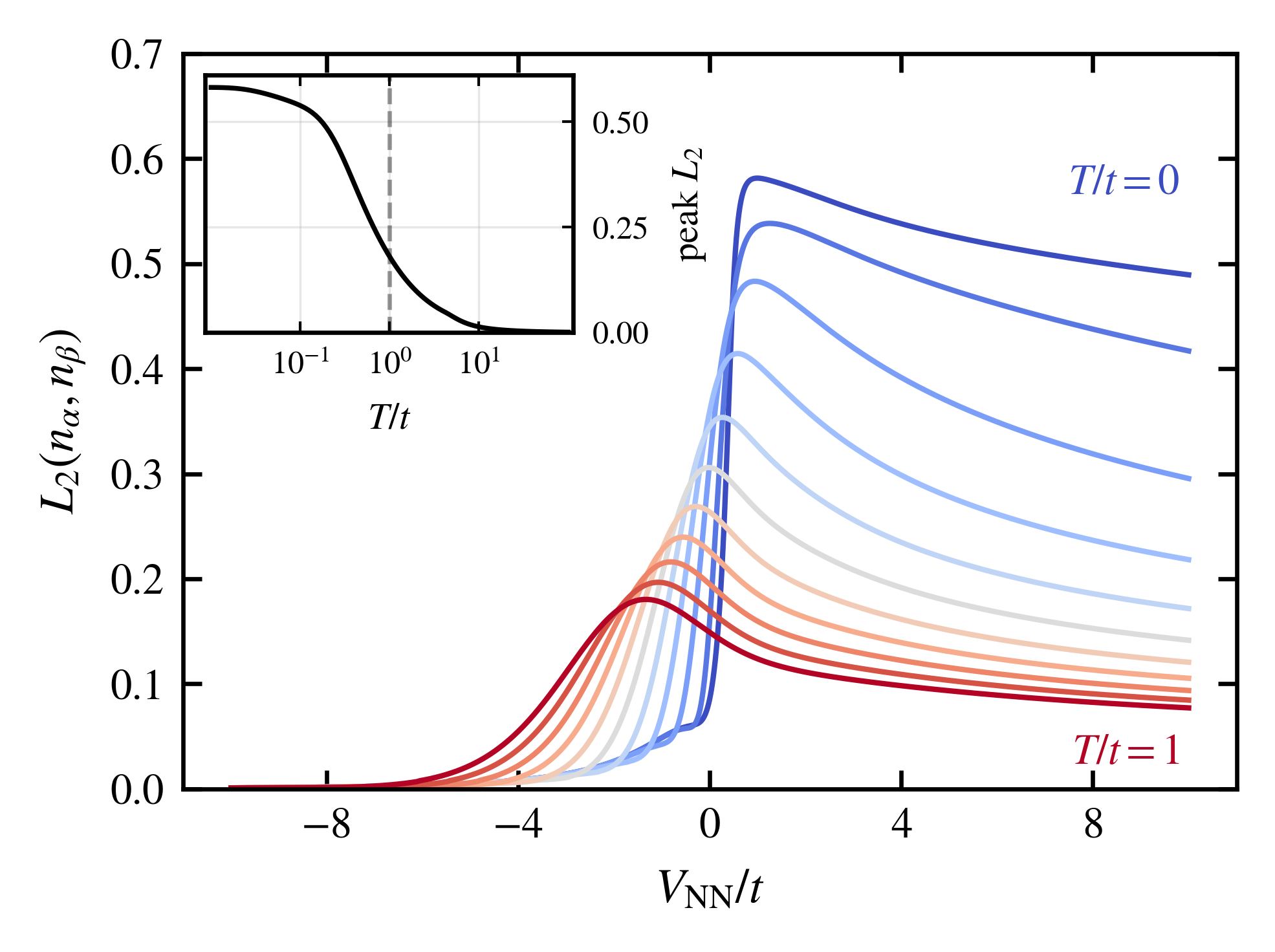}
    \caption{\textbf{The interaction microscope at finite temperatures.} The $L_2$ distance between the thermal relative density profiles for the pair of effective inter-hole potentials $\mathcal{V}_{\mathrm{eff}}^{(1)}$ and $\mathcal{V}_{\mathrm{eff}}^{(2)}$ at finite temperatures (increments of $T/t = 0.1$), showing the enhancement survives to temperatures relevant for quantum simulation ($T/t \lesssim 1$). Inset: the peak $L_2$ distance as a function of temperature. Results are for a 20-site lattice and potential coefficients $\mathcal{W}_\mathrm{eff}^{(1-4)}/t = 4$.}
    \label{fig:finite_T_interaction_microscope}
\end{figure} 

\begin{figure}
    \centering
    \includegraphics[width=\linewidth]{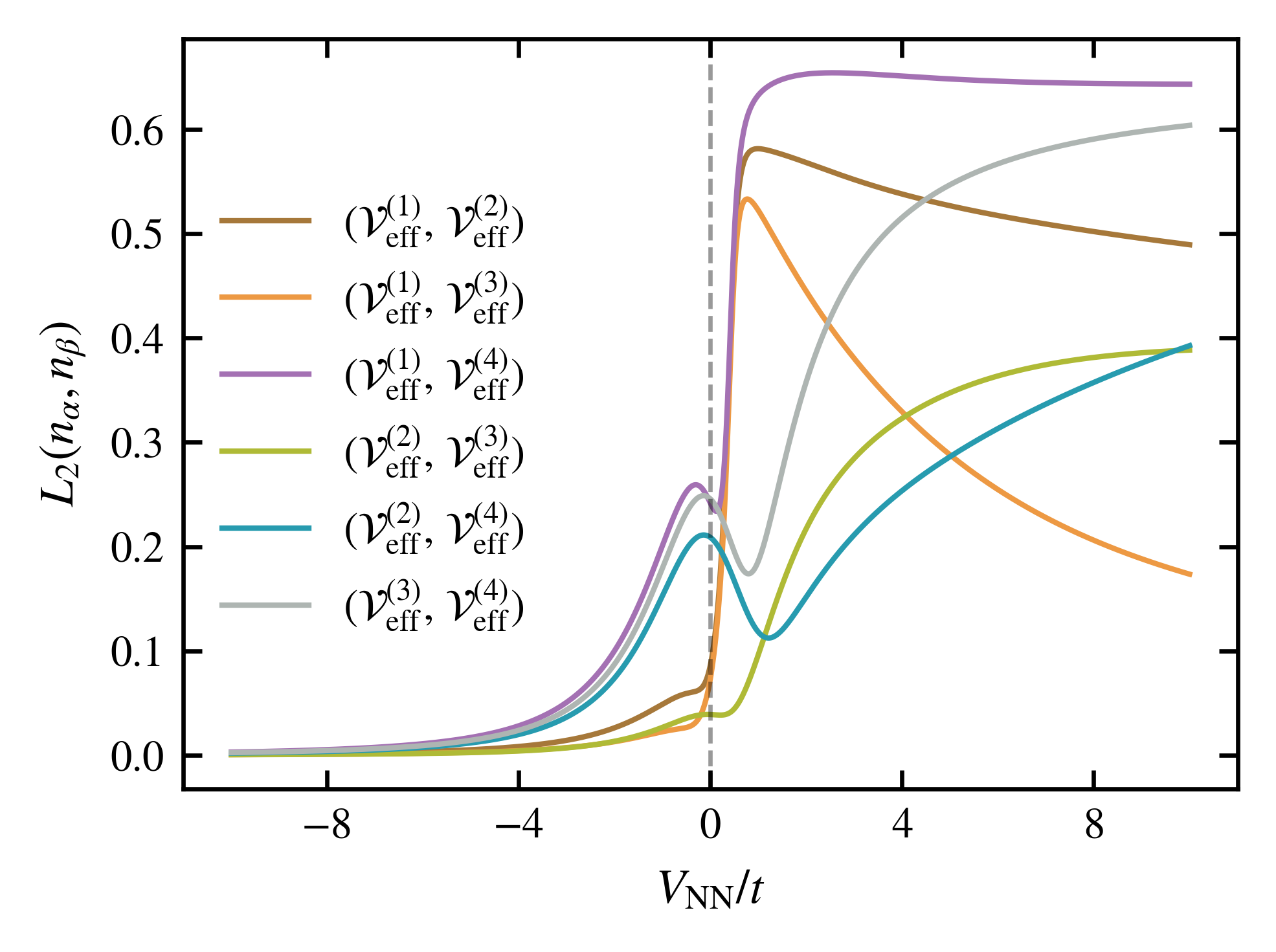}
    \caption{\textbf{Robustness of the interaction microscope for different pairs of effective potentials.} The $L_2$ distance between the respective ground state relative density profiles for each pair of toy effective inter-hole potentials we consider. In particular, note that the peak distinguishability for all pairs occurs at $V_{\text{NN}}/t \neq 0$. Results are for a 20-site lattice and potential coefficients $\mathcal{W}_\mathrm{eff}^{(1-4)}/t = 4$. } 
    \label{fig:pairs_interaction_microscope}
\end{figure}

\subsection{Using the $\ell_2$norm  to measure  distinguishability}

We use  $L_2 (n_\alpha, n_\beta) = \sqrt{\sum_j \big[  n_\alpha(r_j) -n_\beta(r_j) \big]^2 }$ to characterize the difference of two $n_{\alpha}$,
where $n_{\alpha}(r_j)$ is the probability that two holes are separated by $j$ sites, for potentials  $V_{\mathrm{eff}}^{(\alpha)}$.  This is a natural measure of distinguishability, as it provides, for example,   the $\chi^2$ statistic used in hypothesis testing: the $p$-value is a function of  $\chi^2 \propto L_2(n_\alpha,n_\beta)$ and gives the probability of observing a discrepancy at least as large as $L_2$ if $n_\alpha$ and $n_\beta$ arose from the same underlying effective potential.

\subsection{Temperature-dependence of the interaction microscope}

The interaction microscope requires low enough temperatures for the inter-dopant potential to affect hole pairs. Here we investigate the temperature required. Fig.~\ref{fig:finite_T_interaction_microscope} shows $L_2$ versus $V_{\mathrm{NN}}/t$ computed numerically at different temperatures  for our potentials. The $n_\alpha(r)$ are computed as thermal averages $\langle O\rangle_\beta = \text{Tr}[O\exp(-\beta H)] = \sum_{n} \langle n | O | n \rangle \exp(-\beta E_n)$ with $\beta = 1/T$ (we set $k_B=1$). For $T \gg t$, the interaction microscope clearly has no effect, since the inter-hole potential does not affect the physics. 

Figure~\ref{fig:finite_T_interaction_microscope} shows that the interaction microscope remains easily able to distinguish the pair of effective potentials $\mathcal{V}_{\mathrm{eff}}^{(1)}$ and $\mathcal{V}_{\mathrm{eff}}^{(2)}$ up to $T=t$ (the same pair as compared in the main text). As temperature is increased, the maximum of the $L_2$ decreases somewhat but still attains a sizable maximum (at least 0.25) at a non-zero value of $V_{\text{NN}}$. The inset of Fig.~\ref{fig:finite_T_interaction_microscope} shows the maximal value of $\ell_2$ versus $T/t$ on a logarithmic scale. An $L_2$ greater than about 0.1 persists to $T$ several times $t$.
Similar conclusions hold for all pairs of effective potentials considered, as shown in the next section. 

\subsection{Interaction microscope robustness for different effective potential pairs \label{app:robustness}}

A natural question to pose is whether the interaction microscope's capabilities illustrated in Fig. 4 and the previous section are fine-tuned to the pair of potentials presented, or are broadly useful.

Fig. \ref{fig:pairs_interaction_microscope} shows the $L_2$ versus $V_{\mathrm{NN}}/t$  for all six pairs.
This illustrates that all potentials are made significantly more distinguishable by adding $V_{\mathrm{NN}} \neq 0$. Note that the maximal $\ell^2$ distance for all six pairs of potentials occurs on the positive $V_{\mathrm{NN}}$ side. This is because a key mechanism driving the distinguishability is that a maximum in distinguishability is attained when the dipolar interaction causes a bound-unbound transition in one of the potentials being compared.

\end{document}

%% file: u_r6_3d.tex
% GNUPLOT: LaTeX picture with Postscript
\begingroup
  \makeatletter
  \providecommand\color[2][]{%
    \GenericError{(gnuplot) \space\space\space\@spaces}{%
      Package color not loaded in conjunction with
      terminal option `colourtext'%
    }{See the gnuplot documentation for explanation.%
    }{Either use 'blacktext' in gnuplot or load the package
      color.sty in LaTeX.}%
    \renewcommand\color[2][]{}%
  }%
  \providecommand\includegraphics[2][]{%
    \GenericError{(gnuplot) \space\space\space\@spaces}{%
      Package graphicx or graphics not loaded%
    }{See the gnuplot documentation for explanation.%
    }{The gnuplot epslatex terminal needs graphicx.sty or graphics.sty.}%
    \renewcommand\includegraphics[2][]{}%
  }%
  \providecommand\rotatebox[2]{#2}%
  \@ifundefined{ifGPcolor}{%
    \newif\ifGPcolor
    \GPcolortrue
  }{}%
  \@ifundefined{ifGPblacktext}{%
    \newif\ifGPblacktext
    \GPblacktextfalse
  }{}%
  % define a \g@addto@macro without @ in the name:
  \let\gplgaddtomacro\g@addto@macro
  % define empty templates for all commands taking text:
  \gdef\gplbacktext{}%
  \gdef\gplfronttext{}%
  \makeatother
  \ifGPblacktext
    % no textcolor at all
    \def\colorrgb#1{}%
    \def\colorgray#1{}%
  \else
    % gray or color?
    \ifGPcolor
      \def\colorrgb#1{\color[rgb]{#1}}%
      \def\colorgray#1{\color[gray]{#1}}%
      \expandafter\def\csname LTw\endcsname{\color{white}}%
      \expandafter\def\csname LTb\endcsname{\color{black}}%
      \expandafter\def\csname LTa\endcsname{\color{black}}%
      \expandafter\def\csname LT0\endcsname{\color[rgb]{1,0,0}}%
      \expandafter\def\csname LT1\endcsname{\color[rgb]{0,1,0}}%
      \expandafter\def\csname LT2\endcsname{\color[rgb]{0,0,1}}%
      \expandafter\def\csname LT3\endcsname{\color[rgb]{1,0,1}}%
      \expandafter\def\csname LT4\endcsname{\color[rgb]{0,1,1}}%
      \expandafter\def\csname LT5\endcsname{\color[rgb]{1,1,0}}%
      \expandafter\def\csname LT6\endcsname{\color[rgb]{0,0,0}}%
      \expandafter\def\csname LT7\endcsname{\color[rgb]{1,0.3,0}}%
      \expandafter\def\csname LT8\endcsname{\color[rgb]{0.5,0.5,0.5}}%
    \else
      % gray
      \def\colorrgb#1{\color{black}}%
      \def\colorgray#1{\color[gray]{#1}}%
      \expandafter\def\csname LTw\endcsname{\color{white}}%
      \expandafter\def\csname LTb\endcsname{\color{black}}%
      \expandafter\def\csname LTa\endcsname{\color{black}}%
      \expandafter\def\csname LT0\endcsname{\color{black}}%
      \expandafter\def\csname LT1\endcsname{\color{black}}%
      \expandafter\def\csname LT2\endcsname{\color{black}}%
      \expandafter\def\csname LT3\endcsname{\color{black}}%
      \expandafter\def\csname LT4\endcsname{\color{black}}%
      \expandafter\def\csname LT5\endcsname{\color{black}}%
      \expandafter\def\csname LT6\endcsname{\color{black}}%
      \expandafter\def\csname LT7\endcsname{\color{black}}%
      \expandafter\def\csname LT8\endcsname{\color{black}}%
    \fi
  \fi
    \setlength{\unitlength}{0.0500bp}%
    \ifx\gptboxheight\undefined%
      \newlength{\gptboxheight}%
      \newlength{\gptboxwidth}%
      \newsavebox{\gptboxtext}%
    \fi%
    \setlength{\fboxrule}{0.5pt}%
    \setlength{\fboxsep}{1pt}%
    \definecolor{tbcol}{rgb}{1,1,1}%
\begin{picture}(4176.00,4464.00)%
    \gplgaddtomacro\gplbacktext{%
      \csname LTb\endcsname%%
      \put(327,2589){\makebox(0,0)[r]{\strut{}0}}%
      \put(327,2975){\makebox(0,0)[r]{\strut{}}}%
      \put(327,3360){\makebox(0,0)[r]{\strut{}5}}%
      \put(327,3746){\makebox(0,0)[r]{\strut{}}}%
      \put(327,4132){\makebox(0,0)[r]{\strut{}10}}%
      \put(1656,2413){\makebox(0,0){\strut{}\normalsize 0.4}}%
      \put(2852,2413){\makebox(0,0){\strut{}\normalsize 0.8}}%
      \put(4049,2413){\makebox(0,0){\strut{}\normalsize 1.2}}%
      \put(603,4181){\makebox(0,0)[l]{\strut{}\normalsize \textbf{a}}}%
    }%
    \gplgaddtomacro\gplfronttext{%
      \csname LTb\endcsname%%
      \put(-146,3514){\rotatebox{-270}{\makebox(0,0){\strut{}\normalsize $U\ (E_{r})$}}}%
      \put(2254,2282){\makebox(0,0){\strut{}\normalsize $r_{6}/\lambda$}}%
      \csname LTb\endcsname%%
      \put(3411,3507){\makebox(0,0)[r]{\strut{}CsAg}}%
      \csname LTb\endcsname%%
      \put(3411,3287){\makebox(0,0)[r]{\strut{}KAg}}%
      \csname LTb\endcsname%%
      \put(3411,3067){\makebox(0,0)[r]{\strut{}NaCs}}%
      \csname LTb\endcsname%%
      \put(3411,2847){\makebox(0,0)[r]{\strut{}NaK}}%
    }%
    \gplgaddtomacro\gplbacktext{%
      \csname LTb\endcsname%%
      \put(327,521){\makebox(0,0)[r]{\strut{}\normalsize -2}}%
      \put(327,1116){\makebox(0,0)[r]{\strut{}\normalsize 0}}%
      \put(327,1711){\makebox(0,0)[r]{\strut{}\normalsize 2}}%
      \put(1011,47){\makebox(0,0){\strut{}\normalsize -0.5}}%
      \put(2392,47){\makebox(0,0){\strut{}\normalsize 0}}%
      \put(3773,47){\makebox(0,0){\strut{}\normalsize 0.5}}%
      \csname LTb\endcsname%%
      \put(603,1758){\makebox(0,0)[l]{\strut{}\normalsize \textbf{b}}}%
    }%
    \gplgaddtomacro\gplfronttext{%
      \csname LTb\endcsname%%
      \put(-146,1115){\rotatebox{-270}{\makebox(0,0){\strut{}\normalsize $U\ (E_{r})$}}}%
      \put(2254,-107){\makebox(0,0){\strut{}\normalsize $d_{\mathrm{eff}}^{2}\ \left( \mathrm{Debye}^{2} \right)$}}%
    }%
    \gplbacktext
    \put(0,0){\includegraphics[width={208.80bp},height={223.20bp}]{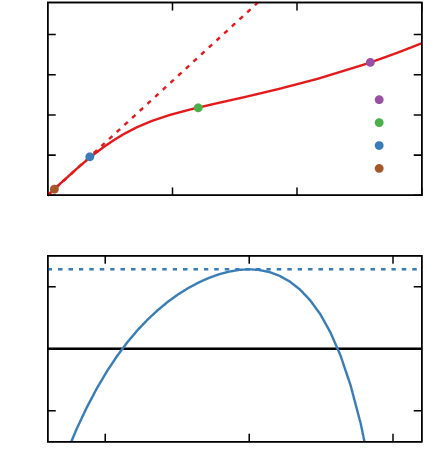}}%
    \gplfronttext
  \end{picture}%
\endgroup

%% file: heatmaps.tex
% GNUPLOT: LaTeX picture with Postscript
\begingroup
  \makeatletter
  \providecommand\color[2][]{%
    \GenericError{(gnuplot) \space\space\space\@spaces}{%
      Package color not loaded in conjunction with
      terminal option `colourtext'%
    }{See the gnuplot documentation for explanation.%
    }{Either use 'blacktext' in gnuplot or load the package
      color.sty in LaTeX.}%
    \renewcommand\color[2][]{}%
  }%
  \providecommand\includegraphics[2][]{%
    \GenericError{(gnuplot) \space\space\space\@spaces}{%
      Package graphicx or graphics not loaded%
    }{See the gnuplot documentation for explanation.%
    }{The gnuplot epslatex terminal needs graphicx.sty or graphics.sty.}%
    \renewcommand\includegraphics[2][]{}%
  }%
  \providecommand\rotatebox[2]{#2}%
  \@ifundefined{ifGPcolor}{%
    \newif\ifGPcolor
    \GPcolortrue
  }{}%
  \@ifundefined{ifGPblacktext}{%
    \newif\ifGPblacktext
    \GPblacktextfalse
  }{}%
  % define a \g@addto@macro without @ in the name:
  \let\gplgaddtomacro\g@addto@macro
  % define empty templates for all commands taking text:
  \gdef\gplbacktext{}%
  \gdef\gplfronttext{}%
  \makeatother
  \ifGPblacktext
    % no textcolor at all
    \def\colorrgb#1{}%
    \def\colorgray#1{}%
  \else
    % gray or color?
    \ifGPcolor
      \def\colorrgb#1{\color[rgb]{#1}}%
      \def\colorgray#1{\color[gray]{#1}}%
      \expandafter\def\csname LTw\endcsname{\color{white}}%
      \expandafter\def\csname LTb\endcsname{\color{black}}%
      \expandafter\def\csname LTa\endcsname{\color{black}}%
      \expandafter\def\csname LT0\endcsname{\color[rgb]{1,0,0}}%
      \expandafter\def\csname LT1\endcsname{\color[rgb]{0,1,0}}%
      \expandafter\def\csname LT2\endcsname{\color[rgb]{0,0,1}}%
      \expandafter\def\csname LT3\endcsname{\color[rgb]{1,0,1}}%
      \expandafter\def\csname LT4\endcsname{\color[rgb]{0,1,1}}%
      \expandafter\def\csname LT5\endcsname{\color[rgb]{1,1,0}}%
      \expandafter\def\csname LT6\endcsname{\color[rgb]{0,0,0}}%
      \expandafter\def\csname LT7\endcsname{\color[rgb]{1,0.3,0}}%
      \expandafter\def\csname LT8\endcsname{\color[rgb]{0.5,0.5,0.5}}%
    \else
      % gray
      \def\colorrgb#1{\color{black}}%
      \def\colorgray#1{\color[gray]{#1}}%
      \expandafter\def\csname LTw\endcsname{\color{white}}%
      \expandafter\def\csname LTb\endcsname{\color{black}}%
      \expandafter\def\csname LTa\endcsname{\color{black}}%
      \expandafter\def\csname LT0\endcsname{\color{black}}%
      \expandafter\def\csname LT1\endcsname{\color{black}}%
      \expandafter\def\csname LT2\endcsname{\color{black}}%
      \expandafter\def\csname LT3\endcsname{\color{black}}%
      \expandafter\def\csname LT4\endcsname{\color{black}}%
      \expandafter\def\csname LT5\endcsname{\color{black}}%
      \expandafter\def\csname LT6\endcsname{\color{black}}%
      \expandafter\def\csname LT7\endcsname{\color{black}}%
      \expandafter\def\csname LT8\endcsname{\color{black}}%
    \fi
  \fi
    \setlength{\unitlength}{0.0500bp}%
    \ifx\gptboxheight\undefined%
      \newlength{\gptboxheight}%
      \newlength{\gptboxwidth}%
      \newsavebox{\gptboxtext}%
    \fi%
    \setlength{\fboxrule}{0.5pt}%
    \setlength{\fboxsep}{1pt}%
\begin{picture}(10080.00,3456.00)%
    \gplgaddtomacro\gplbacktext{%
    }%
    \gplgaddtomacro\gplfronttext{%
      \csname LTb\endcsname%%
      \put(449,363){\makebox(0,0){\strut{}\normalsize 1}}%
      \put(1581,363){\makebox(0,0){\strut{}\normalsize 3}}%
      \put(2713,363){\makebox(0,0){\strut{}\normalsize 5}}%
      \put(1864,77){\makebox(0,0){\strut{}\normalsize $V_{0}\ (E_{r})$}}%
      \put(241,877){\makebox(0,0)[r]{\strut{}\normalsize -0.3}}%
      \put(241,1745){\makebox(0,0)[r]{\strut{}\normalsize 0}}%
      \put(241,2615){\makebox(0,0)[r]{\strut{}\normalsize 0.3}}%
      \put(-221,1630){\rotatebox{-270}{\makebox(0,0){\strut{}\normalsize $d_{\mathrm{eff}}^{2}\ \left( \mathrm{Debye}^{2} \right)$}}}%
      \put(732,3079){\makebox(0,0){\strut{}\small -10}}%
      \put(1299,3079){\makebox(0,0){\strut{}\small -5}}%
      \put(1866,3079){\makebox(0,0){\strut{}\small 0}}%
      \put(2433,3079){\makebox(0,0){\strut{}\small 5}}%
      \put(3000,3079){\makebox(0,0){\strut{}\small 10}}%
      \put(1866,3365){\makebox(0,0){\strut{}\normalsize $U/t$}}%
      \put(789,2627){\makebox(0,0)[l]{\strut{}\normalsize \textbf{a}}}%
    }%
    \gplgaddtomacro\gplbacktext{%
    }%
    \gplgaddtomacro\gplfronttext{%
      \csname LTb\endcsname%%
      \put(3372,363){\makebox(0,0){\strut{}\normalsize 1}}%
      \put(4504,363){\makebox(0,0){\strut{}\normalsize 3}}%
      \put(5636,363){\makebox(0,0){\strut{}\normalsize 5}}%
      \put(4787,77){\makebox(0,0){\strut{}\normalsize $V_{0}\ (E_{r})$}}%
      \put(3164,877){\makebox(0,0)[r]{\strut{}}}%
      \put(3164,1745){\makebox(0,0)[r]{\strut{}}}%
      \put(3164,2615){\makebox(0,0)[r]{\strut{}}}%
      \put(3655,3079){\makebox(0,0){\strut{}\small -10}}%
      \put(4222,3079){\makebox(0,0){\strut{}\small -5}}%
      \put(4789,3079){\makebox(0,0){\strut{}\small 0}}%
      \put(5356,3079){\makebox(0,0){\strut{}\small 5}}%
      \put(5923,3079){\makebox(0,0){\strut{}\small 10}}%
      \put(4789,3365){\makebox(0,0){\strut{}\normalsize $V_{\mathrm{NN}}/t$}}%
      \put(3712,2627){\makebox(0,0)[l]{\strut{}\normalsize \textbf{b}}}%
    }%
    \gplgaddtomacro\gplbacktext{%
      \csname LTb\endcsname%%
      \put(6874,459){\makebox(0,0)[r]{\strut{}\normalsize -10}}%
      \put(6874,1623){\makebox(0,0)[r]{\strut{}\normalsize 0}}%
      \put(6874,2786){\makebox(0,0)[r]{\strut{}\normalsize 10}}%
      \put(7006,327){\makebox(0,0){\strut{}\normalsize 1}}%
      \put(8097,327){\makebox(0,0){\strut{}\normalsize 3}}%
      \put(9188,327){\makebox(0,0){\strut{}\normalsize 5}}%
      \colorrgb{0.22,0.49,0.72}%%
      \put(7033,2344){\makebox(0,0)[l]{\strut{}\normalsize $\frac{V_{\mathrm{NN}}}{t} = 2$ }}%
      \put(8506,1623){\makebox(0,0)[l]{\strut{}\normalsize $\frac{V_{\mathrm{NN}}}{t} = 5$ }}%
      \colorrgb{1.00,0.50,0.00}%%
      \put(7115,924){\makebox(0,0)[l]{\strut{}\normalsize $\frac{U}{t} = 0$ }}%
      \colorrgb{0.60,0.31,0.64}%%
      \put(8833,2228){\makebox(0,0)[l]{\strut{}\normalsize $\frac{U}{t} = 0$ }}%
    }%
    \gplgaddtomacro\gplfronttext{%
      \csname LTb\endcsname%%
      \put(6559,1622){\rotatebox{-270}{\makebox(0,0){\strut{}\normalsize $U/t, V_{\mathrm{NN}}/t$}}}%
      \put(8369,85){\makebox(0,0){\strut{}\normalsize $V_{0}\ (E_{r})$}}%
      \csname LTb\endcsname%%
      \put(7333,2623){\makebox(0,0)[l]{\strut{}\normalsize \textbf{c}}}%
    }%
    \gplbacktext
    \put(0,0){\includegraphics{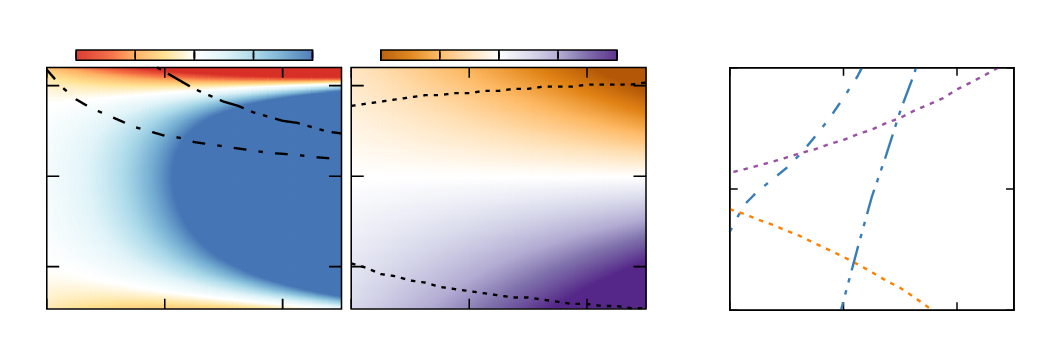}}%
    \gplfronttext
  \end{picture}%
\endgroup

%% file: uharm_r6.tex
% GNUPLOT: LaTeX picture with Postscript
\begingroup
  \makeatletter
  \providecommand\color[2][]{%
    \GenericError{(gnuplot) \space\space\space\@spaces}{%
      Package color not loaded in conjunction with
      terminal option `colourtext'%
    }{See the gnuplot documentation for explanation.%
    }{Either use 'blacktext' in gnuplot or load the package
      color.sty in LaTeX.}%
    \renewcommand\color[2][]{}%
  }%
  \providecommand\includegraphics[2][]{%
    \GenericError{(gnuplot) \space\space\space\@spaces}{%
      Package graphicx or graphics not loaded%
    }{See the gnuplot documentation for explanation.%
    }{The gnuplot epslatex terminal needs graphicx.sty or graphics.sty.}%
    \renewcommand\includegraphics[2][]{}%
  }%
  \providecommand\rotatebox[2]{#2}%
  \@ifundefined{ifGPcolor}{%
    \newif\ifGPcolor
    \GPcolortrue
  }{}%
  \@ifundefined{ifGPblacktext}{%
    \newif\ifGPblacktext
    \GPblacktextfalse
  }{}%
  % define a \g@addto@macro without @ in the name:
  \let\gplgaddtomacro\g@addto@macro
  % define empty templates for all commands taking text:
  \gdef\gplbacktext{}%
  \gdef\gplfronttext{}%
  \makeatother
  \ifGPblacktext
    % no textcolor at all
    \def\colorrgb#1{}%
    \def\colorgray#1{}%
  \else
    % gray or color?
    \ifGPcolor
      \def\colorrgb#1{\color[rgb]{#1}}%
      \def\colorgray#1{\color[gray]{#1}}%
      \expandafter\def\csname LTw\endcsname{\color{white}}%
      \expandafter\def\csname LTb\endcsname{\color{black}}%
      \expandafter\def\csname LTa\endcsname{\color{black}}%
      \expandafter\def\csname LT0\endcsname{\color[rgb]{1,0,0}}%
      \expandafter\def\csname LT1\endcsname{\color[rgb]{0,1,0}}%
      \expandafter\def\csname LT2\endcsname{\color[rgb]{0,0,1}}%
      \expandafter\def\csname LT3\endcsname{\color[rgb]{1,0,1}}%
      \expandafter\def\csname LT4\endcsname{\color[rgb]{0,1,1}}%
      \expandafter\def\csname LT5\endcsname{\color[rgb]{1,1,0}}%
      \expandafter\def\csname LT6\endcsname{\color[rgb]{0,0,0}}%
      \expandafter\def\csname LT7\endcsname{\color[rgb]{1,0.3,0}}%
      \expandafter\def\csname LT8\endcsname{\color[rgb]{0.5,0.5,0.5}}%
    \else
      % gray
      \def\colorrgb#1{\color{black}}%
      \def\colorgray#1{\color[gray]{#1}}%
      \expandafter\def\csname LTw\endcsname{\color{white}}%
      \expandafter\def\csname LTb\endcsname{\color{black}}%
      \expandafter\def\csname LTa\endcsname{\color{black}}%
      \expandafter\def\csname LT0\endcsname{\color{black}}%
      \expandafter\def\csname LT1\endcsname{\color{black}}%
      \expandafter\def\csname LT2\endcsname{\color{black}}%
      \expandafter\def\csname LT3\endcsname{\color{black}}%
      \expandafter\def\csname LT4\endcsname{\color{black}}%
      \expandafter\def\csname LT5\endcsname{\color{black}}%
      \expandafter\def\csname LT6\endcsname{\color{black}}%
      \expandafter\def\csname LT7\endcsname{\color{black}}%
      \expandafter\def\csname LT8\endcsname{\color{black}}%
    \fi
  \fi
    \setlength{\unitlength}{0.0500bp}%
    \ifx\gptboxheight\undefined%
      \newlength{\gptboxheight}%
      \newlength{\gptboxwidth}%
      \newsavebox{\gptboxtext}%
    \fi%
    \setlength{\fboxrule}{0.5pt}%
    \setlength{\fboxsep}{1pt}%
    \definecolor{tbcol}{rgb}{1,1,1}%
\begin{picture}(4176.00,3456.00)%
    \gplgaddtomacro\gplbacktext{%
      \csname LTb\endcsname%%
      \put(264,1257){\makebox(0,0)[r]{\strut{}\normalsize 1}}%
      \put(264,1963){\makebox(0,0)[r]{\strut{}\normalsize 2}}%
      \put(264,2670){\makebox(0,0)[r]{\strut{}\normalsize 3}}%
      \put(396,330){\makebox(0,0){\strut{}0}}%
      \put(1004,330){\makebox(0,0){\strut{}}}%
      \put(1612,330){\makebox(0,0){\strut{}2}}%
      \put(2219,330){\makebox(0,0){\strut{}}}%
      \put(2827,330){\makebox(0,0){\strut{}4}}%
      \put(3435,330){\makebox(0,0){\strut{}}}%
      \put(4043,330){\makebox(0,0){\strut{}6}}%
      \put(578,3047){\makebox(0,0)[l]{\strut{}\normalsize \textbf{a}}}%
    }%
    \gplgaddtomacro\gplfronttext{%
      \csname LTb\endcsname%%
      \put(-275,1892){\rotatebox{-270}{\makebox(0,0){\strut{}\normalsize $U\ (\hbar\omega)$}}}%
      \put(2219,110){\makebox(0,0){\strut{}\normalsize $r_{6}/a_{h}$}}%
      \csname LTb\endcsname%%
      \put(2201,2937){\makebox(0,0)[r]{\strut{}vdW,exact}}%
      \csname LTb\endcsname%%
      \put(2201,2717){\makebox(0,0)[r]{\strut{}Contact, peturbative}}%
      \csname LTb\endcsname%%
      \put(2201,2497){\makebox(0,0)[r]{\strut{}Contact, exact}}%
    }%
    \gplbacktext
    \put(0,0){\includegraphics[width={208.80bp},height={172.80bp}]{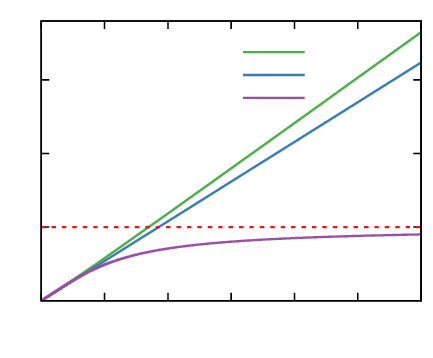}}%
    \gplfronttext
  \end{picture}%
\endgroup

%% file: delta_E_harmonic.tex
% GNUPLOT: LaTeX picture with Postscript
\begingroup
  \makeatletter
  \providecommand\color[2][]{%
    \GenericError{(gnuplot) \space\space\space\@spaces}{%
      Package color not loaded in conjunction with
      terminal option `colourtext'%
    }{See the gnuplot documentation for explanation.%
    }{Either use 'blacktext' in gnuplot or load the package
      color.sty in LaTeX.}%
    \renewcommand\color[2][]{}%
  }%
  \providecommand\includegraphics[2][]{%
    \GenericError{(gnuplot) \space\space\space\@spaces}{%
      Package graphicx or graphics not loaded%
    }{See the gnuplot documentation for explanation.%
    }{The gnuplot epslatex terminal needs graphicx.sty or graphics.sty.}%
    \renewcommand\includegraphics[2][]{}%
  }%
  \providecommand\rotatebox[2]{#2}%
  \@ifundefined{ifGPcolor}{%
    \newif\ifGPcolor
    \GPcolortrue
  }{}%
  \@ifundefined{ifGPblacktext}{%
    \newif\ifGPblacktext
    \GPblacktextfalse
  }{}%
  % define a \g@addto@macro without @ in the name:
  \let\gplgaddtomacro\g@addto@macro
  % define empty templates for all commands taking text:
  \gdef\gplbacktext{}%
  \gdef\gplfronttext{}%
  \makeatother
  \ifGPblacktext
    % no textcolor at all
    \def\colorrgb#1{}%
    \def\colorgray#1{}%
  \else
    % gray or color?
    \ifGPcolor
      \def\colorrgb#1{\color[rgb]{#1}}%
      \def\colorgray#1{\color[gray]{#1}}%
      \expandafter\def\csname LTw\endcsname{\color{white}}%
      \expandafter\def\csname LTb\endcsname{\color{black}}%
      \expandafter\def\csname LTa\endcsname{\color{black}}%
      \expandafter\def\csname LT0\endcsname{\color[rgb]{1,0,0}}%
      \expandafter\def\csname LT1\endcsname{\color[rgb]{0,1,0}}%
      \expandafter\def\csname LT2\endcsname{\color[rgb]{0,0,1}}%
      \expandafter\def\csname LT3\endcsname{\color[rgb]{1,0,1}}%
      \expandafter\def\csname LT4\endcsname{\color[rgb]{0,1,1}}%
      \expandafter\def\csname LT5\endcsname{\color[rgb]{1,1,0}}%
      \expandafter\def\csname LT6\endcsname{\color[rgb]{0,0,0}}%
      \expandafter\def\csname LT7\endcsname{\color[rgb]{1,0.3,0}}%
      \expandafter\def\csname LT8\endcsname{\color[rgb]{0.5,0.5,0.5}}%
    \else
      % gray
      \def\colorrgb#1{\color{black}}%
      \def\colorgray#1{\color[gray]{#1}}%
      \expandafter\def\csname LTw\endcsname{\color{white}}%
      \expandafter\def\csname LTb\endcsname{\color{black}}%
      \expandafter\def\csname LTa\endcsname{\color{black}}%
      \expandafter\def\csname LT0\endcsname{\color{black}}%
      \expandafter\def\csname LT1\endcsname{\color{black}}%
      \expandafter\def\csname LT2\endcsname{\color{black}}%
      \expandafter\def\csname LT3\endcsname{\color{black}}%
      \expandafter\def\csname LT4\endcsname{\color{black}}%
      \expandafter\def\csname LT5\endcsname{\color{black}}%
      \expandafter\def\csname LT6\endcsname{\color{black}}%
      \expandafter\def\csname LT7\endcsname{\color{black}}%
      \expandafter\def\csname LT8\endcsname{\color{black}}%
    \fi
  \fi
    \setlength{\unitlength}{0.0500bp}%
    \ifx\gptboxheight\undefined%
      \newlength{\gptboxheight}%
      \newlength{\gptboxwidth}%
      \newsavebox{\gptboxtext}%
    \fi%
    \setlength{\fboxrule}{0.5pt}%
    \setlength{\fboxsep}{1pt}%
    \definecolor{tbcol}{rgb}{1,1,1}%
\begin{picture}(4176.00,3456.00)%
    \gplgaddtomacro\gplbacktext{%
      \csname LTb\endcsname%%
      \put(264,621){\makebox(0,0)[r]{\strut{}$2$}}%
      \put(264,1783){\makebox(0,0)[r]{\strut{}$2.4$}}%
      \put(264,2944){\makebox(0,0)[r]{\strut{}$2.8$}}%
      \put(396,110){\makebox(0,0){\strut{}$0.01$}}%
      \put(1308,110){\makebox(0,0){\strut{}$0.1$}}%
      \put(2220,110){\makebox(0,0){\strut{}$1$}}%
      \put(3131,110){\makebox(0,0){\strut{}$10$}}%
      \put(4043,110){\makebox(0,0){\strut{}$100$}}%
      \put(578,3032){\makebox(0,0)[l]{\strut{}\normalsize \textbf{b}}}%
    }%
    \gplgaddtomacro\gplfronttext{%
      \csname LTb\endcsname%%
      \put(-275,1782){\rotatebox{-270}{\makebox(0,0){\strut{}\normalsize $\Delta E\ (\hbar\omega)$}}}%
      \put(2219,-110){\makebox(0,0){\strut{}\normalsize $r_{6}/a_{h}$}}%
    }%
    \gplbacktext
    \put(0,0){\includegraphics[width={208.80bp},height={172.80bp}]{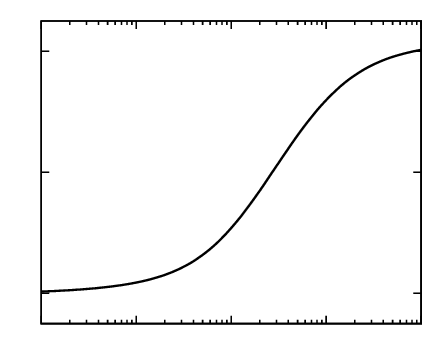}}%
    \gplfronttext
  \end{picture}%
\endgroup

%% file: scan_convergence_latex.tex
% GNUPLOT: LaTeX picture with Postscript
\begingroup
  \makeatletter
  \providecommand\color[2][]{%
    \GenericError{(gnuplot) \space\space\space\@spaces}{%
      Package color not loaded in conjunction with
      terminal option `colourtext'%
    }{See the gnuplot documentation for explanation.%
    }{Either use 'blacktext' in gnuplot or load the package
      color.sty in LaTeX.}%
    \renewcommand\color[2][]{}%
  }%
  \providecommand\includegraphics[2][]{%
    \GenericError{(gnuplot) \space\space\space\@spaces}{%
      Package graphicx or graphics not loaded%
    }{See the gnuplot documentation for explanation.%
    }{The gnuplot epslatex terminal needs graphicx.sty or graphics.sty.}%
    \renewcommand\includegraphics[2][]{}%
  }%
  \providecommand\rotatebox[2]{#2}%
  \@ifundefined{ifGPcolor}{%
    \newif\ifGPcolor
    \GPcolortrue
  }{}%
  \@ifundefined{ifGPblacktext}{%
    \newif\ifGPblacktext
    \GPblacktextfalse
  }{}%
  % define a \g@addto@macro without @ in the name:
  \let\gplgaddtomacro\g@addto@macro
  % define empty templates for all commands taking text:
  \gdef\gplbacktext{}%
  \gdef\gplfronttext{}%
  \makeatother
  \ifGPblacktext
    % no textcolor at all
    \def\colorrgb#1{}%
    \def\colorgray#1{}%
  \else
    % gray or color?
    \ifGPcolor
      \def\colorrgb#1{\color[rgb]{#1}}%
      \def\colorgray#1{\color[gray]{#1}}%
      \expandafter\def\csname LTw\endcsname{\color{white}}%
      \expandafter\def\csname LTb\endcsname{\color{black}}%
      \expandafter\def\csname LTa\endcsname{\color{black}}%
      \expandafter\def\csname LT0\endcsname{\color[rgb]{1,0,0}}%
      \expandafter\def\csname LT1\endcsname{\color[rgb]{0,1,0}}%
      \expandafter\def\csname LT2\endcsname{\color[rgb]{0,0,1}}%
      \expandafter\def\csname LT3\endcsname{\color[rgb]{1,0,1}}%
      \expandafter\def\csname LT4\endcsname{\color[rgb]{0,1,1}}%
      \expandafter\def\csname LT5\endcsname{\color[rgb]{1,1,0}}%
      \expandafter\def\csname LT6\endcsname{\color[rgb]{0,0,0}}%
      \expandafter\def\csname LT7\endcsname{\color[rgb]{1,0.3,0}}%
      \expandafter\def\csname LT8\endcsname{\color[rgb]{0.5,0.5,0.5}}%
    \else
      % gray
      \def\colorrgb#1{\color{black}}%
      \def\colorgray#1{\color[gray]{#1}}%
      \expandafter\def\csname LTw\endcsname{\color{white}}%
      \expandafter\def\csname LTb\endcsname{\color{black}}%
      \expandafter\def\csname LTa\endcsname{\color{black}}%
      \expandafter\def\csname LT0\endcsname{\color{black}}%
      \expandafter\def\csname LT1\endcsname{\color{black}}%
      \expandafter\def\csname LT2\endcsname{\color{black}}%
      \expandafter\def\csname LT3\endcsname{\color{black}}%
      \expandafter\def\csname LT4\endcsname{\color{black}}%
      \expandafter\def\csname LT5\endcsname{\color{black}}%
      \expandafter\def\csname LT6\endcsname{\color{black}}%
      \expandafter\def\csname LT7\endcsname{\color{black}}%
      \expandafter\def\csname LT8\endcsname{\color{black}}%
    \fi
  \fi
    \setlength{\unitlength}{0.0500bp}%
    \ifx\gptboxheight\undefined%
      \newlength{\gptboxheight}%
      \newlength{\gptboxwidth}%
      \newsavebox{\gptboxtext}%
    \fi%
    \setlength{\fboxrule}{0.5pt}%
    \setlength{\fboxsep}{1pt}%
    \definecolor{tbcol}{rgb}{1,1,1}%
\begin{picture}(4176.00,3456.00)%
    \gplgaddtomacro\gplbacktext{%
      \csname LTb\endcsname%%
      \put(435,751){\makebox(0,0)[r]{\strut{}\normalsize 50}}%
      \put(435,1917){\makebox(0,0)[r]{\strut{}\normalsize 100}}%
      \put(435,3083){\makebox(0,0)[r]{\strut{}\normalsize 150}}%
      \put(501,408){\makebox(0,0){\strut{}\normalsize 3}}%
      \put(1252,408){\makebox(0,0){\strut{}\normalsize 6}}%
      \put(2004,408){\makebox(0,0){\strut{}\normalsize 9}}%
      \put(641,3120){\makebox(0,0)[l]{\strut{}\normalsize \textbf{a}}}%
    }%
    \gplgaddtomacro\gplfronttext{%
      \csname LTb\endcsname%%
      \put(-66,1917){\rotatebox{-270}{\makebox(0,0){\strut{}\normalsize $\sqrt{\langle z^{2} \rangle}~(\mathrm{nm})$}}}%
      \put(1377,243){\makebox(0,0){\strut{}\normalsize $r_{\mathrm{max}}~ (a_{h})$}}%
    }%
    \gplgaddtomacro\gplbacktext{%
      \csname LTb\endcsname%%
      \put(2314,751){\makebox(0,0)[r]{\strut{}}}%
      \put(2314,1917){\makebox(0,0)[r]{\strut{}}}%
      \put(2314,3083){\makebox(0,0)[r]{\strut{}}}%
      \put(2380,408){\makebox(0,0){\strut{}\normalsize 3}}%
      \put(2964,408){\makebox(0,0){\strut{}\normalsize 6}}%
      \put(3549,408){\makebox(0,0){\strut{}\normalsize 9}}%
      \put(4133,408){\makebox(0,0){\strut{}\normalsize 12}}%
      \put(2520,3120){\makebox(0,0)[l]{\strut{}\normalsize \textbf{b}}}%
    }%
    \gplgaddtomacro\gplfronttext{%
      \csname LTb\endcsname%%
      \put(3256,243){\makebox(0,0){\strut{}\normalsize $r_{\mathrm{max}}~ (a_{h})$}}%
      \put(3749,3036){\makebox(0,0){\strut{}$V_{0}~(E_{r})$}}%
      \csname LTb\endcsname%%
      \put(3586,2858){\makebox(0,0)[r]{\strut{}5}}%
      \csname LTb\endcsname%%
      \put(3586,2704){\makebox(0,0)[r]{\strut{}10}}%
      \csname LTb\endcsname%%
      \put(3586,2550){\makebox(0,0)[r]{\strut{}15}}%
      \csname LTb\endcsname%%
      \put(3586,2396){\makebox(0,0)[r]{\strut{}20}}%
    }%
    \gplbacktext
    \put(0,0){\includegraphics[width={208.80bp},height={172.80bp}]{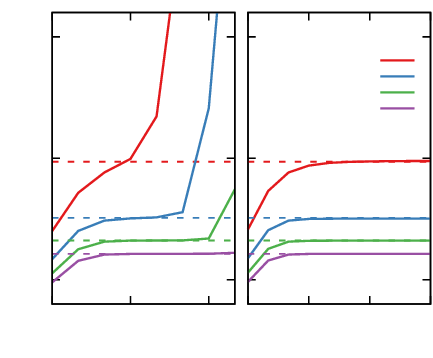}}%
    \gplfronttext
  \end{picture}%
\endgroup

%% file: scan_interactions_convergence.tex
% GNUPLOT: LaTeX picture with Postscript
\begingroup
  \makeatletter
  \providecommand\color[2][]{%
    \GenericError{(gnuplot) \space\space\space\@spaces}{%
      Package color not loaded in conjunction with
      terminal option `colourtext'%
    }{See the gnuplot documentation for explanation.%
    }{Either use 'blacktext' in gnuplot or load the package
      color.sty in LaTeX.}%
    \renewcommand\color[2][]{}%
  }%
  \providecommand\includegraphics[2][]{%
    \GenericError{(gnuplot) \space\space\space\@spaces}{%
      Package graphicx or graphics not loaded%
    }{See the gnuplot documentation for explanation.%
    }{The gnuplot epslatex terminal needs graphicx.sty or graphics.sty.}%
    \renewcommand\includegraphics[2][]{}%
  }%
  \providecommand\rotatebox[2]{#2}%
  \@ifundefined{ifGPcolor}{%
    \newif\ifGPcolor
    \GPcolortrue
  }{}%
  \@ifundefined{ifGPblacktext}{%
    \newif\ifGPblacktext
    \GPblacktextfalse
  }{}%
  % define a \g@addto@macro without @ in the name:
  \let\gplgaddtomacro\g@addto@macro
  % define empty templates for all commands taking text:
  \gdef\gplbacktext{}%
  \gdef\gplfronttext{}%
  \makeatother
  \ifGPblacktext
    % no textcolor at all
    \def\colorrgb#1{}%
    \def\colorgray#1{}%
  \else
    % gray or color?
    \ifGPcolor
      \def\colorrgb#1{\color[rgb]{#1}}%
      \def\colorgray#1{\color[gray]{#1}}%
      \expandafter\def\csname LTw\endcsname{\color{white}}%
      \expandafter\def\csname LTb\endcsname{\color{black}}%
      \expandafter\def\csname LTa\endcsname{\color{black}}%
      \expandafter\def\csname LT0\endcsname{\color[rgb]{1,0,0}}%
      \expandafter\def\csname LT1\endcsname{\color[rgb]{0,1,0}}%
      \expandafter\def\csname LT2\endcsname{\color[rgb]{0,0,1}}%
      \expandafter\def\csname LT3\endcsname{\color[rgb]{1,0,1}}%
      \expandafter\def\csname LT4\endcsname{\color[rgb]{0,1,1}}%
      \expandafter\def\csname LT5\endcsname{\color[rgb]{1,1,0}}%
      \expandafter\def\csname LT6\endcsname{\color[rgb]{0,0,0}}%
      \expandafter\def\csname LT7\endcsname{\color[rgb]{1,0.3,0}}%
      \expandafter\def\csname LT8\endcsname{\color[rgb]{0.5,0.5,0.5}}%
    \else
      % gray
      \def\colorrgb#1{\color{black}}%
      \def\colorgray#1{\color[gray]{#1}}%
      \expandafter\def\csname LTw\endcsname{\color{white}}%
      \expandafter\def\csname LTb\endcsname{\color{black}}%
      \expandafter\def\csname LTa\endcsname{\color{black}}%
      \expandafter\def\csname LT0\endcsname{\color{black}}%
      \expandafter\def\csname LT1\endcsname{\color{black}}%
      \expandafter\def\csname LT2\endcsname{\color{black}}%
      \expandafter\def\csname LT3\endcsname{\color{black}}%
      \expandafter\def\csname LT4\endcsname{\color{black}}%
      \expandafter\def\csname LT5\endcsname{\color{black}}%
      \expandafter\def\csname LT6\endcsname{\color{black}}%
      \expandafter\def\csname LT7\endcsname{\color{black}}%
      \expandafter\def\csname LT8\endcsname{\color{black}}%
    \fi
  \fi
    \setlength{\unitlength}{0.0500bp}%
    \ifx\gptboxheight\undefined%
      \newlength{\gptboxheight}%
      \newlength{\gptboxwidth}%
      \newsavebox{\gptboxtext}%
    \fi%
    \setlength{\fboxrule}{0.5pt}%
    \setlength{\fboxsep}{1pt}%
    \definecolor{tbcol}{rgb}{1,1,1}%
\begin{picture}(4176.00,3456.00)%
    \gplgaddtomacro\gplbacktext{%
      \csname LTb\endcsname%%
      \put(118,714){\makebox(0,0)[r]{\strut{}\normalsize 2}}%
      \put(118,2049){\makebox(0,0)[r]{\strut{}\normalsize 6}}%
      \put(118,3385){\makebox(0,0)[r]{\strut{}\normalsize 10}}%
      \put(250,160){\makebox(0,0){\strut{}\normalsize 3}}%
      \put(697,160){\makebox(0,0){\strut{}\normalsize 5}}%
      \put(1145,160){\makebox(0,0){\strut{}\normalsize 7}}%
      \put(1592,160){\makebox(0,0){\strut{}\normalsize 9}}%
      \put(454,3175){\makebox(0,0)[l]{\strut{}\normalsize \textbf{a}}}%
    }%
    \gplgaddtomacro\gplfronttext{%
      \csname LTb\endcsname%%
      \put(-355,1882){\rotatebox{-270}{\makebox(0,0){\strut{}\normalsize $U/g\ \left(a_{h}^{3} \right)$}}}%
      \put(1033,-60){\makebox(0,0){\strut{}\normalsize $r_{\mathrm{max}}~(a_{h})$}}%
    }%
    \gplgaddtomacro\gplbacktext{%
      \csname LTb\endcsname%%
      \put(2436,630){\makebox(0,0)[r]{\strut{}\normalsize 0.5}}%
      \put(2436,1882){\makebox(0,0)[r]{\strut{}\normalsize 1}}%
      \put(2436,3135){\makebox(0,0)[r]{\strut{}\normalsize 1.5}}%
      \put(2568,160){\makebox(0,0){\strut{}\normalsize 3}}%
      \put(3090,160){\makebox(0,0){\strut{}\normalsize 7}}%
      \put(3611,160){\makebox(0,0){\strut{}\normalsize 11}}%
      \put(4133,160){\makebox(0,0){\strut{}\normalsize 15}}%
      \put(2771,3175){\makebox(0,0)[l]{\strut{}\normalsize \textbf{b}}}%
    }%
    \gplgaddtomacro\gplfronttext{%
      \csname LTb\endcsname%%
      \put(2081,1882){\rotatebox{-270}{\makebox(0,0){\strut{}$U\ (\hbar \omega)$}}}%
      \put(3350,-60){\makebox(0,0){\strut{}\normalsize $r_{\mathrm{max}}~(a_{h})$}}%
      \put(3622,3054){\makebox(0,0){\strut{}$V_{0}~(E_{r})$}}%
      \csname LTb\endcsname%%
      \put(3406,2816){\makebox(0,0)[r]{\strut{}5}}%
      \csname LTb\endcsname%%
      \put(3406,2640){\makebox(0,0)[r]{\strut{}10}}%
      \csname LTb\endcsname%%
      \put(3406,2464){\makebox(0,0)[r]{\strut{}20}}%
      \csname LTb\endcsname%%
      \put(3406,2288){\makebox(0,0)[r]{\strut{}30}}%
    }%
    \gplbacktext
    \put(0,0){\includegraphics[width={208.80bp},height={172.80bp}]{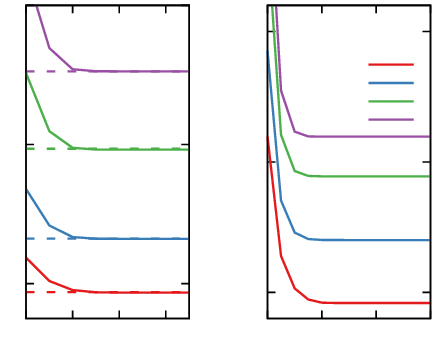}}%
    \gplfronttext
  \end{picture}%
\endgroup

%% file: u_r6_3d_supplement.tex
% GNUPLOT: LaTeX picture with Postscript
\begingroup
  \makeatletter
  \providecommand\color[2][]{%
    \GenericError{(gnuplot) \space\space\space\@spaces}{%
      Package color not loaded in conjunction with
      terminal option `colourtext'%
    }{See the gnuplot documentation for explanation.%
    }{Either use 'blacktext' in gnuplot or load the package
      color.sty in LaTeX.}%
    \renewcommand\color[2][]{}%
  }%
  \providecommand\includegraphics[2][]{%
    \GenericError{(gnuplot) \space\space\space\@spaces}{%
      Package graphicx or graphics not loaded%
    }{See the gnuplot documentation for explanation.%
    }{The gnuplot epslatex terminal needs graphicx.sty or graphics.sty.}%
    \renewcommand\includegraphics[2][]{}%
  }%
  \providecommand\rotatebox[2]{#2}%
  \@ifundefined{ifGPcolor}{%
    \newif\ifGPcolor
    \GPcolortrue
  }{}%
  \@ifundefined{ifGPblacktext}{%
    \newif\ifGPblacktext
    \GPblacktextfalse
  }{}%
  % define a \g@addto@macro without @ in the name:
  \let\gplgaddtomacro\g@addto@macro
  % define empty templates for all commands taking text:
  \gdef\gplbacktext{}%
  \gdef\gplfronttext{}%
  \makeatother
  \ifGPblacktext
    % no textcolor at all
    \def\colorrgb#1{}%
    \def\colorgray#1{}%
  \else
    % gray or color?
    \ifGPcolor
      \def\colorrgb#1{\color[rgb]{#1}}%
      \def\colorgray#1{\color[gray]{#1}}%
      \expandafter\def\csname LTw\endcsname{\color{white}}%
      \expandafter\def\csname LTb\endcsname{\color{black}}%
      \expandafter\def\csname LTa\endcsname{\color{black}}%
      \expandafter\def\csname LT0\endcsname{\color[rgb]{1,0,0}}%
      \expandafter\def\csname LT1\endcsname{\color[rgb]{0,1,0}}%
      \expandafter\def\csname LT2\endcsname{\color[rgb]{0,0,1}}%
      \expandafter\def\csname LT3\endcsname{\color[rgb]{1,0,1}}%
      \expandafter\def\csname LT4\endcsname{\color[rgb]{0,1,1}}%
      \expandafter\def\csname LT5\endcsname{\color[rgb]{1,1,0}}%
      \expandafter\def\csname LT6\endcsname{\color[rgb]{0,0,0}}%
      \expandafter\def\csname LT7\endcsname{\color[rgb]{1,0.3,0}}%
      \expandafter\def\csname LT8\endcsname{\color[rgb]{0.5,0.5,0.5}}%
    \else
      % gray
      \def\colorrgb#1{\color{black}}%
      \def\colorgray#1{\color[gray]{#1}}%
      \expandafter\def\csname LTw\endcsname{\color{white}}%
      \expandafter\def\csname LTb\endcsname{\color{black}}%
      \expandafter\def\csname LTa\endcsname{\color{black}}%
      \expandafter\def\csname LT0\endcsname{\color{black}}%
      \expandafter\def\csname LT1\endcsname{\color{black}}%
      \expandafter\def\csname LT2\endcsname{\color{black}}%
      \expandafter\def\csname LT3\endcsname{\color{black}}%
      \expandafter\def\csname LT4\endcsname{\color{black}}%
      \expandafter\def\csname LT5\endcsname{\color{black}}%
      \expandafter\def\csname LT6\endcsname{\color{black}}%
      \expandafter\def\csname LT7\endcsname{\color{black}}%
      \expandafter\def\csname LT8\endcsname{\color{black}}%
    \fi
  \fi
    \setlength{\unitlength}{0.0500bp}%
    \ifx\gptboxheight\undefined%
      \newlength{\gptboxheight}%
      \newlength{\gptboxwidth}%
      \newsavebox{\gptboxtext}%
    \fi%
    \setlength{\fboxrule}{0.5pt}%
    \setlength{\fboxsep}{1pt}%
    \definecolor{tbcol}{rgb}{1,1,1}%
\begin{picture}(4176.00,3456.00)%
    \gplgaddtomacro\gplbacktext{%
      \csname LTb\endcsname%%
      \put(330,1364){\makebox(0,0)[r]{\strut{}\normalsize 1}}%
      \put(330,2069){\makebox(0,0)[r]{\strut{}\normalsize 2}}%
      \put(330,2773){\makebox(0,0)[r]{\strut{}\normalsize 3}}%
      \put(462,440){\makebox(0,0){\strut{}\normalsize 0}}%
      \put(1503,440){\makebox(0,0){\strut{}\normalsize 4}}%
      \put(2544,440){\makebox(0,0){\strut{}\normalsize 8}}%
      \put(3585,440){\makebox(0,0){\strut{}\normalsize 12}}%
    }%
    \gplgaddtomacro\gplfronttext{%
      \csname LTb\endcsname%%
      \put(-143,1892){\rotatebox{-270}{\makebox(0,0){\strut{}\normalsize $U\ (\hbar\omega)$}}}%
      \put(2153,220){\makebox(0,0){\strut{}\normalsize $r_{6}/a_{h}$}}%
      \csname LTb\endcsname%%
      \put(1129,2878){\makebox(0,0)[r]{\strut{}CsAg}}%
      \csname LTb\endcsname%%
      \put(1129,2702){\makebox(0,0)[r]{\strut{}KAg}}%
      \csname LTb\endcsname%%
      \put(1129,2526){\makebox(0,0)[r]{\strut{}NaCs}}%
      \csname LTb\endcsname%%
      \put(1129,2350){\makebox(0,0)[r]{\strut{}NaK}}%
    }%
    \gplbacktext
    \put(0,0){\includegraphics[width={208.80bp},height={172.80bp}]{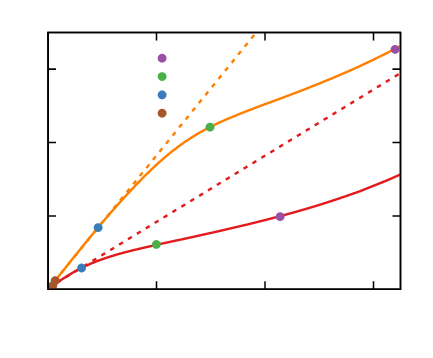}}%
    \gplfronttext
  \end{picture}%
\endgroup

%% file: U_NaCs.tex
% GNUPLOT: LaTeX picture with Postscript
\begingroup
  \makeatletter
  \providecommand\color[2][]{%
    \GenericError{(gnuplot) \space\space\space\@spaces}{%
      Package color not loaded in conjunction with
      terminal option `colourtext'%
    }{See the gnuplot documentation for explanation.%
    }{Either use 'blacktext' in gnuplot or load the package
      color.sty in LaTeX.}%
    \renewcommand\color[2][]{}%
  }%
  \providecommand\includegraphics[2][]{%
    \GenericError{(gnuplot) \space\space\space\@spaces}{%
      Package graphicx or graphics not loaded%
    }{See the gnuplot documentation for explanation.%
    }{The gnuplot epslatex terminal needs graphicx.sty or graphics.sty.}%
    \renewcommand\includegraphics[2][]{}%
  }%
  \providecommand\rotatebox[2]{#2}%
  \@ifundefined{ifGPcolor}{%
    \newif\ifGPcolor
    \GPcolortrue
  }{}%
  \@ifundefined{ifGPblacktext}{%
    \newif\ifGPblacktext
    \GPblacktextfalse
  }{}%
  % define a \g@addto@macro without @ in the name:
  \let\gplgaddtomacro\g@addto@macro
  % define empty templates for all commands taking text:
  \gdef\gplbacktext{}%
  \gdef\gplfronttext{}%
  \makeatother
  \ifGPblacktext
    % no textcolor at all
    \def\colorrgb#1{}%
    \def\colorgray#1{}%
  \else
    % gray or color?
    \ifGPcolor
      \def\colorrgb#1{\color[rgb]{#1}}%
      \def\colorgray#1{\color[gray]{#1}}%
      \expandafter\def\csname LTw\endcsname{\color{white}}%
      \expandafter\def\csname LTb\endcsname{\color{black}}%
      \expandafter\def\csname LTa\endcsname{\color{black}}%
      \expandafter\def\csname LT0\endcsname{\color[rgb]{1,0,0}}%
      \expandafter\def\csname LT1\endcsname{\color[rgb]{0,1,0}}%
      \expandafter\def\csname LT2\endcsname{\color[rgb]{0,0,1}}%
      \expandafter\def\csname LT3\endcsname{\color[rgb]{1,0,1}}%
      \expandafter\def\csname LT4\endcsname{\color[rgb]{0,1,1}}%
      \expandafter\def\csname LT5\endcsname{\color[rgb]{1,1,0}}%
      \expandafter\def\csname LT6\endcsname{\color[rgb]{0,0,0}}%
      \expandafter\def\csname LT7\endcsname{\color[rgb]{1,0.3,0}}%
      \expandafter\def\csname LT8\endcsname{\color[rgb]{0.5,0.5,0.5}}%
    \else
      % gray
      \def\colorrgb#1{\color{black}}%
      \def\colorgray#1{\color[gray]{#1}}%
      \expandafter\def\csname LTw\endcsname{\color{white}}%
      \expandafter\def\csname LTb\endcsname{\color{black}}%
      \expandafter\def\csname LTa\endcsname{\color{black}}%
      \expandafter\def\csname LT0\endcsname{\color{black}}%
      \expandafter\def\csname LT1\endcsname{\color{black}}%
      \expandafter\def\csname LT2\endcsname{\color{black}}%
      \expandafter\def\csname LT3\endcsname{\color{black}}%
      \expandafter\def\csname LT4\endcsname{\color{black}}%
      \expandafter\def\csname LT5\endcsname{\color{black}}%
      \expandafter\def\csname LT6\endcsname{\color{black}}%
      \expandafter\def\csname LT7\endcsname{\color{black}}%
      \expandafter\def\csname LT8\endcsname{\color{black}}%
    \fi
  \fi
    \setlength{\unitlength}{0.0500bp}%
    \ifx\gptboxheight\undefined%
      \newlength{\gptboxheight}%
      \newlength{\gptboxwidth}%
      \newsavebox{\gptboxtext}%
    \fi%
    \setlength{\fboxrule}{0.5pt}%
    \setlength{\fboxsep}{1pt}%
    \definecolor{tbcol}{rgb}{1,1,1}%
\begin{picture}(4176.00,3456.00)%
    \gplgaddtomacro\gplbacktext{%
      \csname LTb\endcsname%%
      \put(252,1284){\makebox(0,0)[r]{\strut{}\normalsize 2}}%
      \put(252,1938){\makebox(0,0)[r]{\strut{}\normalsize 4}}%
      \put(252,2591){\makebox(0,0)[r]{\strut{}\normalsize 6}}%
      \put(252,3245){\makebox(0,0)[r]{\strut{}\normalsize 8}}%
      \put(336,490){\makebox(0,0){\strut{}\normalsize 0}}%
      \put(1344,490){\makebox(0,0){\strut{}\normalsize 5}}%
      \put(2352,490){\makebox(0,0){\strut{}\normalsize 10}}%
      \put(3360,490){\makebox(0,0){\strut{}\normalsize 15}}%
      \put(4049,630){\makebox(0,0)[l]{\strut{}0}}%
      \put(4049,1693){\makebox(0,0)[l]{\strut{}4}}%
      \put(4049,2727){\makebox(0,0)[l]{\strut{}8}}%
    }%
    \gplgaddtomacro\gplfronttext{%
      \csname LTb\endcsname%%
      \put(-301,1937){\rotatebox{-270}{\makebox(0,0){\strut{}\normalsize $U\ (E_{r})$}}}%
      \put(4371,1937){\rotatebox{-270}{\makebox(0,0){\strut{}\normalsize $U/h\ (\mathrm{kHz})$}}}%
      \put(2150,140){\makebox(0,0){\strut{}\normalsize $V_{0}\ (E_{r})$}}%
    }%
    \gplbacktext
    \put(0,0){\includegraphics[width={208.80bp},height={172.80bp}]{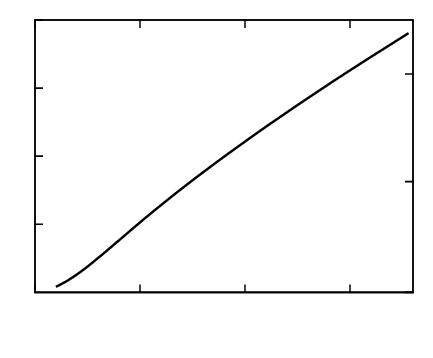}}%
    \gplfronttext
  \end{picture}%
\endgroup

%% file: u_t_3D.tex
% GNUPLOT: LaTeX picture with Postscript
\begingroup
  \makeatletter
  \providecommand\color[2][]{%
    \GenericError{(gnuplot) \space\space\space\@spaces}{%
      Package color not loaded in conjunction with
      terminal option `colourtext'%
    }{See the gnuplot documentation for explanation.%
    }{Either use 'blacktext' in gnuplot or load the package
      color.sty in LaTeX.}%
    \renewcommand\color[2][]{}%
  }%
  \providecommand\includegraphics[2][]{%
    \GenericError{(gnuplot) \space\space\space\@spaces}{%
      Package graphicx or graphics not loaded%
    }{See the gnuplot documentation for explanation.%
    }{The gnuplot epslatex terminal needs graphicx.sty or graphics.sty.}%
    \renewcommand\includegraphics[2][]{}%
  }%
  \providecommand\rotatebox[2]{#2}%
  \@ifundefined{ifGPcolor}{%
    \newif\ifGPcolor
    \GPcolortrue
  }{}%
  \@ifundefined{ifGPblacktext}{%
    \newif\ifGPblacktext
    \GPblacktextfalse
  }{}%
  % define a \g@addto@macro without @ in the name:
  \let\gplgaddtomacro\g@addto@macro
  % define empty templates for all commands taking text:
  \gdef\gplbacktext{}%
  \gdef\gplfronttext{}%
  \makeatother
  \ifGPblacktext
    % no textcolor at all
    \def\colorrgb#1{}%
    \def\colorgray#1{}%
  \else
    % gray or color?
    \ifGPcolor
      \def\colorrgb#1{\color[rgb]{#1}}%
      \def\colorgray#1{\color[gray]{#1}}%
      \expandafter\def\csname LTw\endcsname{\color{white}}%
      \expandafter\def\csname LTb\endcsname{\color{black}}%
      \expandafter\def\csname LTa\endcsname{\color{black}}%
      \expandafter\def\csname LT0\endcsname{\color[rgb]{1,0,0}}%
      \expandafter\def\csname LT1\endcsname{\color[rgb]{0,1,0}}%
      \expandafter\def\csname LT2\endcsname{\color[rgb]{0,0,1}}%
      \expandafter\def\csname LT3\endcsname{\color[rgb]{1,0,1}}%
      \expandafter\def\csname LT4\endcsname{\color[rgb]{0,1,1}}%
      \expandafter\def\csname LT5\endcsname{\color[rgb]{1,1,0}}%
      \expandafter\def\csname LT6\endcsname{\color[rgb]{0,0,0}}%
      \expandafter\def\csname LT7\endcsname{\color[rgb]{1,0.3,0}}%
      \expandafter\def\csname LT8\endcsname{\color[rgb]{0.5,0.5,0.5}}%
    \else
      % gray
      \def\colorrgb#1{\color{black}}%
      \def\colorgray#1{\color[gray]{#1}}%
      \expandafter\def\csname LTw\endcsname{\color{white}}%
      \expandafter\def\csname LTb\endcsname{\color{black}}%
      \expandafter\def\csname LTa\endcsname{\color{black}}%
      \expandafter\def\csname LT0\endcsname{\color{black}}%
      \expandafter\def\csname LT1\endcsname{\color{black}}%
      \expandafter\def\csname LT2\endcsname{\color{black}}%
      \expandafter\def\csname LT3\endcsname{\color{black}}%
      \expandafter\def\csname LT4\endcsname{\color{black}}%
      \expandafter\def\csname LT5\endcsname{\color{black}}%
      \expandafter\def\csname LT6\endcsname{\color{black}}%
      \expandafter\def\csname LT7\endcsname{\color{black}}%
      \expandafter\def\csname LT8\endcsname{\color{black}}%
    \fi
  \fi
    \setlength{\unitlength}{0.0500bp}%
    \ifx\gptboxheight\undefined%
      \newlength{\gptboxheight}%
      \newlength{\gptboxwidth}%
      \newsavebox{\gptboxtext}%
    \fi%
    \setlength{\fboxrule}{0.5pt}%
    \setlength{\fboxsep}{1pt}%
    \definecolor{tbcol}{rgb}{1,1,1}%
\begin{picture}(4176.00,3456.00)%
    \gplgaddtomacro\gplbacktext{%
      \csname LTb\endcsname%%
      \put(330,660){\makebox(0,0)[r]{\strut{}\normalsize 0.1}}%
      \put(330,1175){\makebox(0,0)[r]{\strut{}\normalsize 1}}%
      \put(330,1690){\makebox(0,0)[r]{\strut{}\normalsize 10}}%
      \put(330,2205){\makebox(0,0)[r]{\strut{}\normalsize 100}}%
      \put(330,2720){\makebox(0,0)[r]{\strut{}\normalsize 1000}}%
      \put(330,3235){\makebox(0,0)[r]{\strut{}\normalsize 10000}}%
      \put(462,440){\makebox(0,0){\strut{}\normalsize 1}}%
      \put(1515,440){\makebox(0,0){\strut{}\normalsize 6}}%
      \put(2568,440){\makebox(0,0){\strut{}\normalsize 11}}%
      \put(3622,440){\makebox(0,0){\strut{}\normalsize 16}}%
    }%
    \gplgaddtomacro\gplfronttext{%
      \csname LTb\endcsname%%
      \put(-302,1947){\rotatebox{-270}{\makebox(0,0){\strut{}\normalsize $U/t $}}}%
      \put(2252,110){\makebox(0,0){\strut{}\normalsize $V_{0}\ (E_{r})$}}%
      \csname LTb\endcsname%%
      \put(2089,1515){\makebox(0,0)[r]{\strut{}CsAg}}%
      \csname LTb\endcsname%%
      \put(2089,1339){\makebox(0,0)[r]{\strut{}KAg}}%
      \csname LTb\endcsname%%
      \put(2089,1163){\makebox(0,0)[r]{\strut{}NaCs}}%
      \csname LTb\endcsname%%
      \put(2089,987){\makebox(0,0)[r]{\strut{}LiCu}}%
      \csname LTb\endcsname%%
      \put(2089,811){\makebox(0,0)[r]{\strut{}NaK}}%
    }%
    \gplbacktext
    \put(0,0){\includegraphics[width={208.80bp},height={172.80bp}]{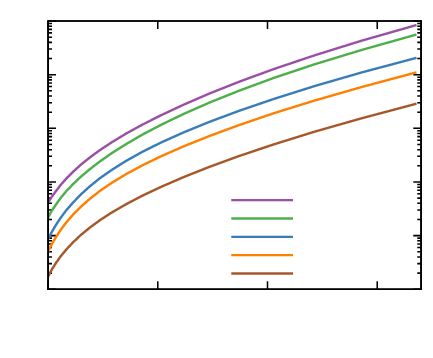}}%
    \gplfronttext
  \end{picture}%
\endgroup

%% file: dip_conv_comparison.tex
% GNUPLOT: LaTeX picture with Postscript
\begingroup
  \makeatletter
  \providecommand\color[2][]{%
    \GenericError{(gnuplot) \space\space\space\@spaces}{%
      Package color not loaded in conjunction with
      terminal option `colourtext'%
    }{See the gnuplot documentation for explanation.%
    }{Either use 'blacktext' in gnuplot or load the package
      color.sty in LaTeX.}%
    \renewcommand\color[2][]{}%
  }%
  \providecommand\includegraphics[2][]{%
    \GenericError{(gnuplot) \space\space\space\@spaces}{%
      Package graphicx or graphics not loaded%
    }{See the gnuplot documentation for explanation.%
    }{The gnuplot epslatex terminal needs graphicx.sty or graphics.sty.}%
    \renewcommand\includegraphics[2][]{}%
  }%
  \providecommand\rotatebox[2]{#2}%
  \@ifundefined{ifGPcolor}{%
    \newif\ifGPcolor
    \GPcolortrue
  }{}%
  \@ifundefined{ifGPblacktext}{%
    \newif\ifGPblacktext
    \GPblacktextfalse
  }{}%
  % define a \g@addto@macro without @ in the name:
  \let\gplgaddtomacro\g@addto@macro
  % define empty templates for all commands taking text:
  \gdef\gplbacktext{}%
  \gdef\gplfronttext{}%
  \makeatother
  \ifGPblacktext
    % no textcolor at all
    \def\colorrgb#1{}%
    \def\colorgray#1{}%
  \else
    % gray or color?
    \ifGPcolor
      \def\colorrgb#1{\color[rgb]{#1}}%
      \def\colorgray#1{\color[gray]{#1}}%
      \expandafter\def\csname LTw\endcsname{\color{white}}%
      \expandafter\def\csname LTb\endcsname{\color{black}}%
      \expandafter\def\csname LTa\endcsname{\color{black}}%
      \expandafter\def\csname LT0\endcsname{\color[rgb]{1,0,0}}%
      \expandafter\def\csname LT1\endcsname{\color[rgb]{0,1,0}}%
      \expandafter\def\csname LT2\endcsname{\color[rgb]{0,0,1}}%
      \expandafter\def\csname LT3\endcsname{\color[rgb]{1,0,1}}%
      \expandafter\def\csname LT4\endcsname{\color[rgb]{0,1,1}}%
      \expandafter\def\csname LT5\endcsname{\color[rgb]{1,1,0}}%
      \expandafter\def\csname LT6\endcsname{\color[rgb]{0,0,0}}%
      \expandafter\def\csname LT7\endcsname{\color[rgb]{1,0.3,0}}%
      \expandafter\def\csname LT8\endcsname{\color[rgb]{0.5,0.5,0.5}}%
    \else
      % gray
      \def\colorrgb#1{\color{black}}%
      \def\colorgray#1{\color[gray]{#1}}%
      \expandafter\def\csname LTw\endcsname{\color{white}}%
      \expandafter\def\csname LTb\endcsname{\color{black}}%
      \expandafter\def\csname LTa\endcsname{\color{black}}%
      \expandafter\def\csname LT0\endcsname{\color{black}}%
      \expandafter\def\csname LT1\endcsname{\color{black}}%
      \expandafter\def\csname LT2\endcsname{\color{black}}%
      \expandafter\def\csname LT3\endcsname{\color{black}}%
      \expandafter\def\csname LT4\endcsname{\color{black}}%
      \expandafter\def\csname LT5\endcsname{\color{black}}%
      \expandafter\def\csname LT6\endcsname{\color{black}}%
      \expandafter\def\csname LT7\endcsname{\color{black}}%
      \expandafter\def\csname LT8\endcsname{\color{black}}%
    \fi
  \fi
    \setlength{\unitlength}{0.0500bp}%
    \ifx\gptboxheight\undefined%
      \newlength{\gptboxheight}%
      \newlength{\gptboxwidth}%
      \newsavebox{\gptboxtext}%
    \fi%
    \setlength{\fboxrule}{0.5pt}%
    \setlength{\fboxsep}{1pt}%
    \definecolor{tbcol}{rgb}{1,1,1}%
\begin{picture}(4176.00,3456.00)%
    \gplgaddtomacro\gplbacktext{%
      \csname LTb\endcsname%%
      \put(202,769){\makebox(0,0)[r]{\strut{}\normalsize -2.8}}%
      \put(202,1523){\makebox(0,0)[r]{\strut{}\normalsize -2.2}}%
      \put(202,2277){\makebox(0,0)[r]{\strut{}\normalsize -1.6}}%
      \put(202,3031){\makebox(0,0)[r]{\strut{}\normalsize -1}}%
      \put(640,298){\makebox(0,0){\strut{}\normalsize 4}}%
      \put(1253,298){\makebox(0,0){\strut{}\normalsize 6}}%
      \put(1865,298){\makebox(0,0){\strut{}\normalsize 8}}%
      \put(554,3089){\makebox(0,0)[l]{\strut{}\normalsize \textbf{a}}}%
    }%
    \gplgaddtomacro\gplfronttext{%
      \csname LTb\endcsname%%
      \put(-403,1900){\rotatebox{-270}{\makebox(0,0){\strut{}\normalsize $U\ (\hbar\omega)$}}}%
      \put(1252,-32){\makebox(0,0){\strut{}\normalsize $r_{\mathrm{max}}\ (a_{h})$}}%
    }%
    \gplgaddtomacro\gplbacktext{%
      \csname LTb\endcsname%%
      \put(2164,769){\makebox(0,0)[r]{\strut{}}}%
      \put(2164,1523){\makebox(0,0)[r]{\strut{}}}%
      \put(2164,2277){\makebox(0,0)[r]{\strut{}}}%
      \put(2164,3031){\makebox(0,0)[r]{\strut{}}}%
      \put(2602,298){\makebox(0,0){\strut{}\normalsize 4}}%
      \put(3215,298){\makebox(0,0){\strut{}\normalsize 6}}%
      \put(3827,298){\makebox(0,0){\strut{}\normalsize 8}}%
      \put(2516,3089){\makebox(0,0)[l]{\strut{}\normalsize \textbf{b}}}%
    }%
    \gplgaddtomacro\gplfronttext{%
      \csname LTb\endcsname%%
      \put(3214,-32){\makebox(0,0){\strut{}\normalsize $r_{\mathrm{max}}\ (a_{h})$}}%
      \put(3566,3061){\makebox(0,0){\strut{}n,N}}%
      \csname LTb\endcsname%%
      \put(3205,2867){\makebox(0,0)[r]{\strut{}2}}%
      \csname LTb\endcsname%%
      \put(3205,2691){\makebox(0,0)[r]{\strut{}4}}%
      \csname LTb\endcsname%%
      \put(3205,2515){\makebox(0,0)[r]{\strut{}5}}%
      \csname LTb\endcsname%%
      \put(3205,2339){\makebox(0,0)[r]{\strut{}6}}%
    }%
    \gplbacktext
    \put(0,0){\includegraphics[width={208.80bp},height={172.80bp}]{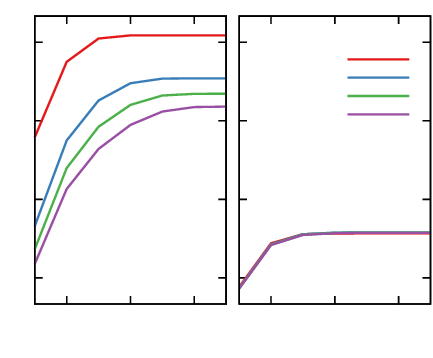}}%
    \gplfronttext
  \end{picture}%
\endgroup

%% file: deff_v10.tex
% GNUPLOT: LaTeX picture with Postscript
\begingroup
  \makeatletter
  \providecommand\color[2][]{%
    \GenericError{(gnuplot) \space\space\space\@spaces}{%
      Package color not loaded in conjunction with
      terminal option `colourtext'%
    }{See the gnuplot documentation for explanation.%
    }{Either use 'blacktext' in gnuplot or load the package
      color.sty in LaTeX.}%
    \renewcommand\color[2][]{}%
  }%
  \providecommand\includegraphics[2][]{%
    \GenericError{(gnuplot) \space\space\space\@spaces}{%
      Package graphicx or graphics not loaded%
    }{See the gnuplot documentation for explanation.%
    }{The gnuplot epslatex terminal needs graphicx.sty or graphics.sty.}%
    \renewcommand\includegraphics[2][]{}%
  }%
  \providecommand\rotatebox[2]{#2}%
  \@ifundefined{ifGPcolor}{%
    \newif\ifGPcolor
    \GPcolortrue
  }{}%
  \@ifundefined{ifGPblacktext}{%
    \newif\ifGPblacktext
    \GPblacktextfalse
  }{}%
  % define a \g@addto@macro without @ in the name:
  \let\gplgaddtomacro\g@addto@macro
  % define empty templates for all commands taking text:
  \gdef\gplbacktext{}%
  \gdef\gplfronttext{}%
  \makeatother
  \ifGPblacktext
    % no textcolor at all
    \def\colorrgb#1{}%
    \def\colorgray#1{}%
  \else
    % gray or color?
    \ifGPcolor
      \def\colorrgb#1{\color[rgb]{#1}}%
      \def\colorgray#1{\color[gray]{#1}}%
      \expandafter\def\csname LTw\endcsname{\color{white}}%
      \expandafter\def\csname LTb\endcsname{\color{black}}%
      \expandafter\def\csname LTa\endcsname{\color{black}}%
      \expandafter\def\csname LT0\endcsname{\color[rgb]{1,0,0}}%
      \expandafter\def\csname LT1\endcsname{\color[rgb]{0,1,0}}%
      \expandafter\def\csname LT2\endcsname{\color[rgb]{0,0,1}}%
      \expandafter\def\csname LT3\endcsname{\color[rgb]{1,0,1}}%
      \expandafter\def\csname LT4\endcsname{\color[rgb]{0,1,1}}%
      \expandafter\def\csname LT5\endcsname{\color[rgb]{1,1,0}}%
      \expandafter\def\csname LT6\endcsname{\color[rgb]{0,0,0}}%
      \expandafter\def\csname LT7\endcsname{\color[rgb]{1,0.3,0}}%
      \expandafter\def\csname LT8\endcsname{\color[rgb]{0.5,0.5,0.5}}%
    \else
      % gray
      \def\colorrgb#1{\color{black}}%
      \def\colorgray#1{\color[gray]{#1}}%
      \expandafter\def\csname LTw\endcsname{\color{white}}%
      \expandafter\def\csname LTb\endcsname{\color{black}}%
      \expandafter\def\csname LTa\endcsname{\color{black}}%
      \expandafter\def\csname LT0\endcsname{\color{black}}%
      \expandafter\def\csname LT1\endcsname{\color{black}}%
      \expandafter\def\csname LT2\endcsname{\color{black}}%
      \expandafter\def\csname LT3\endcsname{\color{black}}%
      \expandafter\def\csname LT4\endcsname{\color{black}}%
      \expandafter\def\csname LT5\endcsname{\color{black}}%
      \expandafter\def\csname LT6\endcsname{\color{black}}%
      \expandafter\def\csname LT7\endcsname{\color{black}}%
      \expandafter\def\csname LT8\endcsname{\color{black}}%
    \fi
  \fi
    \setlength{\unitlength}{0.0500bp}%
    \ifx\gptboxheight\undefined%
      \newlength{\gptboxheight}%
      \newlength{\gptboxwidth}%
      \newsavebox{\gptboxtext}%
    \fi%
    \setlength{\fboxrule}{0.5pt}%
    \setlength{\fboxsep}{1pt}%
    \definecolor{tbcol}{rgb}{1,1,1}%
\begin{picture}(4176.00,3456.00)%
    \gplgaddtomacro\gplbacktext{%
      \csname LTb\endcsname%%
      \put(330,708){\makebox(0,0)[r]{\strut{}\normalsize -0.5}}%
      \put(330,1498){\makebox(0,0)[r]{\strut{}\normalsize 0}}%
      \put(330,2287){\makebox(0,0)[r]{\strut{}\normalsize 0.5}}%
      \put(330,3077){\makebox(0,0)[r]{\strut{}\normalsize 1}}%
      \put(1013,330){\makebox(0,0){\strut{}\normalsize -0.5}}%
      \put(2390,330){\makebox(0,0){\strut{}\normalsize 0}}%
      \put(3768,330){\makebox(0,0){\strut{}\normalsize 0.5}}%
    }%
    \gplgaddtomacro\gplfronttext{%
      \csname LTb\endcsname%%
      \put(-143,1892){\rotatebox{-270}{\makebox(0,0){\strut{}\normalsize $U\ (\hbar \omega)$}}}%
      \put(2252,110){\makebox(0,0){\strut{}\normalsize $d_{\mathrm{eff}}^{2} \left( \mathrm{Debye}^{2} \right)$}}%
      \csname LTb\endcsname%%
      \put(2353,1383){\makebox(0,0)[r]{\strut{}Harmonic}}%
      \csname LTb\endcsname%%
      \put(2353,1163){\makebox(0,0)[r]{\strut{}$1\mathrm{D}_{\parallel}$}}%
      \csname LTb\endcsname%%
      \put(2353,943){\makebox(0,0)[r]{\strut{}$1\mathrm{D}_{\perp}$}}%
      \csname LTb\endcsname%%
      \put(2353,723){\makebox(0,0)[r]{\strut{}$3\mathrm{D}$}}%
    }%
    \gplbacktext
    \put(0,0){\includegraphics[width={208.80bp},height={172.80bp}]{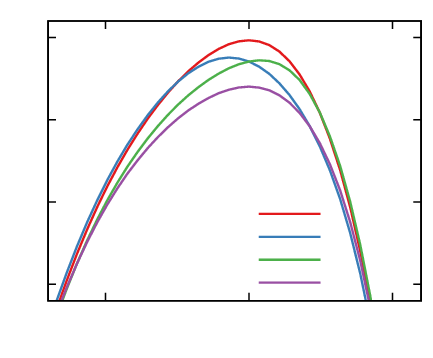}}%
    \gplfronttext
  \end{picture}%
\endgroup

%% file: U_with_harm.tex
% GNUPLOT: LaTeX picture with Postscript
\begingroup
  \makeatletter
  \providecommand\color[2][]{%
    \GenericError{(gnuplot) \space\space\space\@spaces}{%
      Package color not loaded in conjunction with
      terminal option `colourtext'%
    }{See the gnuplot documentation for explanation.%
    }{Either use 'blacktext' in gnuplot or load the package
      color.sty in LaTeX.}%
    \renewcommand\color[2][]{}%
  }%
  \providecommand\includegraphics[2][]{%
    \GenericError{(gnuplot) \space\space\space\@spaces}{%
      Package graphicx or graphics not loaded%
    }{See the gnuplot documentation for explanation.%
    }{The gnuplot epslatex terminal needs graphicx.sty or graphics.sty.}%
    \renewcommand\includegraphics[2][]{}%
  }%
  \providecommand\rotatebox[2]{#2}%
  \@ifundefined{ifGPcolor}{%
    \newif\ifGPcolor
    \GPcolortrue
  }{}%
  \@ifundefined{ifGPblacktext}{%
    \newif\ifGPblacktext
    \GPblacktextfalse
  }{}%
  % define a \g@addto@macro without @ in the name:
  \let\gplgaddtomacro\g@addto@macro
  % define empty templates for all commands taking text:
  \gdef\gplbacktext{}%
  \gdef\gplfronttext{}%
  \makeatother
  \ifGPblacktext
    % no textcolor at all
    \def\colorrgb#1{}%
    \def\colorgray#1{}%
  \else
    % gray or color?
    \ifGPcolor
      \def\colorrgb#1{\color[rgb]{#1}}%
      \def\colorgray#1{\color[gray]{#1}}%
      \expandafter\def\csname LTw\endcsname{\color{white}}%
      \expandafter\def\csname LTb\endcsname{\color{black}}%
      \expandafter\def\csname LTa\endcsname{\color{black}}%
      \expandafter\def\csname LT0\endcsname{\color[rgb]{1,0,0}}%
      \expandafter\def\csname LT1\endcsname{\color[rgb]{0,1,0}}%
      \expandafter\def\csname LT2\endcsname{\color[rgb]{0,0,1}}%
      \expandafter\def\csname LT3\endcsname{\color[rgb]{1,0,1}}%
      \expandafter\def\csname LT4\endcsname{\color[rgb]{0,1,1}}%
      \expandafter\def\csname LT5\endcsname{\color[rgb]{1,1,0}}%
      \expandafter\def\csname LT6\endcsname{\color[rgb]{0,0,0}}%
      \expandafter\def\csname LT7\endcsname{\color[rgb]{1,0.3,0}}%
      \expandafter\def\csname LT8\endcsname{\color[rgb]{0.5,0.5,0.5}}%
    \else
      % gray
      \def\colorrgb#1{\color{black}}%
      \def\colorgray#1{\color[gray]{#1}}%
      \expandafter\def\csname LTw\endcsname{\color{white}}%
      \expandafter\def\csname LTb\endcsname{\color{black}}%
      \expandafter\def\csname LTa\endcsname{\color{black}}%
      \expandafter\def\csname LT0\endcsname{\color{black}}%
      \expandafter\def\csname LT1\endcsname{\color{black}}%
      \expandafter\def\csname LT2\endcsname{\color{black}}%
      \expandafter\def\csname LT3\endcsname{\color{black}}%
      \expandafter\def\csname LT4\endcsname{\color{black}}%
      \expandafter\def\csname LT5\endcsname{\color{black}}%
      \expandafter\def\csname LT6\endcsname{\color{black}}%
      \expandafter\def\csname LT7\endcsname{\color{black}}%
      \expandafter\def\csname LT8\endcsname{\color{black}}%
    \fi
  \fi
    \setlength{\unitlength}{0.0500bp}%
    \ifx\gptboxheight\undefined%
      \newlength{\gptboxheight}%
      \newlength{\gptboxwidth}%
      \newsavebox{\gptboxtext}%
    \fi%
    \setlength{\fboxrule}{0.5pt}%
    \setlength{\fboxsep}{1pt}%
    \definecolor{tbcol}{rgb}{1,1,1}%
\begin{picture}(4176.00,3456.00)%
    \gplgaddtomacro\gplbacktext{%
      \csname LTb\endcsname%%
      \put(330,844){\makebox(0,0)[r]{\strut{}\normalsize -0.3}}%
      \put(330,1396){\makebox(0,0)[r]{\strut{}\normalsize 0}}%
      \put(330,1948){\makebox(0,0)[r]{\strut{}\normalsize 0.3}}%
      \put(330,2499){\makebox(0,0)[r]{\strut{}\normalsize 0.6}}%
      \put(330,3051){\makebox(0,0)[r]{\strut{}\normalsize 0.9}}%
      \put(462,440){\makebox(0,0){\strut{}\normalsize 1}}%
      \put(1305,440){\makebox(0,0){\strut{}\normalsize 5}}%
      \put(2147,440){\makebox(0,0){\strut{}\normalsize 9}}%
      \put(2990,440){\makebox(0,0){\strut{}\normalsize 13}}%
      \put(3832,440){\makebox(0,0){\strut{}\normalsize 17}}%
    }%
    \gplgaddtomacro\gplfronttext{%
      \csname LTb\endcsname%%
      \put(-275,1947){\rotatebox{-270}{\makebox(0,0){\strut{}\normalsize $U\ (\hbar \omega)$}}}%
      \put(2252,220){\makebox(0,0){\strut{}\normalsize $V_{0}\ (E_{r})$}}%
      \put(3439,2591){\makebox(0,0){\strut{}\normalsize $d_{\mathrm{eff}}^{2} \left( \mathrm{Debye^{2}} \right)$}}%
      \csname LTb\endcsname%%
      \put(2779,2393){\makebox(0,0)[r]{\strut{}-0.45}}%
      \csname LTb\endcsname%%
      \put(2779,2195){\makebox(0,0)[r]{\strut{}0}}%
      \csname LTb\endcsname%%
      \put(2779,1997){\makebox(0,0)[r]{\strut{}0.35}}%
    }%
    \gplbacktext
    \put(0,0){\includegraphics[width={208.80bp},height={172.80bp}]{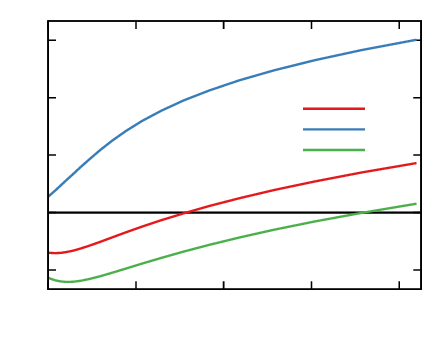}}%
    \gplfronttext
  \end{picture}%
\endgroup

%% file: Vnn_together.tex
% GNUPLOT: LaTeX picture with Postscript
\begingroup
  \makeatletter
  \providecommand\color[2][]{%
    \GenericError{(gnuplot) \space\space\space\@spaces}{%
      Package color not loaded in conjunction with
      terminal option `colourtext'%
    }{See the gnuplot documentation for explanation.%
    }{Either use 'blacktext' in gnuplot or load the package
      color.sty in LaTeX.}%
    \renewcommand\color[2][]{}%
  }%
  \providecommand\includegraphics[2][]{%
    \GenericError{(gnuplot) \space\space\space\@spaces}{%
      Package graphicx or graphics not loaded%
    }{See the gnuplot documentation for explanation.%
    }{The gnuplot epslatex terminal needs graphicx.sty or graphics.sty.}%
    \renewcommand\includegraphics[2][]{}%
  }%
  \providecommand\rotatebox[2]{#2}%
  \@ifundefined{ifGPcolor}{%
    \newif\ifGPcolor
    \GPcolortrue
  }{}%
  \@ifundefined{ifGPblacktext}{%
    \newif\ifGPblacktext
    \GPblacktextfalse
  }{}%
  % define a \g@addto@macro without @ in the name:
  \let\gplgaddtomacro\g@addto@macro
  % define empty templates for all commands taking text:
  \gdef\gplbacktext{}%
  \gdef\gplfronttext{}%
  \makeatother
  \ifGPblacktext
    % no textcolor at all
    \def\colorrgb#1{}%
    \def\colorgray#1{}%
  \else
    % gray or color?
    \ifGPcolor
      \def\colorrgb#1{\color[rgb]{#1}}%
      \def\colorgray#1{\color[gray]{#1}}%
      \expandafter\def\csname LTw\endcsname{\color{white}}%
      \expandafter\def\csname LTb\endcsname{\color{black}}%
      \expandafter\def\csname LTa\endcsname{\color{black}}%
      \expandafter\def\csname LT0\endcsname{\color[rgb]{1,0,0}}%
      \expandafter\def\csname LT1\endcsname{\color[rgb]{0,1,0}}%
      \expandafter\def\csname LT2\endcsname{\color[rgb]{0,0,1}}%
      \expandafter\def\csname LT3\endcsname{\color[rgb]{1,0,1}}%
      \expandafter\def\csname LT4\endcsname{\color[rgb]{0,1,1}}%
      \expandafter\def\csname LT5\endcsname{\color[rgb]{1,1,0}}%
      \expandafter\def\csname LT6\endcsname{\color[rgb]{0,0,0}}%
      \expandafter\def\csname LT7\endcsname{\color[rgb]{1,0.3,0}}%
      \expandafter\def\csname LT8\endcsname{\color[rgb]{0.5,0.5,0.5}}%
    \else
      % gray
      \def\colorrgb#1{\color{black}}%
      \def\colorgray#1{\color[gray]{#1}}%
      \expandafter\def\csname LTw\endcsname{\color{white}}%
      \expandafter\def\csname LTb\endcsname{\color{black}}%
      \expandafter\def\csname LTa\endcsname{\color{black}}%
      \expandafter\def\csname LT0\endcsname{\color{black}}%
      \expandafter\def\csname LT1\endcsname{\color{black}}%
      \expandafter\def\csname LT2\endcsname{\color{black}}%
      \expandafter\def\csname LT3\endcsname{\color{black}}%
      \expandafter\def\csname LT4\endcsname{\color{black}}%
      \expandafter\def\csname LT5\endcsname{\color{black}}%
      \expandafter\def\csname LT6\endcsname{\color{black}}%
      \expandafter\def\csname LT7\endcsname{\color{black}}%
      \expandafter\def\csname LT8\endcsname{\color{black}}%
    \fi
  \fi
    \setlength{\unitlength}{0.0500bp}%
    \ifx\gptboxheight\undefined%
      \newlength{\gptboxheight}%
      \newlength{\gptboxwidth}%
      \newsavebox{\gptboxtext}%
    \fi%
    \setlength{\fboxrule}{0.5pt}%
    \setlength{\fboxsep}{1pt}%
    \definecolor{tbcol}{rgb}{1,1,1}%
\begin{picture}(4176.00,3456.00)%
    \gplgaddtomacro\gplbacktext{%
      \csname LTb\endcsname%%
      \put(118,462){\makebox(0,0)[r]{\strut{}\normalsize -1.2}}%
      \put(118,1163){\makebox(0,0)[r]{\strut{}\normalsize -0.6}}%
      \put(118,1865){\makebox(0,0)[r]{\strut{}\normalsize 0}}%
      \put(118,2567){\makebox(0,0)[r]{\strut{}\normalsize 0.6}}%
      \put(118,3268){\makebox(0,0)[r]{\strut{}\normalsize 1.2}}%
      \put(250,125){\makebox(0,0){\strut{}\normalsize 1}}%
      \put(1002,125){\makebox(0,0){\strut{}\normalsize 3}}%
      \put(1753,125){\makebox(0,0){\strut{}\normalsize 5}}%
      \put(475,3172){\makebox(0,0)[l]{\strut{}\normalsize \textbf{a}}}%
    }%
    \gplgaddtomacro\gplfronttext{%
      \csname LTb\endcsname%%
      \put(-355,1865){\rotatebox{-270}{\makebox(0,0){\strut{}\normalsize $V_{\mathrm{NN}}\ (E_{r})$}}}%
      \put(1189,-95){\makebox(0,0){\strut{}\normalsize $V_{0} (E_{r})$}}%
      \put(1271,2169){\makebox(0,0){\strut{}$d^2_{\mathrm{eff}}(\mathrm{Debye^2})$}}%
      \csname LTb\endcsname%%
      \put(1073,1878){\makebox(0,0)[r]{\strut{}-0.45}}%
      \csname LTb\endcsname%%
      \put(1073,1614){\makebox(0,0)[r]{\strut{}0.35}}%
    }%
    \gplgaddtomacro\gplbacktext{%
      \csname LTb\endcsname%%
      \put(2123,462){\makebox(0,0)[r]{\strut{}}}%
      \put(2123,1163){\makebox(0,0)[r]{\strut{}}}%
      \put(2123,1865){\makebox(0,0)[r]{\strut{}}}%
      \put(2123,2567){\makebox(0,0)[r]{\strut{}}}%
      \put(2123,3268){\makebox(0,0)[r]{\strut{}}}%
      \put(2544,125){\makebox(0,0){\strut{}\normalsize -0.5}}%
      \put(3266,125){\makebox(0,0){\strut{}\normalsize 0}}%
      \put(3989,125){\makebox(0,0){\strut{}\normalsize 0.5}}%
      \put(2480,3172){\makebox(0,0)[l]{\strut{}\normalsize \textbf{b}}}%
    }%
    \gplgaddtomacro\gplfronttext{%
      \csname LTb\endcsname%%
      \put(3194,-95){\makebox(0,0){\strut{}\normalsize $d_{\mathrm{eff}}^{2}\  \left( \mathrm{Debye}^{2} \right)$}}%
      \put(2948,1561){\makebox(0,0){\strut{}$V_{0}\ (E_{r})$}}%
      \csname LTb\endcsname%%
      \put(2552,1343){\makebox(0,0)[r]{\strut{}3}}%
      \csname LTb\endcsname%%
      \put(2552,1145){\makebox(0,0)[r]{\strut{}6}}%
    }%
    \gplbacktext
    \put(0,0){\includegraphics[width={208.80bp},height={172.80bp}]{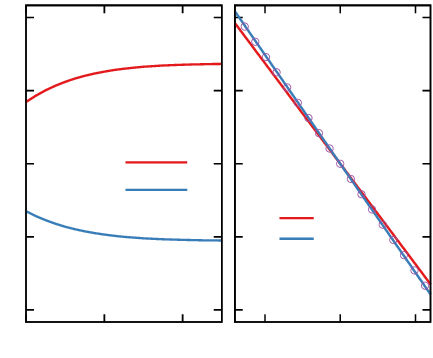}}%
    \gplfronttext
  \end{picture}%
\endgroup

%% file: Vnn_r6.tex
% GNUPLOT: LaTeX picture with Postscript
\begingroup
  \makeatletter
  \providecommand\color[2][]{%
    \GenericError{(gnuplot) \space\space\space\@spaces}{%
      Package color not loaded in conjunction with
      terminal option `colourtext'%
    }{See the gnuplot documentation for explanation.%
    }{Either use 'blacktext' in gnuplot or load the package
      color.sty in LaTeX.}%
    \renewcommand\color[2][]{}%
  }%
  \providecommand\includegraphics[2][]{%
    \GenericError{(gnuplot) \space\space\space\@spaces}{%
      Package graphicx or graphics not loaded%
    }{See the gnuplot documentation for explanation.%
    }{The gnuplot epslatex terminal needs graphicx.sty or graphics.sty.}%
    \renewcommand\includegraphics[2][]{}%
  }%
  \providecommand\rotatebox[2]{#2}%
  \@ifundefined{ifGPcolor}{%
    \newif\ifGPcolor
    \GPcolortrue
  }{}%
  \@ifundefined{ifGPblacktext}{%
    \newif\ifGPblacktext
    \GPblacktextfalse
  }{}%
  % define a \g@addto@macro without @ in the name:
  \let\gplgaddtomacro\g@addto@macro
  % define empty templates for all commands taking text:
  \gdef\gplbacktext{}%
  \gdef\gplfronttext{}%
  \makeatother
  \ifGPblacktext
    % no textcolor at all
    \def\colorrgb#1{}%
    \def\colorgray#1{}%
  \else
    % gray or color?
    \ifGPcolor
      \def\colorrgb#1{\color[rgb]{#1}}%
      \def\colorgray#1{\color[gray]{#1}}%
      \expandafter\def\csname LTw\endcsname{\color{white}}%
      \expandafter\def\csname LTb\endcsname{\color{black}}%
      \expandafter\def\csname LTa\endcsname{\color{black}}%
      \expandafter\def\csname LT0\endcsname{\color[rgb]{1,0,0}}%
      \expandafter\def\csname LT1\endcsname{\color[rgb]{0,1,0}}%
      \expandafter\def\csname LT2\endcsname{\color[rgb]{0,0,1}}%
      \expandafter\def\csname LT3\endcsname{\color[rgb]{1,0,1}}%
      \expandafter\def\csname LT4\endcsname{\color[rgb]{0,1,1}}%
      \expandafter\def\csname LT5\endcsname{\color[rgb]{1,1,0}}%
      \expandafter\def\csname LT6\endcsname{\color[rgb]{0,0,0}}%
      \expandafter\def\csname LT7\endcsname{\color[rgb]{1,0.3,0}}%
      \expandafter\def\csname LT8\endcsname{\color[rgb]{0.5,0.5,0.5}}%
    \else
      % gray
      \def\colorrgb#1{\color{black}}%
      \def\colorgray#1{\color[gray]{#1}}%
      \expandafter\def\csname LTw\endcsname{\color{white}}%
      \expandafter\def\csname LTb\endcsname{\color{black}}%
      \expandafter\def\csname LTa\endcsname{\color{black}}%
      \expandafter\def\csname LT0\endcsname{\color{black}}%
      \expandafter\def\csname LT1\endcsname{\color{black}}%
      \expandafter\def\csname LT2\endcsname{\color{black}}%
      \expandafter\def\csname LT3\endcsname{\color{black}}%
      \expandafter\def\csname LT4\endcsname{\color{black}}%
      \expandafter\def\csname LT5\endcsname{\color{black}}%
      \expandafter\def\csname LT6\endcsname{\color{black}}%
      \expandafter\def\csname LT7\endcsname{\color{black}}%
      \expandafter\def\csname LT8\endcsname{\color{black}}%
    \fi
  \fi
    \setlength{\unitlength}{0.0500bp}%
    \ifx\gptboxheight\undefined%
      \newlength{\gptboxheight}%
      \newlength{\gptboxwidth}%
      \newsavebox{\gptboxtext}%
    \fi%
    \setlength{\fboxrule}{0.5pt}%
    \setlength{\fboxsep}{1pt}%
    \definecolor{tbcol}{rgb}{1,1,1}%
\begin{picture}(4176.00,3456.00)%
    \gplgaddtomacro\gplbacktext{%
      \csname LTb\endcsname%%
      \put(285,1935){\makebox(0,0)[r]{\strut{}\normalsize 0}}%
      \put(285,2660){\makebox(0,0)[r]{\strut{}\normalsize 0.02}}%
      \put(285,3385){\makebox(0,0)[r]{\strut{}\normalsize 0.04}}%
      \put(417,1715){\makebox(0,0){\strut{}}}%
      \put(1532,1715){\makebox(0,0){\strut{}}}%
      \put(2647,1715){\makebox(0,0){\strut{}}}%
      \put(3761,1715){\makebox(0,0){\strut{}}}%
      \put(603,3284){\makebox(0,0)[l]{\strut{}\normalsize \textbf{a}}}%
    }%
    \gplgaddtomacro\gplfronttext{%
      \csname LTb\endcsname%%
      \put(-320,2660){\rotatebox{-270}{\makebox(0,0){\strut{}\normalsize $V_{\mathrm{NN}}\ (E_{r})$}}}%
    }%
    \gplgaddtomacro\gplbacktext{%
      \csname LTb\endcsname%%
      \put(285,345){\makebox(0,0)[r]{\strut{}\normalsize 0}}%
      \put(285,967){\makebox(0,0)[r]{\strut{}\normalsize 0.3}}%
      \put(285,1589){\makebox(0,0)[r]{\strut{}\normalsize 0.6}}%
      \put(417,125){\makebox(0,0){\strut{}\normalsize 0}}%
      \put(1532,125){\makebox(0,0){\strut{}\normalsize 3}}%
      \put(2647,125){\makebox(0,0){\strut{}\normalsize 6}}%
      \put(3761,125){\makebox(0,0){\strut{}\normalsize 9}}%
      \put(603,1694){\makebox(0,0)[l]{\strut{}\normalsize \textbf{b}}}%
    }%
    \gplgaddtomacro\gplfronttext{%
      \csname LTb\endcsname%%
      \put(-320,1070){\rotatebox{-270}{\makebox(0,0){\strut{}\normalsize $V_{\mathrm{NN}}\ (E_{r})$}}}%
      \put(2275,-95){\makebox(0,0){\strut{}\normalsize Cutoff ($E_{r}$)}}%
      \put(3507,1046){\makebox(0,0){\strut{}$V_{0}\ (E_{r})$}}%
      \csname LTb\endcsname%%
      \put(3146,837){\makebox(0,0)[r]{\strut{}3}}%
      \csname LTb\endcsname%%
      \put(3146,617){\makebox(0,0)[r]{\strut{}6}}%
    }%
    \gplbacktext
    \put(0,0){\includegraphics[width={208.80bp},height={172.80bp}]{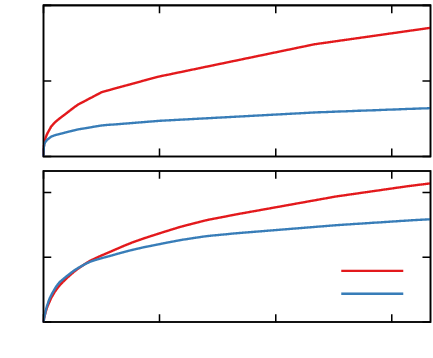}}%
    \gplfronttext
  \end{picture}%
\endgroup

%% file: overlaps_nodipole.tex
% GNUPLOT: LaTeX picture with Postscript
\begingroup
  \makeatletter
  \providecommand\color[2][]{%
    \GenericError{(gnuplot) \space\space\space\@spaces}{%
      Package color not loaded in conjunction with
      terminal option `colourtext'%
    }{See the gnuplot documentation for explanation.%
    }{Either use 'blacktext' in gnuplot or load the package
      color.sty in LaTeX.}%
    \renewcommand\color[2][]{}%
  }%
  \providecommand\includegraphics[2][]{%
    \GenericError{(gnuplot) \space\space\space\@spaces}{%
      Package graphicx or graphics not loaded%
    }{See the gnuplot documentation for explanation.%
    }{The gnuplot epslatex terminal needs graphicx.sty or graphics.sty.}%
    \renewcommand\includegraphics[2][]{}%
  }%
  \providecommand\rotatebox[2]{#2}%
  \@ifundefined{ifGPcolor}{%
    \newif\ifGPcolor
    \GPcolortrue
  }{}%
  \@ifundefined{ifGPblacktext}{%
    \newif\ifGPblacktext
    \GPblacktextfalse
  }{}%
  % define a \g@addto@macro without @ in the name:
  \let\gplgaddtomacro\g@addto@macro
  % define empty templates for all commands taking text:
  \gdef\gplbacktext{}%
  \gdef\gplfronttext{}%
  \makeatother
  \ifGPblacktext
    % no textcolor at all
    \def\colorrgb#1{}%
    \def\colorgray#1{}%
  \else
    % gray or color?
    \ifGPcolor
      \def\colorrgb#1{\color[rgb]{#1}}%
      \def\colorgray#1{\color[gray]{#1}}%
      \expandafter\def\csname LTw\endcsname{\color{white}}%
      \expandafter\def\csname LTb\endcsname{\color{black}}%
      \expandafter\def\csname LTa\endcsname{\color{black}}%
      \expandafter\def\csname LT0\endcsname{\color[rgb]{1,0,0}}%
      \expandafter\def\csname LT1\endcsname{\color[rgb]{0,1,0}}%
      \expandafter\def\csname LT2\endcsname{\color[rgb]{0,0,1}}%
      \expandafter\def\csname LT3\endcsname{\color[rgb]{1,0,1}}%
      \expandafter\def\csname LT4\endcsname{\color[rgb]{0,1,1}}%
      \expandafter\def\csname LT5\endcsname{\color[rgb]{1,1,0}}%
      \expandafter\def\csname LT6\endcsname{\color[rgb]{0,0,0}}%
      \expandafter\def\csname LT7\endcsname{\color[rgb]{1,0.3,0}}%
      \expandafter\def\csname LT8\endcsname{\color[rgb]{0.5,0.5,0.5}}%
    \else
      % gray
      \def\colorrgb#1{\color{black}}%
      \def\colorgray#1{\color[gray]{#1}}%
      \expandafter\def\csname LTw\endcsname{\color{white}}%
      \expandafter\def\csname LTb\endcsname{\color{black}}%
      \expandafter\def\csname LTa\endcsname{\color{black}}%
      \expandafter\def\csname LT0\endcsname{\color{black}}%
      \expandafter\def\csname LT1\endcsname{\color{black}}%
      \expandafter\def\csname LT2\endcsname{\color{black}}%
      \expandafter\def\csname LT3\endcsname{\color{black}}%
      \expandafter\def\csname LT4\endcsname{\color{black}}%
      \expandafter\def\csname LT5\endcsname{\color{black}}%
      \expandafter\def\csname LT6\endcsname{\color{black}}%
      \expandafter\def\csname LT7\endcsname{\color{black}}%
      \expandafter\def\csname LT8\endcsname{\color{black}}%
    \fi
  \fi
    \setlength{\unitlength}{0.0500bp}%
    \ifx\gptboxheight\undefined%
      \newlength{\gptboxheight}%
      \newlength{\gptboxwidth}%
      \newsavebox{\gptboxtext}%
    \fi%
    \setlength{\fboxrule}{0.5pt}%
    \setlength{\fboxsep}{1pt}%
    \definecolor{tbcol}{rgb}{1,1,1}%
\begin{picture}(4176.00,3456.00)%
    \gplgaddtomacro\gplbacktext{%
      \csname LTb\endcsname%%
      \put(330,121){\makebox(0,0)[r]{\strut{}\normalsize 0}}%
      \put(330,1579){\makebox(0,0)[r]{\strut{}\normalsize 0.5}}%
      \put(330,3037){\makebox(0,0)[r]{\strut{}\normalsize 1}}%
      \put(462,-121){\makebox(0,0){\strut{}}}%
      \put(1894,-121){\makebox(0,0){\strut{}}}%
      \put(3327,-121){\makebox(0,0){\strut{}}}%
      \put(534,2607){\makebox(0,0)[l]{\strut{}\normalsize \textbf{a}}}%
    }%
    \gplgaddtomacro\gplfronttext{%
      \csname LTb\endcsname%%
      \put(-275,1557){\rotatebox{-270}{\makebox(0,0){\strut{}\large $\langle \omega_{i}|\Psi_{0} \rangle ^{2}$}}}%
      \put(3621,2140){\makebox(0,0){\strut{}i}}%
      \csname LTb\endcsname%%
      \put(3260,1849){\makebox(0,0)[r]{\strut{}0}}%
      \csname LTb\endcsname%%
      \put(3260,1607){\makebox(0,0)[r]{\strut{}1}}%
      \csname LTb\endcsname%%
      \put(3260,1365){\makebox(0,0)[r]{\strut{}2}}%
    }%
    \gplgaddtomacro\gplbacktext{%
      \csname LTb\endcsname%%
      \put(897,1097){\makebox(0,0)[r]{\strut{}\normalsize 0.01}}%
      \put(897,1710){\makebox(0,0)[r]{\strut{}\normalsize 0.03}}%
      \put(897,2324){\makebox(0,0)[r]{\strut{}\normalsize 0.05}}%
      \put(963,680){\makebox(0,0){\strut{}\normalsize 1}}%
      \put(1727,680){\makebox(0,0){\strut{}\normalsize 3}}%
      \put(2492,680){\makebox(0,0){\strut{}\normalsize 5}}%
    }%
    \gplgaddtomacro\gplfronttext{%
    }%
    \gplbacktext
    \put(0,0){\includegraphics[width={208.80bp},height={172.80bp}]{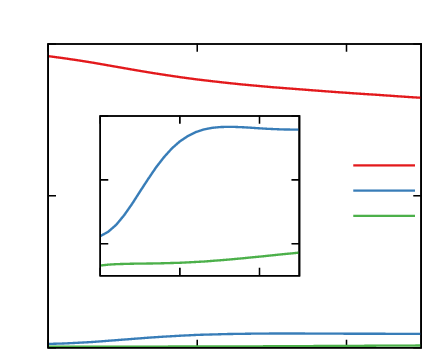}}%
    \gplfronttext
  \end{picture}%
\endgroup

%% file: overlaps_with_dipole.tex
% GNUPLOT: LaTeX picture with Postscript
\begingroup
  \makeatletter
  \providecommand\color[2][]{%
    \GenericError{(gnuplot) \space\space\space\@spaces}{%
      Package color not loaded in conjunction with
      terminal option `colourtext'%
    }{See the gnuplot documentation for explanation.%
    }{Either use 'blacktext' in gnuplot or load the package
      color.sty in LaTeX.}%
    \renewcommand\color[2][]{}%
  }%
  \providecommand\includegraphics[2][]{%
    \GenericError{(gnuplot) \space\space\space\@spaces}{%
      Package graphicx or graphics not loaded%
    }{See the gnuplot documentation for explanation.%
    }{The gnuplot epslatex terminal needs graphicx.sty or graphics.sty.}%
    \renewcommand\includegraphics[2][]{}%
  }%
  \providecommand\rotatebox[2]{#2}%
  \@ifundefined{ifGPcolor}{%
    \newif\ifGPcolor
    \GPcolortrue
  }{}%
  \@ifundefined{ifGPblacktext}{%
    \newif\ifGPblacktext
    \GPblacktextfalse
  }{}%
  % define a \g@addto@macro without @ in the name:
  \let\gplgaddtomacro\g@addto@macro
  % define empty templates for all commands taking text:
  \gdef\gplbacktext{}%
  \gdef\gplfronttext{}%
  \makeatother
  \ifGPblacktext
    % no textcolor at all
    \def\colorrgb#1{}%
    \def\colorgray#1{}%
  \else
    % gray or color?
    \ifGPcolor
      \def\colorrgb#1{\color[rgb]{#1}}%
      \def\colorgray#1{\color[gray]{#1}}%
      \expandafter\def\csname LTw\endcsname{\color{white}}%
      \expandafter\def\csname LTb\endcsname{\color{black}}%
      \expandafter\def\csname LTa\endcsname{\color{black}}%
      \expandafter\def\csname LT0\endcsname{\color[rgb]{1,0,0}}%
      \expandafter\def\csname LT1\endcsname{\color[rgb]{0,1,0}}%
      \expandafter\def\csname LT2\endcsname{\color[rgb]{0,0,1}}%
      \expandafter\def\csname LT3\endcsname{\color[rgb]{1,0,1}}%
      \expandafter\def\csname LT4\endcsname{\color[rgb]{0,1,1}}%
      \expandafter\def\csname LT5\endcsname{\color[rgb]{1,1,0}}%
      \expandafter\def\csname LT6\endcsname{\color[rgb]{0,0,0}}%
      \expandafter\def\csname LT7\endcsname{\color[rgb]{1,0.3,0}}%
      \expandafter\def\csname LT8\endcsname{\color[rgb]{0.5,0.5,0.5}}%
    \else
      % gray
      \def\colorrgb#1{\color{black}}%
      \def\colorgray#1{\color[gray]{#1}}%
      \expandafter\def\csname LTw\endcsname{\color{white}}%
      \expandafter\def\csname LTb\endcsname{\color{black}}%
      \expandafter\def\csname LTa\endcsname{\color{black}}%
      \expandafter\def\csname LT0\endcsname{\color{black}}%
      \expandafter\def\csname LT1\endcsname{\color{black}}%
      \expandafter\def\csname LT2\endcsname{\color{black}}%
      \expandafter\def\csname LT3\endcsname{\color{black}}%
      \expandafter\def\csname LT4\endcsname{\color{black}}%
      \expandafter\def\csname LT5\endcsname{\color{black}}%
      \expandafter\def\csname LT6\endcsname{\color{black}}%
      \expandafter\def\csname LT7\endcsname{\color{black}}%
      \expandafter\def\csname LT8\endcsname{\color{black}}%
    \fi
  \fi
    \setlength{\unitlength}{0.0500bp}%
    \ifx\gptboxheight\undefined%
      \newlength{\gptboxheight}%
      \newlength{\gptboxwidth}%
      \newsavebox{\gptboxtext}%
    \fi%
    \setlength{\fboxrule}{0.5pt}%
    \setlength{\fboxsep}{1pt}%
    \definecolor{tbcol}{rgb}{1,1,1}%
\begin{picture}(4176.00,3456.00)%
    \gplgaddtomacro\gplbacktext{%
      \csname LTb\endcsname%%
      \put(330,660){\makebox(0,0)[r]{\strut{}\normalsize 0}}%
      \put(330,2036){\makebox(0,0)[r]{\strut{}\normalsize 0.5}}%
      \put(330,3411){\makebox(0,0)[r]{\strut{}\normalsize 1}}%
      \put(462,440){\makebox(0,0){\strut{}\normalsize 1}}%
      \put(1894,440){\makebox(0,0){\strut{}\normalsize 3}}%
      \put(3327,440){\makebox(0,0){\strut{}\normalsize 5}}%
      \put(534,3026){\makebox(0,0)[l]{\strut{}\normalsize \textbf{b}}}%
    }%
    \gplgaddtomacro\gplfronttext{%
      \csname LTb\endcsname%%
      \put(-275,2035){\rotatebox{-270}{\makebox(0,0){\strut{}\large $\langle \omega_{i}|\Psi_{0} \rangle ^{2}$}}}%
      \put(2252,110){\makebox(0,0){\strut{}\normalsize $V_{0}(E_{r})$}}%
      \put(3621,2448){\makebox(0,0){\strut{}i}}%
      \csname LTb\endcsname%%
      \put(3260,2206){\makebox(0,0)[r]{\strut{}0}}%
      \csname LTb\endcsname%%
      \put(3260,1964){\makebox(0,0)[r]{\strut{}1}}%
      \csname LTb\endcsname%%
      \put(3260,1722){\makebox(0,0)[r]{\strut{}2}}%
      \csname LTb\endcsname%%
      \put(3260,1480){\makebox(0,0)[r]{\strut{}3}}%
    }%
    \gplgaddtomacro\gplbacktext{%
      \csname LTb\endcsname%%
      \put(938,1531){\makebox(0,0)[r]{\strut{}\normalsize 0.01}}%
      \put(938,2029){\makebox(0,0)[r]{\strut{}\normalsize 0.03}}%
      \put(938,2527){\makebox(0,0)[r]{\strut{}\normalsize 0.05}}%
      \put(1004,1172){\makebox(0,0){\strut{}\normalsize 1}}%
      \put(1768,1172){\makebox(0,0){\strut{}\normalsize 3}}%
      \put(2533,1172){\makebox(0,0){\strut{}\normalsize 5}}%
    }%
    \gplgaddtomacro\gplfronttext{%
    }%
    \gplbacktext
    \put(0,0){\includegraphics[width={208.80bp},height={172.80bp}]{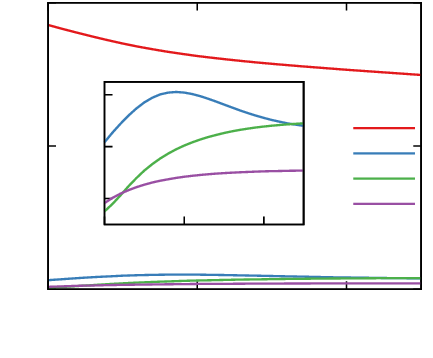}}%
    \gplfronttext
  \end{picture}%
\endgroup

%% file: induce_t.tex
% GNUPLOT: LaTeX picture with Postscript
\begingroup
  \makeatletter
  \providecommand\color[2][]{%
    \GenericError{(gnuplot) \space\space\space\@spaces}{%
      Package color not loaded in conjunction with
      terminal option `colourtext'%
    }{See the gnuplot documentation for explanation.%
    }{Either use 'blacktext' in gnuplot or load the package
      color.sty in LaTeX.}%
    \renewcommand\color[2][]{}%
  }%
  \providecommand\includegraphics[2][]{%
    \GenericError{(gnuplot) \space\space\space\@spaces}{%
      Package graphicx or graphics not loaded%
    }{See the gnuplot documentation for explanation.%
    }{The gnuplot epslatex terminal needs graphicx.sty or graphics.sty.}%
    \renewcommand\includegraphics[2][]{}%
  }%
  \providecommand\rotatebox[2]{#2}%
  \@ifundefined{ifGPcolor}{%
    \newif\ifGPcolor
    \GPcolortrue
  }{}%
  \@ifundefined{ifGPblacktext}{%
    \newif\ifGPblacktext
    \GPblacktextfalse
  }{}%
  % define a \g@addto@macro without @ in the name:
  \let\gplgaddtomacro\g@addto@macro
  % define empty templates for all commands taking text:
  \gdef\gplbacktext{}%
  \gdef\gplfronttext{}%
  \makeatother
  \ifGPblacktext
    % no textcolor at all
    \def\colorrgb#1{}%
    \def\colorgray#1{}%
  \else
    % gray or color?
    \ifGPcolor
      \def\colorrgb#1{\color[rgb]{#1}}%
      \def\colorgray#1{\color[gray]{#1}}%
      \expandafter\def\csname LTw\endcsname{\color{white}}%
      \expandafter\def\csname LTb\endcsname{\color{black}}%
      \expandafter\def\csname LTa\endcsname{\color{black}}%
      \expandafter\def\csname LT0\endcsname{\color[rgb]{1,0,0}}%
      \expandafter\def\csname LT1\endcsname{\color[rgb]{0,1,0}}%
      \expandafter\def\csname LT2\endcsname{\color[rgb]{0,0,1}}%
      \expandafter\def\csname LT3\endcsname{\color[rgb]{1,0,1}}%
      \expandafter\def\csname LT4\endcsname{\color[rgb]{0,1,1}}%
      \expandafter\def\csname LT5\endcsname{\color[rgb]{1,1,0}}%
      \expandafter\def\csname LT6\endcsname{\color[rgb]{0,0,0}}%
      \expandafter\def\csname LT7\endcsname{\color[rgb]{1,0.3,0}}%
      \expandafter\def\csname LT8\endcsname{\color[rgb]{0.5,0.5,0.5}}%
    \else
      % gray
      \def\colorrgb#1{\color{black}}%
      \def\colorgray#1{\color[gray]{#1}}%
      \expandafter\def\csname LTw\endcsname{\color{white}}%
      \expandafter\def\csname LTb\endcsname{\color{black}}%
      \expandafter\def\csname LTa\endcsname{\color{black}}%
      \expandafter\def\csname LT0\endcsname{\color{black}}%
      \expandafter\def\csname LT1\endcsname{\color{black}}%
      \expandafter\def\csname LT2\endcsname{\color{black}}%
      \expandafter\def\csname LT3\endcsname{\color{black}}%
      \expandafter\def\csname LT4\endcsname{\color{black}}%
      \expandafter\def\csname LT5\endcsname{\color{black}}%
      \expandafter\def\csname LT6\endcsname{\color{black}}%
      \expandafter\def\csname LT7\endcsname{\color{black}}%
      \expandafter\def\csname LT8\endcsname{\color{black}}%
    \fi
  \fi
    \setlength{\unitlength}{0.0500bp}%
    \ifx\gptboxheight\undefined%
      \newlength{\gptboxheight}%
      \newlength{\gptboxwidth}%
      \newsavebox{\gptboxtext}%
    \fi%
    \setlength{\fboxrule}{0.5pt}%
    \setlength{\fboxsep}{1pt}%
    \definecolor{tbcol}{rgb}{1,1,1}%
\begin{picture}(4176.00,3456.00)%
    \gplgaddtomacro\gplbacktext{%
      \csname LTb\endcsname%%
      \put(396,550){\makebox(0,0)[r]{\strut{}\normalsize -0.1}}%
      \put(396,1317){\makebox(0,0)[r]{\strut{}\normalsize 0}}%
      \put(396,2084){\makebox(0,0)[r]{\strut{}\normalsize 0.1}}%
      \put(396,2851){\makebox(0,0)[r]{\strut{}\normalsize 0.2}}%
      \put(528,330){\makebox(0,0){\strut{}\normalsize 1}}%
      \put(1934,330){\makebox(0,0){\strut{}\normalsize 3}}%
      \put(3340,330){\makebox(0,0){\strut{}\normalsize 5}}%
    }%
    \gplgaddtomacro\gplfronttext{%
      \csname LTb\endcsname%%
      \put(-209,1892){\rotatebox{-270}{\makebox(0,0){\strut{}\normalsize $\mathrm{Energy}~ (E_{r})$}}}%
      \put(2285,0){\makebox(0,0){\strut{}\normalsize $V_{0}~ (E_{r})$}}%
      \csname LTb\endcsname%%
      \put(3056,3062){\makebox(0,0)[r]{\strut{}$t$}}%
      \csname LTb\endcsname%%
      \put(3056,2842){\makebox(0,0)[r]{\strut{}$J_{\mathrm{contact}}$}}%
      \csname LTb\endcsname%%
      \put(3056,2622){\makebox(0,0)[r]{\strut{}$J$}}%
      \csname LTb\endcsname%%
      \put(3056,2402){\makebox(0,0)[r]{\strut{}$J_{\mathrm{dip}}$}}%
    }%
    \gplbacktext
    \put(0,0){\includegraphics[width={208.80bp},height={172.80bp}]{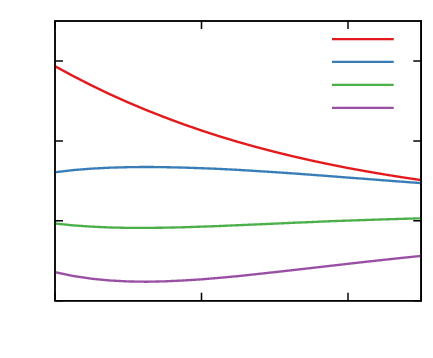}}%
    \gplfronttext
  \end{picture}%
\endgroup